\documentclass[]{aa} 

\usepackage[english]{babel}
\usepackage[T1]{fontenc}
\usepackage{graphicx}
\usepackage{txfonts}
\usepackage{hyperref}
\usepackage{gensymb}
\usepackage{newtxtext,newtxmath}
\usepackage{longtable}
\usepackage{setspace}
\usepackage{booktabs}
\usepackage{wallpaper}
\usepackage{epstopdf}
\usepackage{eso-pic}
\usepackage{eqparbox}
\usepackage{multicol}
\usepackage{dcolumn}

\usepackage[singlelinecheck=off]{caption}
\usepackage{longtable}

\usepackage{ae,aecompl}

\usepackage{amsmath}	
\usepackage{amssymb}	

\def\cratio{$^{12}$C/$^{13}$C}
\def\nratio{$^{14}$N/$^{15}$N}

\newcommand{\asec}{$^{\prime\prime}$}
\newcommand{\GG}[1]{}
\def\fm{HC(O)NH$_{2}$}
\def\ia{HNCO}
\def\am{CH$_{3}$C(O)NH$_{2}$}
\def\mi{CH$_{3}$NCO}
\def\mieq{{\rm CH}_{3}{\rm NCO}}
\def\nm{CH$_{3}$NHCHO}
\def\ei{CH$_{3}$CH$_{2}$NCO}
\def\ur{NH$_{2}$C(O)NH$_{2}$}
\def\gly{HOCH$_{2}$C(O)NH$_{2}$}
\def\cfm{NH$_{2}$C(O)CN}

\newcommand{\mcl}[3]{\multicolumn{#1}{#2}{#3}}
\newcolumntype{.}{D{.}{.}{-1}}

\begin{document}

   \title{The GUAPOS project\\ II. A comprehensive study of peptide-like bond molecules}


  \author{L. Colzi\inst{1, 2}
         \and
         V. M. Rivilla\inst{1, 2}
         \and
          M. T. Beltrán\inst{2}
         \and
          I. Jiménez-Serra\inst{1}
          \and
           C. Mininni\inst{2,3,4}
          \and
           M. Melosso\inst{5} 
           \and
           R. Cesaroni\inst{2}  
           \and
          F. Fontani\inst{2}  
          \and
          A. Lorenzani\inst{2} 
          \and
            A. Sánchez-Monge\inst{6}
            \and
           S. Viti\inst{7}
            \and
        P. Schilke\inst{6}
          \and
          L. Testi\inst{2, 8}
           \and
           E. R. Alonso\inst{9}
            \and
           L. Kolesniková\inst{9,10}  
			 }
          
   \institute{Centro de Astrobiología (CSIC-INTA), Ctra. de Ajalvir Km. 4, 28850, Torrejón de Ardoz, Madrid, Spain \\
              \email{lcolzi@cab.inta-csic.es}
         \and
            INAF-Osservatorio Astrofisico di Arcetri, Largo E. Fermi 5, I-50125, Florence, Italy
             \and 
                Università degli studi di Firenze, Dipartimento di fisica e Astronomia, Via Sansone 1, 50019 Sesto Fiorentino, Italy
                \and     
                INAF-IAPS, via del Fosso del Cavaliere 100, I-00133 Roma, Italy 
                    \and
            Dipartimento di Chimica "Giacomo Ciamician”, Università di Bologna, Via F. Selmi 2, 40126 Bologna, Italy    
              \and
          I. Physikalisches Institut, Universität zu Köln, Zülpicher Str. 77, 50937 Köln, Germany
                      \and
            University of Leiden, Niels Bohrweg 2, 2333 CA, Leiden, The Netherlands
            \and
           European Southern Observatory, Karl-Schwarzschild-Strasse 2, 85748 Garching bei München, Germany
           \and 
	Grupo de Espectroscopia Molecular (GEM), Edificio Quifima, Área de  Química-Física, Laboratorios de Espectroscopia y
Bioespectroscopia, Parque Científico UVa, Unidad Asociada CSIC,  Universidad de Valladolid, 47011 Valladolid, Spain.
		\and
		Department of Analytical Chemistry, University of Chemistry and Technology, Technická 5, 166 28 Prague 6, Czech Republic}

\date{Received 17 June 2021 / Accepted 23 July 2021}


\abstract 
{Peptide-like molecules, which can take part to the formation of proteins in a primitive Earth environment, have been detected up to now only towards a few hot cores and hot corinos.}
{We present a study of \ia, \fm, \mi, \am, \nm, \ei, \ur, \cfm, and \gly\;towards the hot core G31.41+0.31. The aim of this work is to study these species together to allow a consistent study among them.} 
{We have used the spectrum obtained from the ALMA 3mm spectral survey GUAPOS, with a spectral resolution of $\sim$0.488 MHz ($\sim$1.3--1.7 km s$^{-1}$) and an angular resolution of 1\farcs2$\times$1\farcs2 ($\sim$4500 au), to derive column densities of all the molecular species presented in this work, together with 0\farcs2$\times$0\farcs2 ($\sim$750 au) ALMA observations from another project to study the morphology of \ia, \fm\;and \am.}
{We have detected \ia, \fm, \mi, \am, and \nm, but no \ei, \ur, \cfm, and \gly. This is the first time that these molecules have been detected all together outside the Galactic center. We have obtained molecular fractional abundances with respect to H$_{2}$ from 10$^{-7}$ down to a few 10$^{-9}$ and abundances with respect to CH$_{3}$OH from 10$^{-3}$ to $\sim$4$\times$10$^{-2}$, and their emission is found to be compact ($\sim$2\asec, i.e. $\sim$7500 au). From the comparison with other sources, we find that regions in an earlier stage of evolution, such as pre-stellar cores, show abundances at least two orders of magnitude lower than those in hot cores, hot corinos or shocked regions. Moreover, molecular abundance ratios towards different sources are found to be consistent between them within $\sim$1 order of magnitude, regardless of the physical properties (e.g. different masses and luminosities), or the source position throughout the Galaxy. Correlations have also been found between \ia\;and \fm, and \mi\;and \ia\;abundances, and for the first time between \mi\;and \fm, \am\;and \ia, and \am\;and \fm\;abundances.
These results suggest that all these species are formed on grain surfaces in early evolutionary stages of molecular clouds, and that they are subsequently released back to the gas-phase through thermal desorption or shock-triggered desorption.}
  {}
  

   \maketitle
   \titlerunning{Peptide-like molecules}
   \authorrunning{L. Colzi et al.}
%

\section{Introduction}

\begin{figure*}
\centering
\includegraphics[width=37pc]{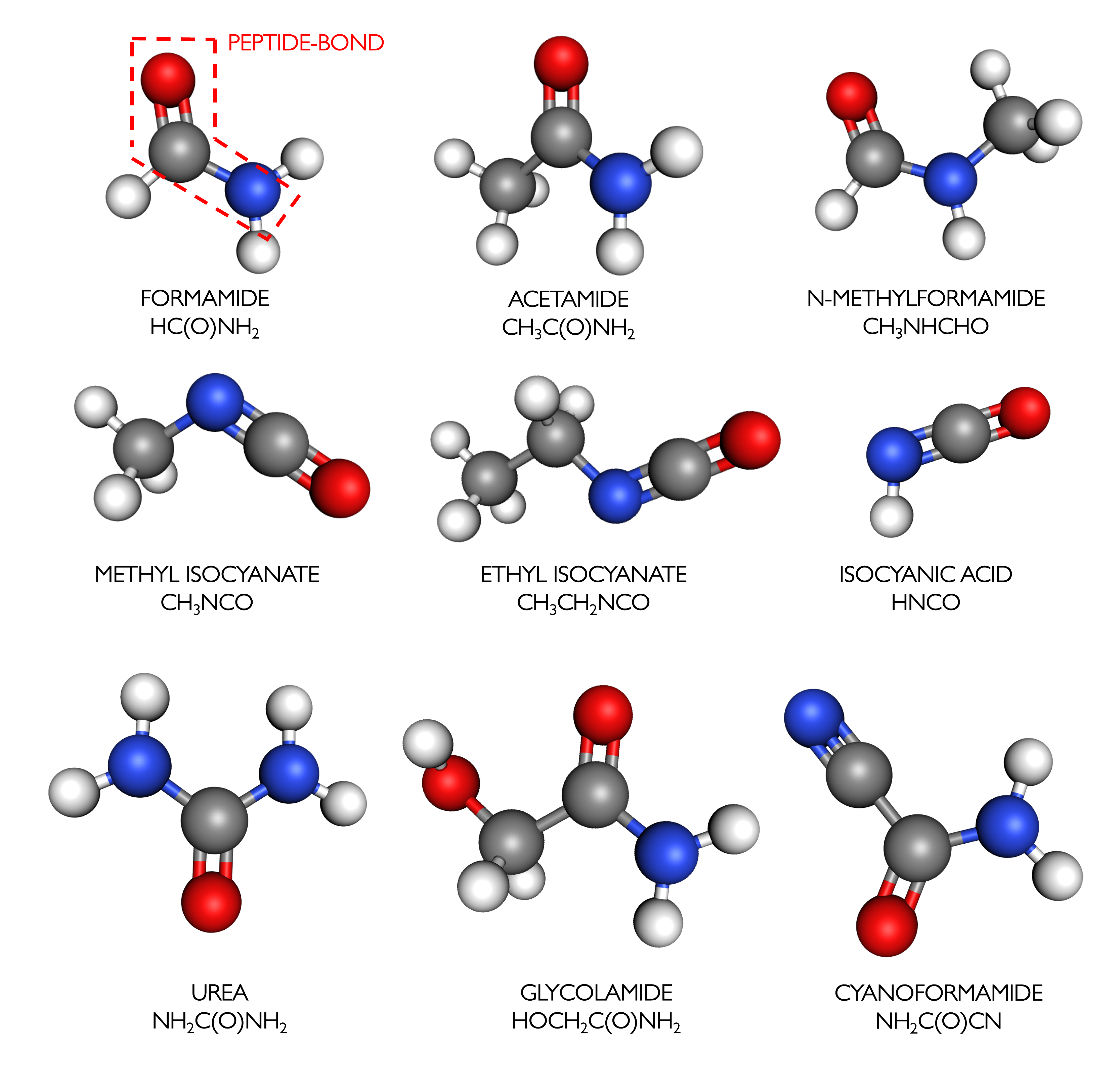}
\caption{Chemical structure of the peptide-like bond molecules studied in this paper. White, grey, red and blue spheres correspond to H, C, O, and N atoms, respectively.}
\label{fig-molecules}
\end{figure*}

The study of the origin of life on Earth is one of the main challenges among biologists, chemists, geologists, and, in recent years, also astrophysicists. In fact, the advent of more sensitive and higher spatial and spectral resolution astronomical instruments, like the Atacama Large Millimeter Array (ALMA), allowed the detection of $\sim$240 molecular species in the interstellar medium (ISM), about 100 of which  are complex organic molecules (COMs), i.e. molecules containing carbon with 6 or more atoms (\citealt{mcguire2018}). In particular, COMs have been detected ubiquitously in the ISM
towards high-mass and low-mass star-forming regions (e.g. \citealt{hollis2004}; \citealt{beltran2009}; \citealt{belloche2013}; \citealt{jorgensen2012}), protostellar molecular outflows (e.g. \citealt{arce2008}; \citealt{codella2020}), photon-dominated regions (e.g. \citealt{guzman2013}; \citealt{cuadrado2017}), dark clouds cores and pre-stellar cores (e.g. \citealt{marcelino2007}; \citealt{bacmann2012}; \citealt{jimenez-serra2016}), and Galactic Center (GC) molecular clouds (e.g. \citealt{requena-torres2006}; \citealt{zeng2018}; \citealt{rivilla2019}; \citealt{jimenez-serra2020}; \citealt{rivilla2021}). 

Among prebiotic COMs, those containing peptide-like bonds (NCO backbone) are of great interest because they can participate in the link of amino acids forming proteins (e.g. \citealt{pascal2005}). Peptide-like bond molecules detected to date in the ISM are isocyanic acid (\ia), formamide (\fm), methyl isocyanate (\mi), acetamide (\am) and its isomer N-methylformamide (\nm), and urea (\ur). 
Unlike these species, ethyl isocyanate (\ei), cyanoformamide (\cfm), and glycolamide (\gly), which also contain the peptide-like bond, have never been detected in the ISM. \citet{kolesnikova2018} report the rotational spectrum of \ei\;from 80 to 340 GHz, and searched for it towards the Orion KL and Sgr B2 hot molecular cores (HMCs) without success. \citet{sanz-novo2020} provided experimental frequencies of the rotational lines in the ground vibrational state of \gly\;and searched for it towards SgrB2(N) also without success. Finally, \cfm\;searches in the ISM have never been reported.
Isocyanic acid, HNCO, was detected towards HMCs, low-mass protostars, translucent molecular clouds, molecular outflows, and extragalactic regions (e.g. \citealt{snyder1972}; \citealt{turner1999}; \citealt{helmich1997}; \citealt{bisschop2007}; \citealt{zeng2018}; \citealt{rodriguez-fernandez2010}; \citealt{nguyen-q-rieu1991}; \citealt{nazari2021}; \citealt{canelo2021}). Formamide, \fm, has been also detected towards many different high- and low-mass star-forming regions (e.g. \citealt{lopez-sepulcre2019}). Methyl isocyanate, \mi\, was detected towards Sgr B2(N), Orion KL, the low-mass protostars IRAS 16293-2422 A and B, the high-mass protostellar object G328.2551-0.5321, the G10.47+0.03 and G31.41+0.31 HMCs, and the Serpens SMM1 hot corino (\citet{halfen2015}; \citealt{cernicharo2016}; \citealt{martin-domenech2017}; \citealt{ligterink2017}; \citet{ligterink2018}; \citet{manigand2020}; \citealt{csengeri2019}; \citealt{gorai2020,gorai2021}; \citealt{ligterink2021}). 
Acetamide, \am, was detected towards different high-mass star-forming regions (\citealt{hollis2006}; \citealt{halfen2011}; \citealt{cernicharo2016}; \citealt{belloche2017}; \citealt{ligterink2020}) and tentatively detected towards  IRAS 16293-2422 B (\citealt{ligterink2018}). 
Moreover, the second most stable C$_{2}$H$_{5}$NO isomer after \am, N-methylformamide, \nm\;(\citealt{lattelais2010}), was detected towards Sgr B2(N1S) and NGC 6334I and tentatively detected towards Sgr B2(N2) (\citealt{belloche2017}; \citealt{belloche2019}, \citealt{ligterink2020}). Recently, urea, NH$_{2}$C(O)NH$_{2}$, has been detected towards Sgr B2(N1) and the GC molecular cloud G+0.693-0.027 (\citealt{belloche2019}, \citealt{jimenez-serra2020}), and tentatively detected towards NGC 7538 IRS9, Sgr B2(N-LMH), and NGC 6334I (\citealt{raunier2004}; \citealt{remijan2014}; \citealt{ligterink2020}).

These peptide-like bond species have preferentially been detected so far in massive and clustered star-forming regions. In this sense it is worth noting that our Sun is thought to have been formed in a clustered environment in the presence of massive stars (e.g \citealt{adams2010}; \citealt{lichtenberg2019}; \citealt{wallner2020}; \citealt{korschinek2020}). Therefore, the study of the chemical reservoir of the birth environment of massive stars, known as HMCs, can give us important hints about the chemical heritage that our own Solar System received from its natal environment.

For all of these species, different chemical formation and destruction pathways, both on grain surfaces and in gas-phase, have been proposed (e.g. \citealt{agarwal1985}; \citealt{grim1989}; \citealt{garrod2008}; \citealt{jones2011}; \citealt{noble2015}; \citealt{fedoseev2016}; \citealt{belloche2017}; \citealt{quenard2018}).
However, chemical networks still need important new inputs from observations in different astronomical environments to be properly constrained.

In this paper we study molecules with one or more peptide-like bond (see Fig.~\ref{fig-molecules}) towards the HMC G31.41+0.31 (hereafter G31) in the context of the G31 Unbiased ALMA sPectral Observational Survey (GUAPOS, \citealt{mininni2020}). All these molecules are reported for the first time in this paper towards G31, with the exceptions of \fm, recently detected by \citet{coletta2020} using single-dish data, and \mi, recently reported by \citet{gorai2021}. G31 is a HMC located at a distance of 3.75 kpc (\citealt{immer2019}), with a luminosity of $\sim$4.5$\times$10$^{4}$ L$_{\odot}$ (from \citealt{osorio2009}) and a mass of $\sim$70 M$_{\odot}$ (\citealt{cesaroni2019}). The core harbours two free-free continuum sources separated by $\sim$0\farcs2 (\citealt{cesaroni2010}), and new VLA and ALMA observations show that there are at least 4 massive star-forming regions in the core (\citealt{beltran2021}). Moreover, few molecular lines present an inverse P-Cygni profile, indicating that the core is collapsing and rotating with respect to the direction of a magnetic field, revealed by polarization measurements (\citealt{girart2009}; \citealt{beltran2019}). G31 is an excellent source to search for complex molecules since it presents a very rich chemistry, as already shown by previous works (e.g. \citealt{beltran2005}; \citealt{beltran2009}; \citealt{rivilla2017}; \citealt{beltran2018}; \citealt{mininni2020}, \citealt{gorai2021}). This is the second paper from this survey, after the first one which presented the GUAPOS project and the analysis of the three energetically most stable C$_{2}$H$_{4}$O$_{2}$ isomers (\citealt{mininni2020}). 

In Sect.~\ref{observations} and \ref{results} we present the observations and the results, respectively. In Sect.~\ref{discussion} we provide a detailed comparison with previous observations towards other sources and a discussion of the main formation and destruction reactions proposed for these species, giving new inputs for future chemical models. Finally, the conclusions are summarised in Sect.~\ref{conclusions}. 

\section{Observations and data analysis}
\label{observations}

\subsection{ALMA data}
\label{observation-data}

Observations towards the HMC G31 were taken with ALMA during Cycle 5 (project 2017.1.00501.S, P.I.: M. T. Beltrán) obtaining an unbiased spectral survey in Band 3, from 84.05 GHz up to 115.91 GHz. The frequency resolution is 0.49 MHz, corresponding to a velocity resolution of $\sim$1.6 km s$^{-1}$ at 90 GHz. The final angular resolution is $\sim$1\farcs2 ($\sim$4500 au). The root mean square (rms) noise of the maps varies between 0.5 mJy beam$^{-1}$ and 1.9 mJy beam$^{-1}$. The pointing center of the observations is $\alpha_{\rm J2000}$ = 18$^{\rm h}$47$^{\rm m}$34$^{\rm s}$ and $\delta_{\rm J2000}$ = -0.1\degree12$^{\prime}$45\asec. The uncertainty on the flux calibration is of $\sim$5\%. For more details see \citet{mininni2020}. 

In this work, as in \citet{mininni2020}, we have analysed a spectrum extracted inside an area equal to the beam size towards the continuum peak position ($\alpha_{\rm J2000}$ = 18$^{\rm h}$47$^{\rm m}$34$^{\rm s}$.321 and $\delta_{\rm J2000}$ = $-$0.1\degree 12$^{\prime}$45\farcs977). The rms noise of the spectrum varies from 7 mK to 27 mK.
The spectrum of G31 is very line rich, which prevents to subtract the continuum by simply using line-free channels. Thus,  we have applied the corrected sigma clipping method (c-SMC) approach of the Python-based tool STATCONT\footnote{ STATCONT is freely accessible here: \url{https://hera.ph1.uni-koeln.de/~sanchez/statcont}.} (\citealt{sanchez-monge2018}) to subtract the continuum. 
A detailed discussion of the results of this method is presented in Appendix \ref{cont-sub}. The error associated with the final Band 3 spectrum is of $\pm$1.2 K, which corresponds to the 11\% of the continuum level at a reference frequency of 84.579 GHz. This error is included as an additional error on the parameters derived from the fit to the spectrum. 

\subsection{Additional high-angular resolution data}
\label{obs-highres}
Interferometric observations of G31 at higher angular resolution were carried out with ALMA in Cycle 2 in July and September 2015 as part of project 2013.1.00489.S. (P.I.: R.\ Cesaroni). The observations were carried out in Band 6 with the array in an extended configuration. The digital correlator was configured in thirteen spectral windows (SPW), one broad window for the continuum and twelve narrow ones for the lines, covering different bandwidths from $\sim$217\,GHz to $\sim$236.5\,GHz. The phase reference center of the observations is $\alpha_{\rm J2000}$= 18$^{\rm h}$47$^{\rm m}$34$^{\rm s}$.315, $\delta_{\rm J2000}$ = -01\degree 12$^\prime$45\farcs90. The resulting synthesized cleaned beam of the maps is 0\farcs2$\times$0\farcs2 ($\sim$750 au) for the lines analysed in this work. The rms noise of the maps is $\sim$1.3 mJy beam$^{-1}$ at $\sim$217 GHz and 218 GHz, $\sim$1.5 mJy beam$^{-1}$ at $\sim$219 GHz, and $\sim$2 mJy beam$^{-1}$ at $\sim$220 GHz. We refer to \citet{cesaroni2017} and \citet{beltran2018} for detailed information on the observations.

 \subsection{Spectral analysis}
\label{analysis}

The line identification of the molecular species present in the GUAPOS spectrum has been done using the version 01/12/2020 of the SLIM (Spectral Line Identification and Modeling) tool within the MADCUBA package\footnote{Madrid Data Cube Analysis on ImageJ is a software developed at the Center of Astrobiology (CAB) in Madrid; \url{http://cab.intacsic.es/madcuba/Portada.html}.} (\citealt{martin2019}).
SLIM uses the spectroscopic entries from the Cologne Database for Molecular Spectroscopy\footnote{\url{http://cdms.astro.uni-koeln.de/classic/}.} (CDMS, \citealt{muller2001, muller2005}; \citealt{endres2016}), the Jet Propulsion Laboratory\footnote{\url{https://spec.jpl.nasa.gov/ftp/pub/catalog/catdir.html}.} (JPL, \citealt{pickett1998}), and for species whose spectroscopy was not present in the catalogues we have added entries using available spectroscopic works. Then, SLIM generates a synthetic spectrum, assuming local thermodynamic equilibrium (LTE) conditions and taking into account the line opacity. In this study, we focused the analysis on the molecules containing one or more peptide-like bonds (NCO backbone) \ia, \fm, \mi, \am, \nm, \ei, \ur, \cfm, and \gly\;(Fig.~\ref{fig-molecules}). Details on the spectroscopic entries used for these species can be found in Appendix \ref{spectroscopy}. Beside the analysis of the peptide-like bond molecules, a preliminary identification of other molecular species has been done to evaluate the effect of possible line contaminations (e.g. thin red line of Fig.~\ref{fig-res-ia}). For each molecular species we used the MADCUBA-AUTOFIT tool to compare the observed spectrum with the LTE synthetic one. This fitting tool provides the best non-linear least-squared fit to all the transitions taken into account by using the Levenberg-Marquardt algorithm. The free parameters of the fitting are the column density of the molecule, $N$, the excitation temperature, $T_{\rm ex}$, the peak velocity, $\varv_{\rm LSR}$, and the full-width-half-maximum (FWHM), $\Delta \varv$. As described by \citet{mininni2020}, the LTE assumption is well justified because of the high volume density $n$(H$_{2}$) $\sim$ 10$^{8}$ cm$^{-3}$.  
Since the emission of the HMC fills the beam (i.e. the region from which we have extracted the spectrum, see Sect.~\ref{integrated-maps} and \citealt{mininni2020}), we did not consider beam-dilution in the fitting procedure. Moreover, since we have also considered the possible contamination from other molecular species, we have limited the fit to non-contaminated or slightly contaminated transitions. In particular, we have considered (i) non-contamination from a line of another species if its peak is at a distance larger than FWHM from the peak of the considered line, and ii) in case of blending with peak separation $<$FWHM, the contamination of the line of the other species should be $<$ 15\%. Thus, the parameters above are left free to obtain the best-LTE fit whenever the algorithm reaches the convergence. If convergence cannot be reached, FWHM, $\varv_{\rm LSR}$, and/or $T_{\rm ex}$ have been fixed as explained in Sect.~\ref{sect-result-fit}. When the algorithm converges, it also provides the errors on the parameters.
The transitions used for the fit of each molecule are listed in Table \ref{transitions}. 
When the spectroscopic information of vibrationally excited states is present, the total partition function used for the fit ($Q_{\rm vibrot}$) has been derived taking into account the contribution of both the rotational partition function, $Q_{\rm rot}$, and the vibrational one, $Q_{\rm vib}$, $Q_{\rm vibrot}$=$Q_{\rm rot} \times Q_{\rm vib}$ (see Appendix \ref{spectroscopy}). The final relative error on the total column densities has been derived as the square root of the quadratic sum of the relative error given by the fit algorithm and the error on the continuum level (11\%, see Sect.~\ref{observation-data}).
Finally, the molecular abundances ($X$) were derived using the column density of H$_{2}$, obtained from the continuum emission inside the area from which we extracted the spectrum, $N_{\rm H_{2}}$ = (1.0$\pm$0.2)$\times$10$^{25}$~cm$^{-2}$ (see \citealt{mininni2020}).

The analysis of the different vibrational states has been done separately for \ia, \fm, and \mi, because of the difficulty to fit all of the transitions of a molecule with a single $T_{\rm ex}$. In fact, it is well know that the HMC G31 has a temperature gradient, with temperatures of $\sim$100 K towards the outer part (at $\sim$0.03 pc from the center), and of $\sim$450 K towards its center (\citealt{beltran2018}). Moreover, as we discuss in Sects.~\ref{res-ia}, \ref{res-fm} and \ref{res-mi}, the ground vibrational states of these molecules are optically thicker than the vibrationally excited ones.
Thus, we only discuss the results obtained from the $^{13}$C-isotopologues, or from the vibrationally excited states if the $^{13}$C-isotopologues have not been detected.
Conversely, the resulting fit for \am\;was obtained taking into account the $v$=0, and $v_{\rm t}$=1, 2 transitions all together, because we find that they all are optically thin. For \nm, only the ground vibrational state transitions have been detected and analysed, and upper limits will be provided for its excited states transitions. Finally, \ei, \ur, \cfm, and \gly\;have not been detected, and only upper limits of the ground vibrational state will be provided.

\section{Results}
\label{results}

\begin{table*}
\setlength{\tabcolsep}{5pt}
\caption{Line parameters obtained from the best LTE fit and abundance for \ia, \fm, \mi\;(and their $^{13}$C-isotopologues), \am, \nm, \ei, \ur, \cfm, and \gly.}
\centering
  \begin{tabular}{lccccc}
  \hline
     & FWHM & $\varv_{\rm LSR}$  & $T_{\rm ex}$ & $N$ & $X$\\ 
         & (km s$^{-1}$) & (km s$^{-1}$) & (K)  & ($\times$10$^{16}$ cm$^{-2}$) & ($\times$10$^{-9}$)\\
  \hline
  \ia, $v$=0 &8& 97$\pm$17 &  217$\pm$22 &11.5$\pm$1.8  & 11$\pm$3 \\
  \ia, $v_{4}$=1 &8& 97 &  217 & 63$\pm$8\tablefootmark{a}  & 63$\pm$15\tablefootmark{a} \\
\ia, $v_{5}$=1 &8& 97 &  217 & 79$\pm$17\tablefootmark{a}  & 79$\pm$23\tablefootmark{a} \\
\ia, $v_{6}$=1 &8& 97 &  217 & $\le$95\tablefootmark{d}  & $\le$95 \\
HN$^{13}$CO, $v$=0  &8 &97 &  217 & 3.8$\pm$0.5 &  3.8$\pm$0.9 \\
  \ia\tablefootmark{b} && & & 141$\pm$49\tablefootmark{c} &  141$\pm$57\tablefootmark{c} \\
  H$^{15}$NCO, $v$=0 & 8 & 97 & 217 & 0.49$\pm$0.09 \tablefootmark{a} & 0.49$\pm$0.13\tablefootmark{a} \\
  \hline
 \fm, $v$=0 & 8.6$\pm$0.2 & 97$\pm$13 & 150$\pm$26  & 5.4$\pm$1.1 & 5.4$\pm$1.5 \\
  \fm, $v_{12}$=1 & 8.4$\pm$0.2 & 97 & 245$\pm$88 & 15$\pm$2 & 15$\pm$4\\
  H$^{13}$C(O)NH$_{2}$, $v$=0 &8.6 &97 &150 &  0.47$\pm$0.07 & 0.47$\pm$0.11 \\
  \fm\tablefootmark{b} && & & 17$\pm$6\tablefootmark{c} &  17$\pm$7\tablefootmark{c} \\
  \hline
  \mi, $v_{\rm b}$=0 & 7.15$\pm$0.15 & 97$\pm$11 & 122$\pm$7 & 4.3$\pm$0.6 & 4.3$\pm$1.0 \\
   \mi, $v_{\rm b}$=1 & 7.15 & 97& 91$\pm$37 & 12$\pm$3\tablefootmark{c} & 12$\pm$4\tablefootmark{c} \\
   $^{13}$CH$_{3}$NCO, $v_{\rm b}$=0 & 7.15 & 97& 122 & 0.49$\pm$0.09\tablefootmark{a} & 0.49$\pm$0.13\tablefootmark{a} \\
 CH$_{3}$N$^{13}$CO, $v_{\rm b}$=0  &  7.15 & 97  & 122 & $\l\le$0.63\tablefootmark{d} & $<$0.63 \\
   \mi\tablefootmark{b} && & & 18$\pm$7 &  18$\pm$9 \\
  \hline
  \am, $v$=0, $v_{\rm t}$=1, 2 & 6.2$\pm$0.4 & 96.7 & 285$\pm$50 & 8$\pm$4\tablefootmark{c} & 8$\pm$4\tablefootmark{c} \\
  \hline
  \nm, $v$=0 & 7&96.5 &285 &3.7$\pm$1.6\tablefootmark{c} & 3.7$\pm$1.7\tablefootmark{c} \\
  \nm, $v_{\rm t}$=1& 7 &96.5 &285 & $\le$2.3\tablefootmark{d} & $\le$2.3 \\
  \nm, $v_{\rm t}$=2 & 7 &96.5  &285& $\le$2.1\tablefootmark{d} & $\le$2.1 \\
  \hline
  \ei,  $v$=0 &7.15 & 97& 91& $\le$0.5\tablefootmark{d} & $\le$0.5 \\
   \hline
  \ur  $v$=0 &7 & 96.5 & 150  & $\le$0.016\tablefootmark{d} & $\le$0.016 \\
    \hline
 \cfm,  $v$=0 &7 & 96.5 & 150  & $\le$0.3\tablefootmark{d} & $\le$0.3 \\
    \hline
\gly,  $v$=0  &7 & 96.5 & 150  & $\le$0.07\tablefootmark{d} & $\le$0.07 \\
  \hline
  \normalsize
  \label{table-fitresult-v0}
  \end{tabular}
  \tablefoot{$N$ is the average column density inside the beam of 1\farcs2; $X$ is the abundance inside a beam calculated as $N$/$N_{\rm H_{2}}$, where $N_{\rm H_{2}}$ = (1.0$\pm$0.2)$\times$10$^{25}$~cm$^{-2}$. Parameters without errors are fixed in the fitting procedure, as explained in Sect.~\ref{analysis}.\\
 \tablefoottext{a}{Tentative detection}. \tablefoottext{b}{Column density and abundances derived from the $^{13}$C-isotopologue multiplied by \cratio=37$\pm$12 (\citealt{yan2019}). In the case of \mi\;we have used the $^{13}$CH$_{3}$NCO column density.}
\tablefoottext{c}{Value used for the discussion in Sect.~\ref{discussion}.} \tablefoottext{d}{Note that this upper limit should be taken with caution since it has been derived by eye because all of the transitions are contaminated by other species.}
}
      \normalsize
\end{table*}

In this section we show the results obtained from the line fitting procedure performed with MADCUBA-SLIM for the nine species (Sects.~\ref{res-ia}--\ref{res-ei} and Appendix \ref{figure-tentative}). The non-contaminated transitions used to fit the molecular lines are listed in Table \ref{transitions}.  Moreover, we present the integrated emission maps of the molecular species obtained from the GUAPOS survey ($\sim$1\farcs2), and higher angular resolution observations ($\sim$0\farcs2) for \ia, \fm, and \mi \;(Sect.~\ref{integrated-maps}).
The results from the fitting procedure are listed in Table \ref{table-fitresult-v0}, and the total spectrum with the fit to all of the molecules studied in this paper is given in Appendix \ref{spectra}.
This is the first time that \ia, \fm, \mi, \am, and \nm\;have been detected together towards G31 and outside the GC, after the detections in Sgr B2(N2) (\citealt{belloche2017}).

\subsection{LTE fits}
\label{sect-result-fit}

\subsubsection{Isocyanic acid (\ia)}
\label{res-ia}
\begin{figure*}
\centering
\includegraphics[width=35pc]{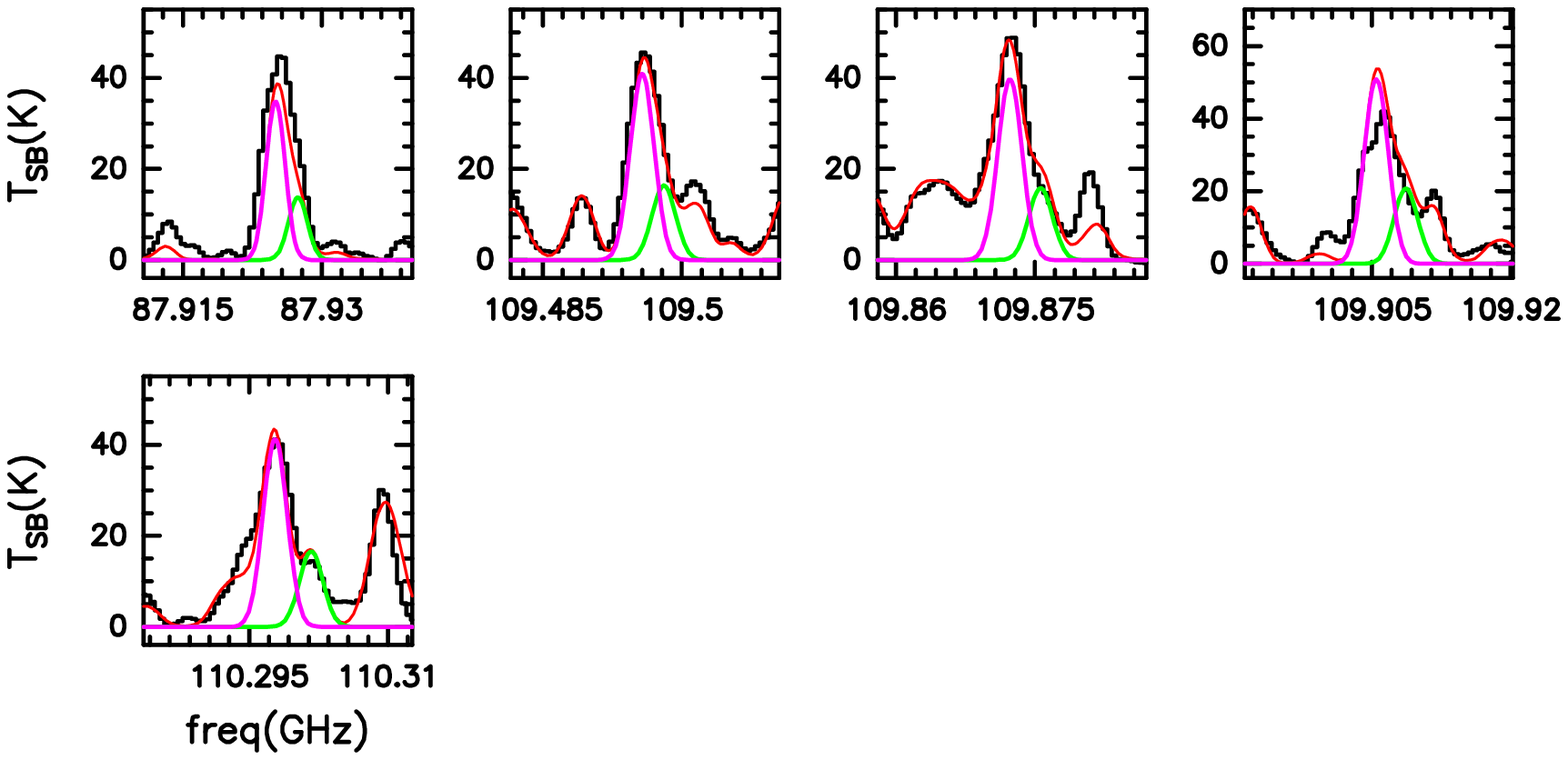}
\caption{Transitions listed in Table~\ref{transitions} and used to fit the $v$=0 state of isocyanic acid (\ia) and its $^{13}$C-isotopologue (HN$^{13}$CO). The magenta and green curves represent the best LTE fits obtained with MADCUBA for HNCO and HN$^{13}$CO, respectively (Table \ref{table-fitresult-v0}). The red curve shows the simulated spectrum taking into account all the species identified so far in the region.}
\label{fig-res-ia}
\end{figure*}

The HNCO and HN$^{13}$CO molecules have a similar center of mass and for this reason the frequencies of their rotational transitions are very close ($\sim$1.8 MHz of difference). Thus, since the typical line width towards G31 is of $\sim$6--8 km s$^{-1}$ ($\sim$2.0--2.6 MHz at 100 GHz, see e.g.~\citealt{mininni2020}) the lines of the two species appear blended. 
We have performed the fit of both species simultaneously (magenta and green curves in Fig.~\ref{fig-res-ia}) using MADCUBA, considering also the contribution from other species already identified in the GUAPOS survey (thin red line in Fig.~\ref{fig-res-ia}). 

To perform the final fit of HNCO and HN$^{13}$CO we have fixed the FWHM to 8 km s$^{-1}$, which best reproduce the observed line profiles (Fig.~\ref{fig-res-ia}). The best fit of HNCO provided $T_{\rm ex}$ = 217$\pm$22 K, $N$ = (1.15$\pm$0.18)$\times$10$^{17}$ cm$^{-2}$, and molecular abundance with respect to H$_{2}$, $X$ = (1.1$\pm$0.3)$\times$10$^{-8}$. To derive the best fit for HN$^{13}$CO we fixed $T_{\rm ex}$ to the same estimated value for HNCO because the fit did not converge leaving it free, and we have obtained $N$ = (3.8$\pm$0.5)$\times$10$^{16}$ cm$^{-2}$, and $X$ = (3.8$\pm$0.9)$\times$10$^{-9}$. 

Moreover, we have also tentatively detected the vibrationally excited states $v_{4}$=1 and $v_{5}$=1, while an upper limit to the $v_{6}$=1 state is given. MADCUBA derives the upper limits to the integrated intensity using the formula 3$\times$rms$\times\Delta$v/$\sqrt{n_{\rm chan}}$, where rms is the root-mean-square measured over a line-free spectral range, and $n_{\rm chan}$ is the number of channels covered by the FWHM, $\Delta \varv$. As shown in Figs.~\ref{fig-res-ia-va} and \ref{fig-res-ia-vb}, only five and three rotational transitions of the $v_{4}$=1 and $v_{5}$=1 states, respectively, have been tentatively detected, since they are partially contaminated with other molecules. The main contaminants of the $v_{4}$=1 transitions are HNCO, $v_{5}$=1 at 110.086 GHz, CH$_{3}$COCH$_{3}$ at 110.089 GHz, HCOOC$_{2}$H$_{5}$ at 110.417 GHz and 110.103 GHz, CH$_{2}$DOH at 110.105 GHz, and CH$_{3}$CHO at 106.792 GHz. The main contaminant of the $v_{5}$=1 transitions is ethylene glycol at 87.739 GHz.
To perform the final fit we have fixed FWHM, $T_{\rm ex}$, and $\varv_{\rm LSR}$ as those of the ground state of HNCO (Table \ref{table-fitresult-v0}). The best fit gave as output $N$ = (6.3$\pm$0.8)$\times$10$^{17}$ cm$^{-2}$, and  $X$ = (6.3$\pm$1.5)$\times$10$^{-8}$ for $v_{4}$=1, and $N$ = (7.9$\pm$1.7)$\times$10$^{17}$ cm$^{-2}$, and $X$ = (7.9$\pm$2.3)$\times$10$^{-8}$ for $v_{5}$=1. Moreover, for $v_{6}$=1 we have derived a rough estimate of the upper limit of the column density using MADCUBA-SLIM given the difficulty of obtaining a precise result due to the blending with lines of other species. Thus, we have assumed $T_{\rm ex}$, $\varv_{\rm LSR}$, and FWHM derived for \ia\;$v_{\rm b}$=0, and have increased $N$ to the maximum value compatible with the observed spectrum. We have obtained $N$ $\le$ 9.5$\times$10$^{17}$ cm$^{-2}$ and an abundance $X$ $\le$ 9.5 $\times$10$^{-8}$. 
We have also found hints of possible emission of three transitions of H$^{15}$NCO, at 106.224, 106.578, and 106.614 GHz, respectively (Fig.~\ref{fig-ia-15N}). The transition at 106.224 GHz is partially contaminated by n-C$_{3}$H$_{7}$CN, $v$=0 (20\% of contamination). Moreover, the transition at 106.614 GHz seems to be contaminated by a non-identified line. Taking into account these transitions, and fixing all the parameters as those estimated for HNCO, we have derived a tentative $N$ = (4.9$\pm$0.9)$\times$10$^{15}$ cm$^{-2}$, and $X$ = (4.9$\pm$1.3)$\times$10$^{-10}$.

From the results of the fit we have derived a \cratio\;ratio of 3.0$\pm$0.6 and a \nratio\;ratio of 23$\pm$6 using HNCO, $v$=0.  The low \cratio\;is probably due to the line opacity, $\tau$, of the optically thick main isotopologue. In fact, from the fit of HN$^{13}$CO we obtain values of $\tau$ as high as 0.04, while the $\tau$ obtained for HNCO, $v$=0 are one order of magnitude higher (see Table \ref{transitions}). Moreover, this is also confirmed by the low \nratio\;ratio (23) measured. This is by far the lowest ratio ever estimated towards massive star-forming regions, for which typical values range between 200 and 1000 (e.g., \citealt{colzi18b}), and it is even lower that the values measured in the pristine material of meteorites (see e.g. \citealt{bonal2009}).

Since HNCO is optically thick, we have derived the column density using the optically thinner $^{13}$C-isotopologue (see Table \ref{transitions}), correcting it by the \cratio\;ratio derived following the galactocentric trend recently obtained by \citet{yan2019}. At the galactocentric distance of G31, $D_{\rm GC}$=5.02 kpc, the \cratio\;ratio is 37$\pm$12. 
Thus, the corrected column density is $N$ = (1.4$\pm$0.5)$\times$10$^{18}$ cm$^{-2}$, and the corrected abundance is $X$=(1.4$\pm$0.6)$\times$10$^{-7}$.  If we simulate the spectrum of the transitions in Fig.~\ref{fig-res-ia} of HNCO, $v$=0 with this corrected column density, the derived line opacities are in the range 1.3--3.4 confirming that HNCO, $v$=0 is optically thick. This column density and abundance will be used for the discussion presented in Sect.~\ref{discussion} and are consistent with those derived from the tentatively detected vibrationally excited state $v_{5}$=1. Moreover, the [HN$^{13}$CO/H$^{15}$NCO]$\times$37 value derived, 287$\pm$66, is consistent with the Galactic \nratio\;value of 340$\pm$90 derived from the linear relation found by \citet{colzi18b}, which suggests that both isotopologues are reasonably optically thin.

\subsubsection{Formamide (\fm)}
\label{res-fm}

\begin{figure*}
\centering
\includegraphics[width=35pc]{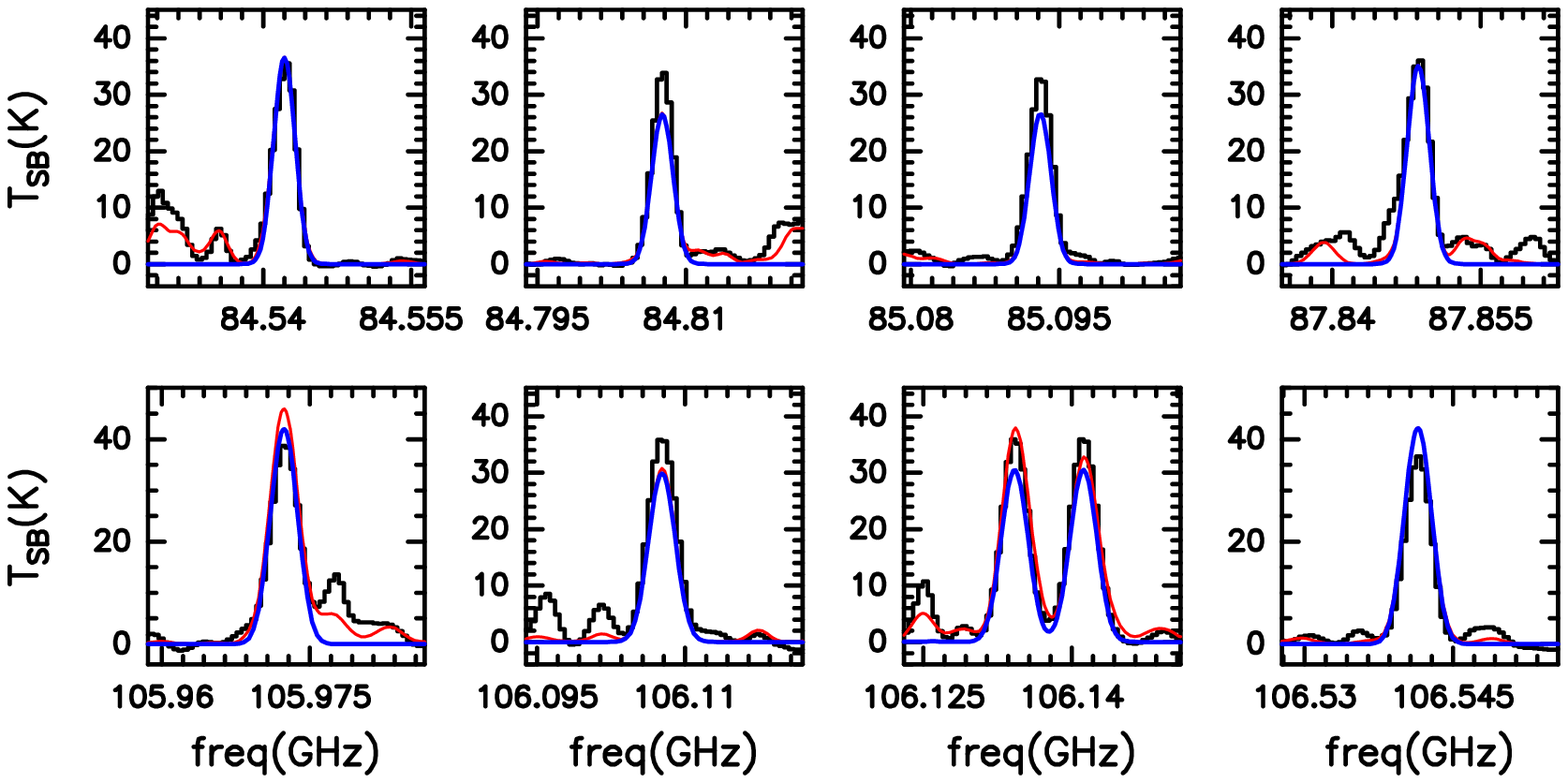}
\caption{Transitions listed in Table~\ref{transitions} and used to fit the $v$=0 state of formamide (\fm). The blue curve represents the best LTE fit obtained with MADCUBA (Table \ref{table-fitresult-v0}). The red curve shows the simulated spectrum taking into account all the species identified so far in the region.}
\label{fig-res-fm}
\end{figure*}

\begin{figure*}
\centering
\includegraphics[width=35pc]{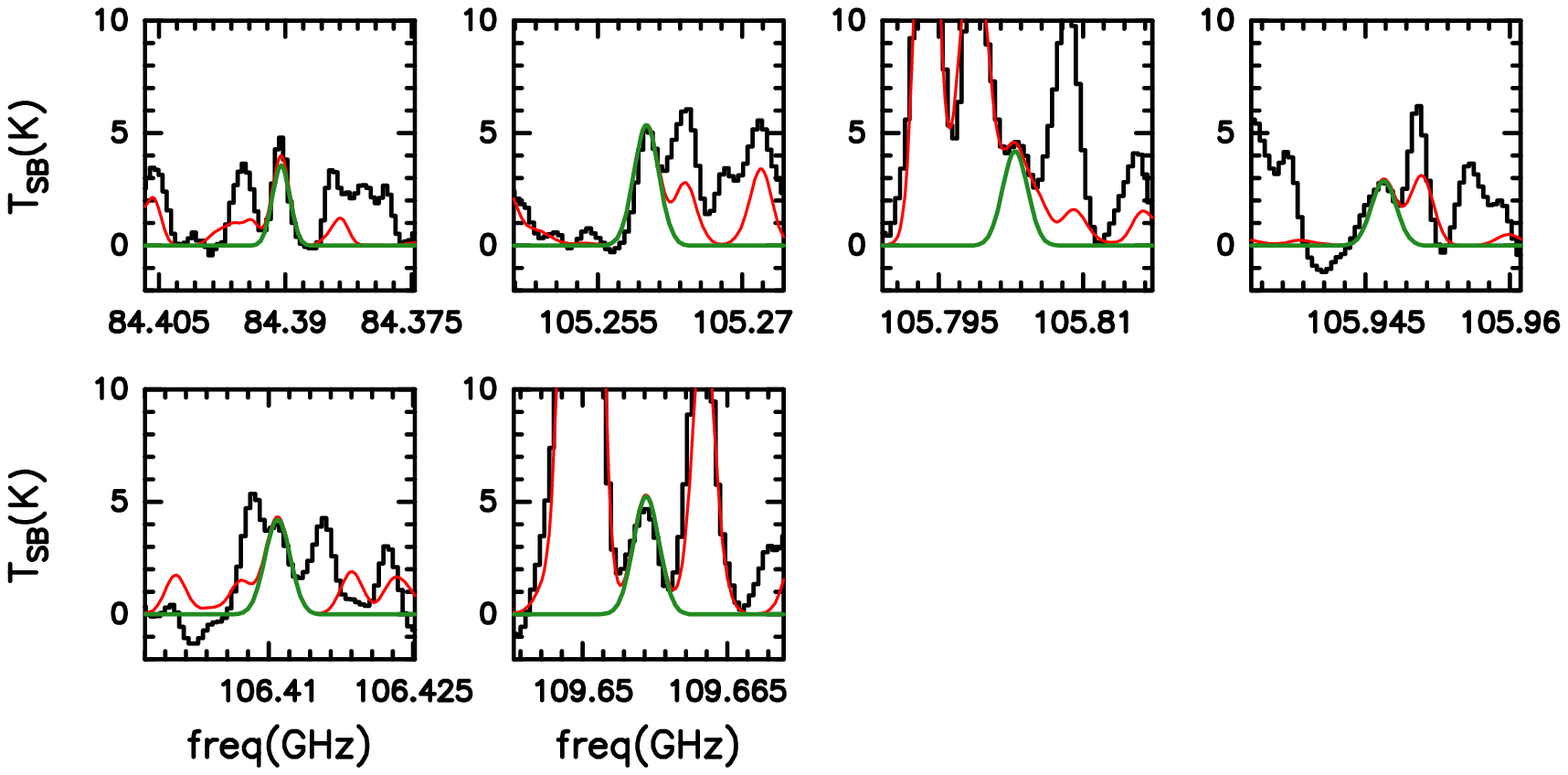}
\caption{Transitions listed in Table~\ref{transitions} and used to fit the $v$=0 state of the $^{13}$C-isotopologue of formamide (H$^{13}$C(O)NH$_{2}$). The dark green curve represents the best LTE fit obtained with MADCUBA (Table \ref{table-fitresult-v0}). The red curve shows the simulated spectrum taking into account all the species identified so far in the region.}
\label{fig-res-13fm}
\end{figure*}

\begin{figure*}
\centering
\includegraphics[width=35pc]{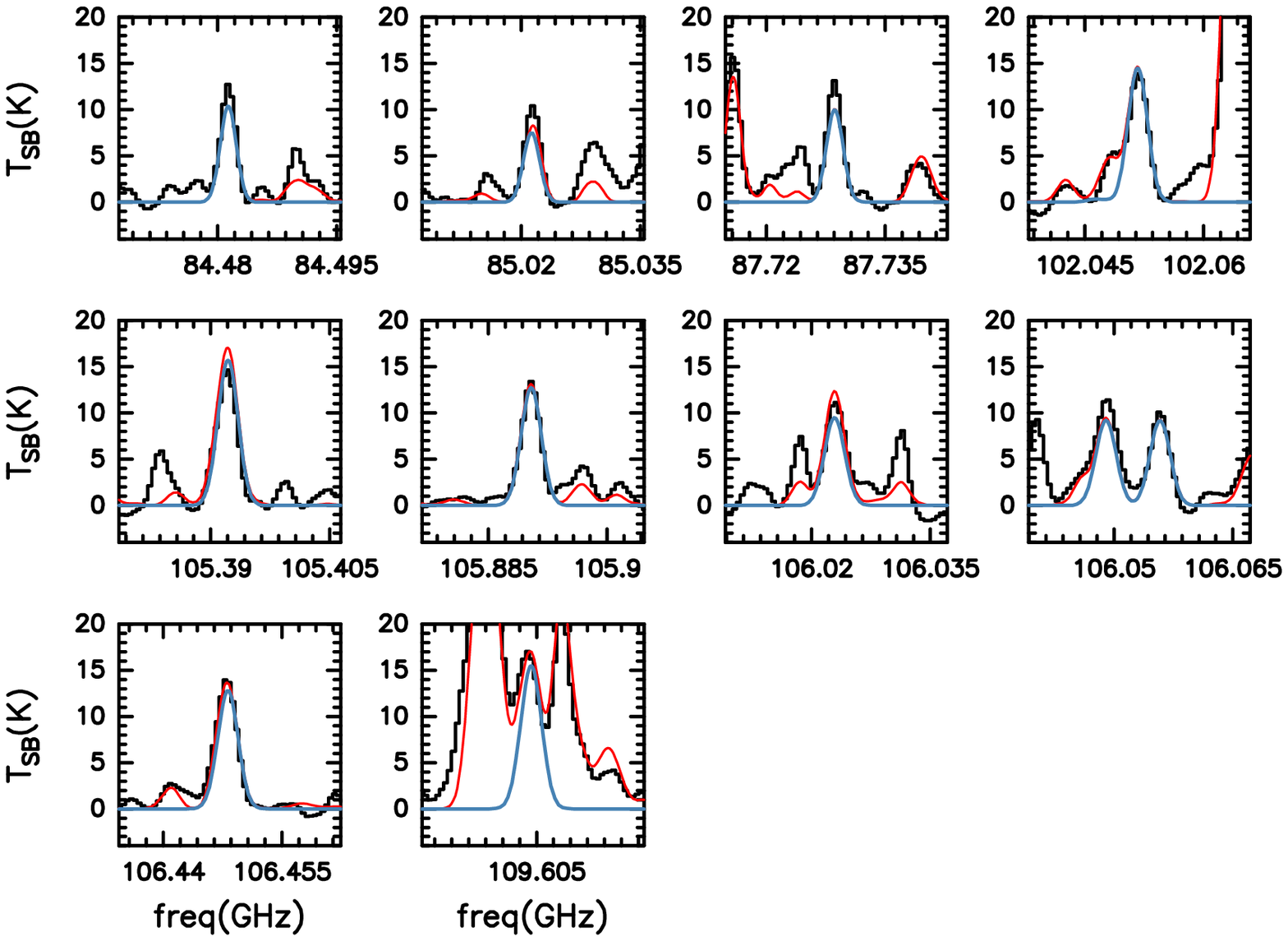}
\caption{Transitions listed in Table~\ref{transitions} and used to fit the $v_{12}$=1 state of formamide (\fm). The steel blue curve represents the best LTE fits obtained with MADCUBA (Table \ref{table-fitresult-v0}). The red curve shows the simulated spectrum taking into account all the species identified so far in the region.}
\label{fig-res-fm-v1}
\end{figure*}

Figures \ref{fig-res-fm} and \ref{fig-res-13fm} show the non-contaminated transitions used to fit \fm\;and H$^{13}$C(O)NH$_{2}$, while Fig.~\ref{fig-res-fm-v1} shows the non-contaminated transitions of the vibrationally excited state $v_{12}$=1.
The best-fit parameters obtained with MADCUBA for the ground state are $T_{\rm ex}$ = 150$\pm$26 K, FWHM = 8.6$\pm$0.2 km s$^{-1}$, $N$ = (5.4$\pm$1.1)$\times$10$^{16}$ cm$^{-2}$, and $X$ = (5.4$\pm$1.5)$\times$10$^{-9}$ (Table \ref{table-fitresult-v0}). For its $^{13}$C-isotopologue, H$^{13}$C(O)NH$_{2}$, we have fixed the FWHM, the $T_{\rm ex}$, and the $\varv_{\rm LSR}$ as those estimated for \fm\;and found $N$ = (4.7$\pm$0.7)$\times$10$^{15}$ cm$^{-2}$ and $X$ = (4.7$\pm$1.1)$\times$10$^{-10}$. 

From the fit results we have derived a \cratio\;ratio of 12$\pm$3. 
As in the case of HNCO, this low \cratio\;ratio is probably affected by opacity effects. As already done for HNCO, we discuss the column density and abundance of \fm\;assuming the \cratio\;ratio from the galactocentric distance dependence and deriving the values from those of the $^{13}$C-isotopologue. 
Thus, the corrected column density is $N$ = (1.7$\pm$0.6)$\times$10$^{17}$ cm$^{-2}$, and the corrected abundance is $X$ = (1.7$\pm$0.7)$\times$10$^{-8}$.
These values will be used for the discussion presented in Sect.~\ref{discussion}. 

The best-fit parameters obtained with MADCUBA for the $v_{\rm 12}$=1 state are $T_{\rm ex}$ = 245$\pm$88 K (consistent within the error with that obtained for $v$=0), FWHM = 8.4$\pm$0.2 km s$^{-1}$, $N$ = (1.5$\pm$0.2)$\times$10$^{17}$ cm$^{-2}$, and $X$ = (1.5$\pm$0.4)$\times$10$^{-8}$. Note that the column density and the abundance derived from the $^{13}$C-isotopologue multiplied by the \cratio\;ratio, are consistent, within the errors, with those derived separately from the \fm\;$v_{\rm 12}$=1 state (see Table \ref{table-fitresult-v0}). This result is also consistent with what was found for HNCO, HN$^{13}$CO and the tentatively detected vibrationally excited states. It should be noted that if we fix $T_{\rm ex}$ of H$^{13}$C(O)NH$_{2}$ to that obtained for the \fm\;$v_{\rm 12}$=1 state (245 K), we obtain $N$ = (9.1$\pm$1.4)$\times$10$^{15}$ cm$^{-2}$. Thus, the $N$ corrected for the \cratio\;ratio would be $N$ = (3.4$\pm$1.2)$\times$10$^{17}$ cm$^{-2}$, a factor of two higher with respect to the result obtained for the $v_{\rm 12}$=1 state of the main species, but consistent within the errors.

Since we have found that rotational transitions of the vibrationally excited state are optically thin, this could be an indication that the $v$=0 state is affected by opacity effects, and that the assumed \cratio\;ratio to obtain the final results is a good approximation. In fact, if we simulate the spectrum of the transitions in Fig.~\ref{fig-res-fm} of \fm, $v$=0 with the column density derived from the $^{13}$C-isotopologue ($N$=1.7$\times$10$^{17}$ cm$^{-2}$), the derived line opacities are in the range 0.3--1.1.

\subsubsection{Methyl isocyanate (\mi)}
\label{res-mi}

\begin{figure*}[h!]
\centering
\includegraphics[width=35pc]{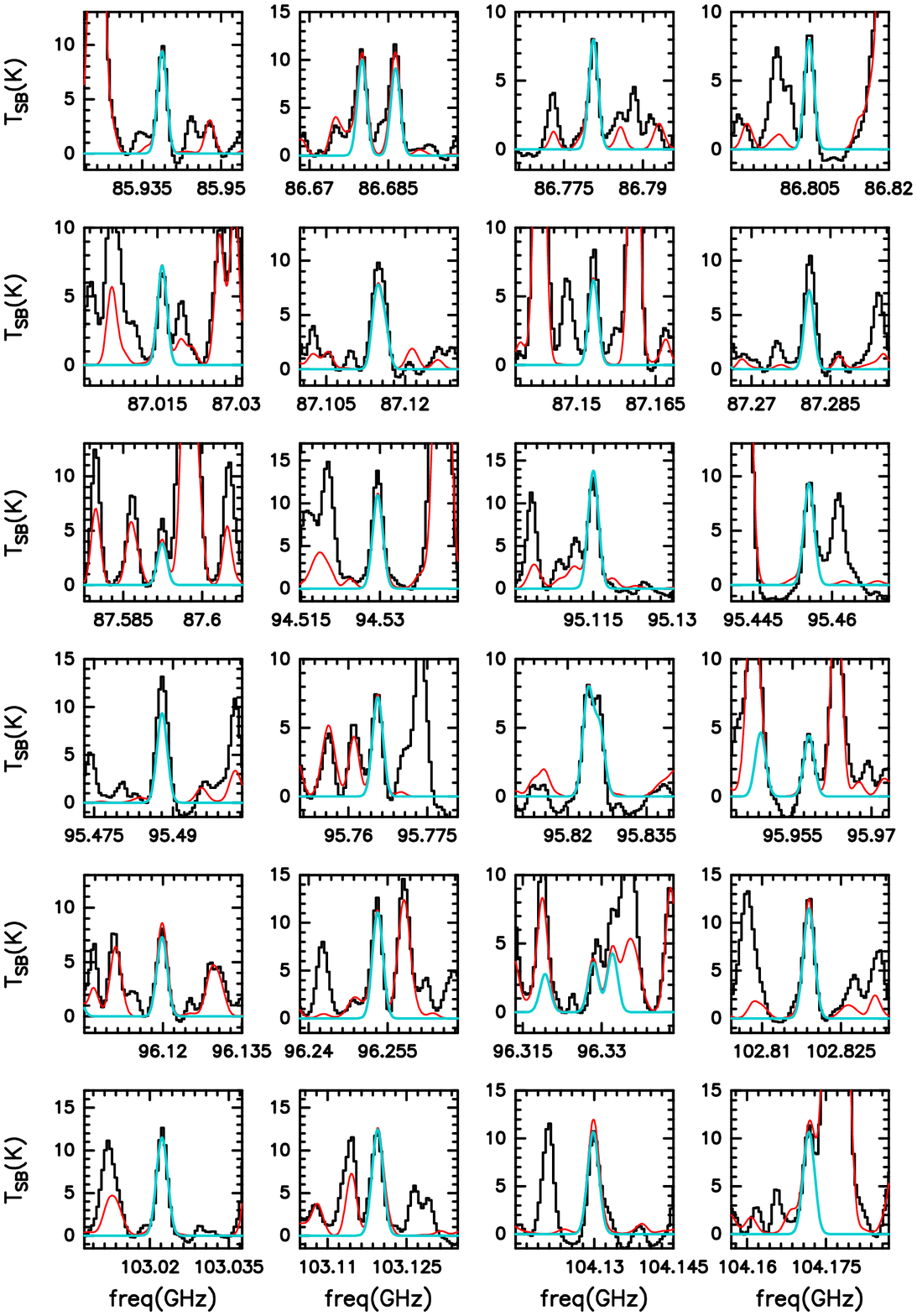}
\caption{Transitions listed in Table~\ref{transitions} and used to fit the $v_{\rm b}$=0 state of methyl isocyanate (\mi). The dark turquoise curve represents the best LTE fit obtained with MADCUBA (Table \ref{table-fitresult-v0}). The red curve shows the simulated spectrum taking into account all the species identified so far in the region.}
\label{fig-res-mi-1}
\end{figure*}

\begin{figure*}
\centering
\addtocounter{figure}{-1}
\includegraphics[width=35pc]{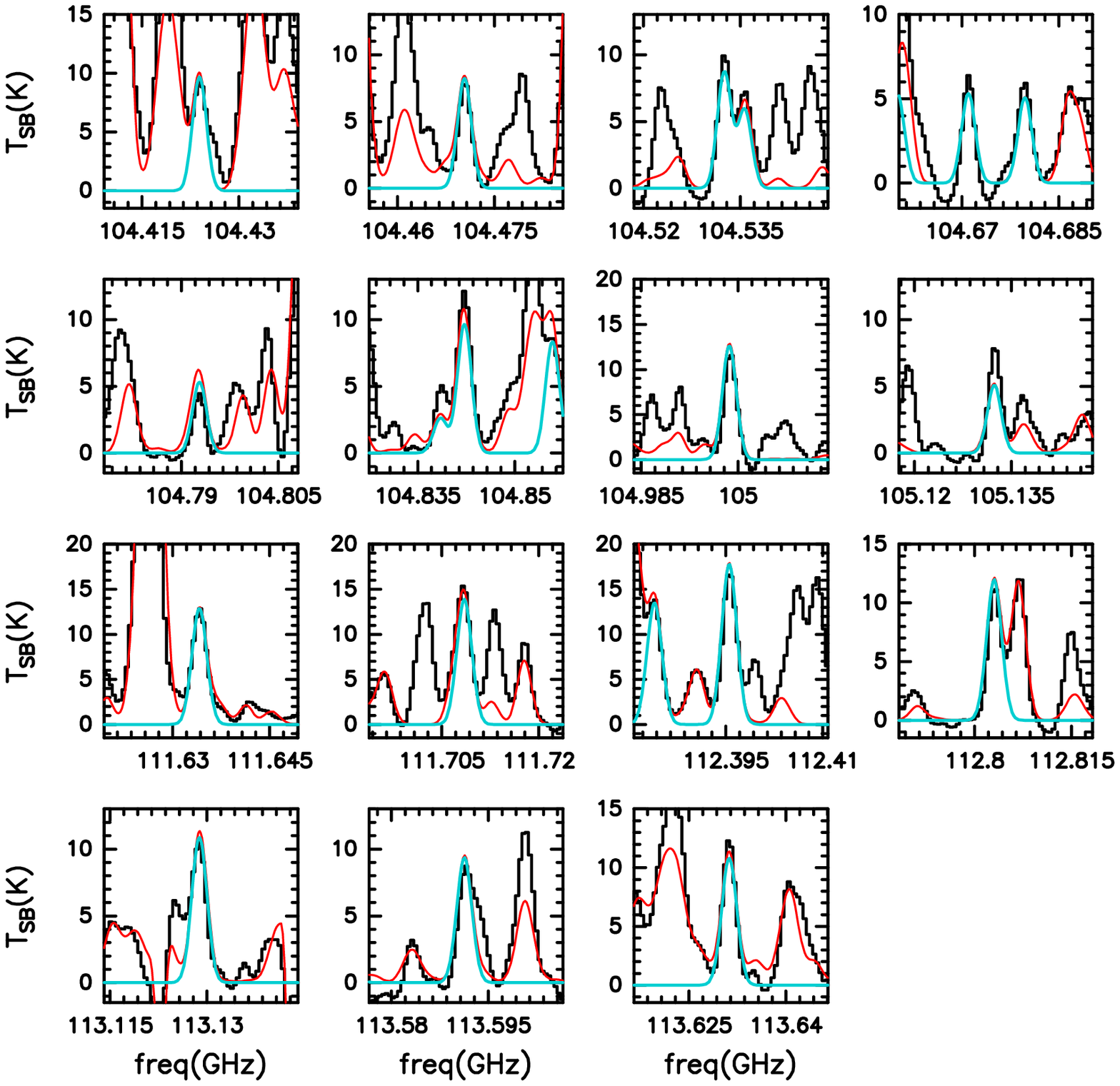}
\caption{Continued.}
\label{fig-res-mi-2}
\end{figure*}

\begin{figure*}
\centering
\includegraphics[width=35pc]{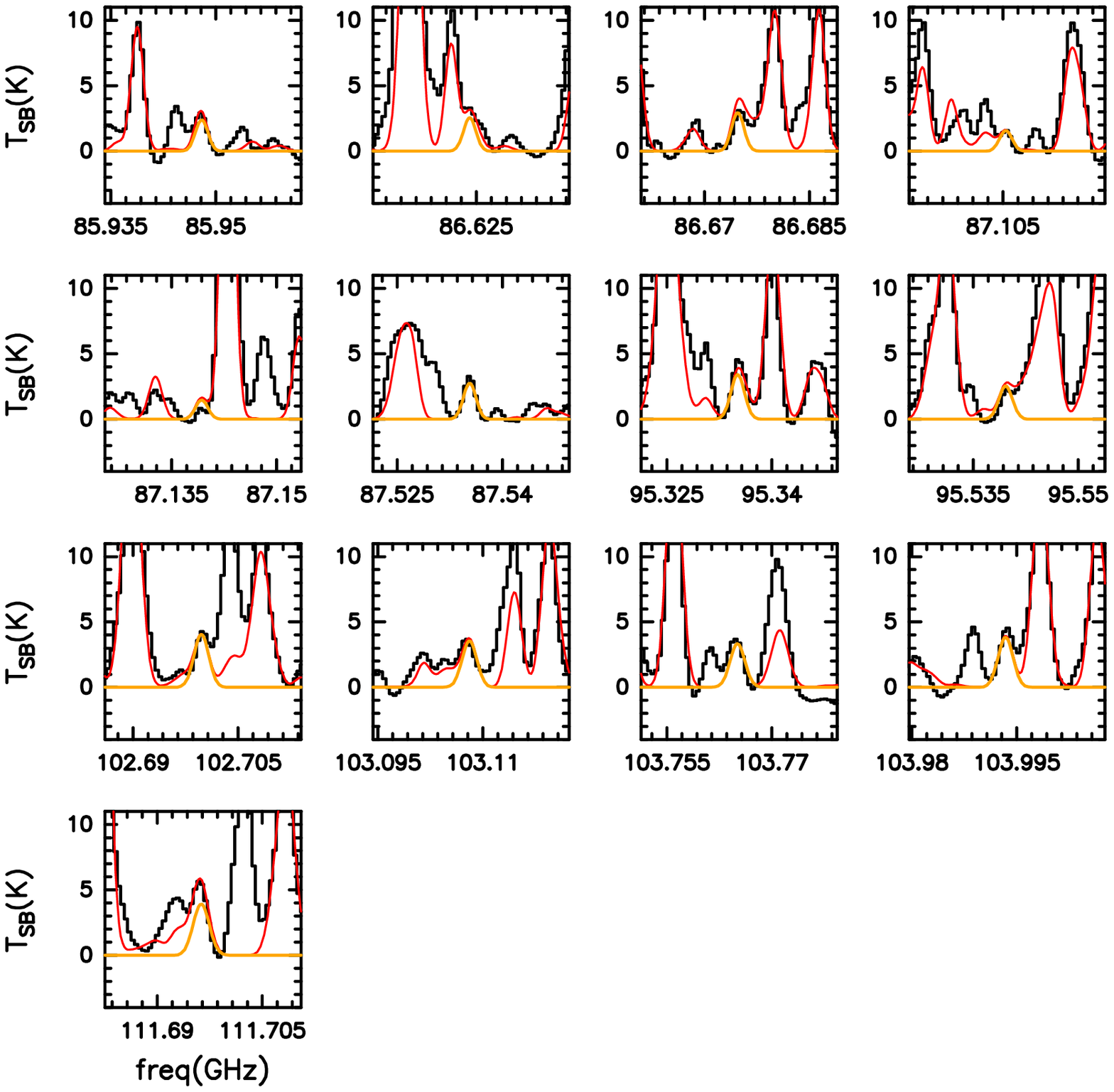}
\caption{Transitions listed in Table~\ref{transitions} and used to fit the $v_{\rm b}$=1 state of methyl isocyanate (\mi). The light orange curve represents the best LTE fits obtained with MADCUBA (Table \ref{table-fitresult-v0}). The red curve shows the simulated spectrum taking into account all the species identified so far in the region.}
\label{fig-res-mi-v1}
\end{figure*}

Figure \ref{fig-res-mi-1} shows the non-contaminated transitions used to fit the ground vibrational state of \mi.
The best-fit parameters obtained with MADCUBA are $T_{\rm ex}$ = 122$\pm$7 K, FWHM = 7.15$\pm$0.15 km s$^{-1}$, $N$ = (4.3$\pm$0.6)$\times$10$^{16}$ cm$^{-2}$, and $X$ = (4.3$\pm$1.0)$\times$10$^{-9}$.

Figure~\ref{fig-res-mi-v1} shows the non-contaminated transitions for the vibrationally excited state $v_{\rm b}$=1. The best fit has been obtained fixing the same FWHM as the ground state, and resulted in $T_{\rm ex}$ = 91$\pm$37 K, $N$ = (1.2$\pm$0.3)$\times$10$^{17}$ cm$^{-2}$, and $X$ = (1.2$\pm$0.4)$\times$10$^{-8}$.

One $^{13}$C-isotopologue, $^{13}$CH$_{3}$NCO, is tentatively detected because most of the transitions are contaminated or partially blended with those of other molecular species. Figure \ref{fig-mi-13C} shows the less contaminated transitions of $^{13}$CH$_{3}$NCO, $v_{\rm b}$=0. In particular, the main contaminants are $^{33}$SO$_{2}$ at 93.070 GHz, 93.071 GHz, and 93.073 GHz, \nm\;at 93.406 GHz, CH$_{3}$C(O)NH$_{2}$ at 93.459 GHz, CH$_{3}$OCHO at 93.457 GHz, CH$_{3}$COOH at 100.203 GHz, and 100.942 GHz, CH$_{3}$CHO at 101.892 GHz, and CH$_{3}$O$^{13}$CHO at 108.576 GHz. Moreover, at 108.406 GHz the baseline derived from STATCONT is slightly high, and the simulated spectra do not match exactly the observed one. However, the peak of the simulated transition is of $\sim$1 K and it is consistent with the error on the derived baseline (see Sect.~\ref{cont-sub}).
To perform the final fit we have fixed the FWHM, the $T_{\rm ex}$, and the $\varv_{\rm LSR}$ as those of the ground state of \mi\;(Sect.~\ref{res-mi}). The best fit gave as $N$ = (4.9$\pm$0.9)$\times$10$^{15}$ cm$^{-2}$ and $X$ = (4.9$\pm$1.3)$\times$10$^{-10}$.
Conversely, CH$_{3}$N$^{13}$CO was not detected and only an upper limit, consistent with what is obtained for $^{13}$CH$_{3}$NCO, can be provided for the column density and abundance (Table \ref{table-fitresult-v0}). As with HNCO, $v_{6}$=1, it was not possible to directly derive the upper limit of the CH$_{3}$N$^{13}$CO column density due to the blending with other species, and we roughly estimated the column density by assuming the same $T_{\rm ex}$, $\varv_{\rm LSR}$, and FWHM of the ground state of \mi.

For this molecular species, the ground vibrational state has $\tau$ up to 0.12, while the $v_{\rm b}$=1 is optically thin ($\tau\le$0.05). Moreover, from the tentative detection of $^{13}$CH$_{3}$NCO, and correcting for the same \cratio\;used above, we obtain $N$ = (1.8$\pm$0.7)$\times$10$^{17}$ cm$^{-2}$, and $X$ = (1.8$\pm$0.8)$\times$10$^{-8}$, consistent with the results obtained for the $v_{\rm b}$=1 state, as already found for HNCO and \fm. If we simulate the spectrum of the transitions in Fig.~\ref{fig-res-mi-1} of \mi, $v_{\rm b}$=0 with the column density obtained from the $v_{\rm b}$=1 states, the derived line opacities are up to 0.4 confirming that \mi, $v_{\rm b}$=0 is partially optically thick. 
Thus, we discuss the results obtained for \mi\;taking into account the best fit of the vibrationally excited state (Sect.~\ref{discussion}).

\mi\;is the molecule that presents the lowest $T_{\rm ex}$ with respect to the other species studied in this work (Table \ref{table-fitresult-v0}). First of all, the $T_{\rm ex}$ found for the ground vibrational state could be affected by the opacity of these rotational transitions. Secondly, the transitions of the $v_{\rm b}$=1 state, from which we take the final result, present a very small range of $E_{\rm up}$ (from 285 K up 350 K). This means that the $T_{\rm ex}$ could not be well constrained. However, even considering a higher $T_{\rm ex}$ of 300 K, the derived $N$ varies only by a factor of 1.2, consistent within the errors with the $N$ derived leaving $T_{\rm ex}$ free.

\subsubsection{Acetamide (\am)}
\label{res-am}

\begin{figure*}[h!]
\centering
\includegraphics[width=32pc]{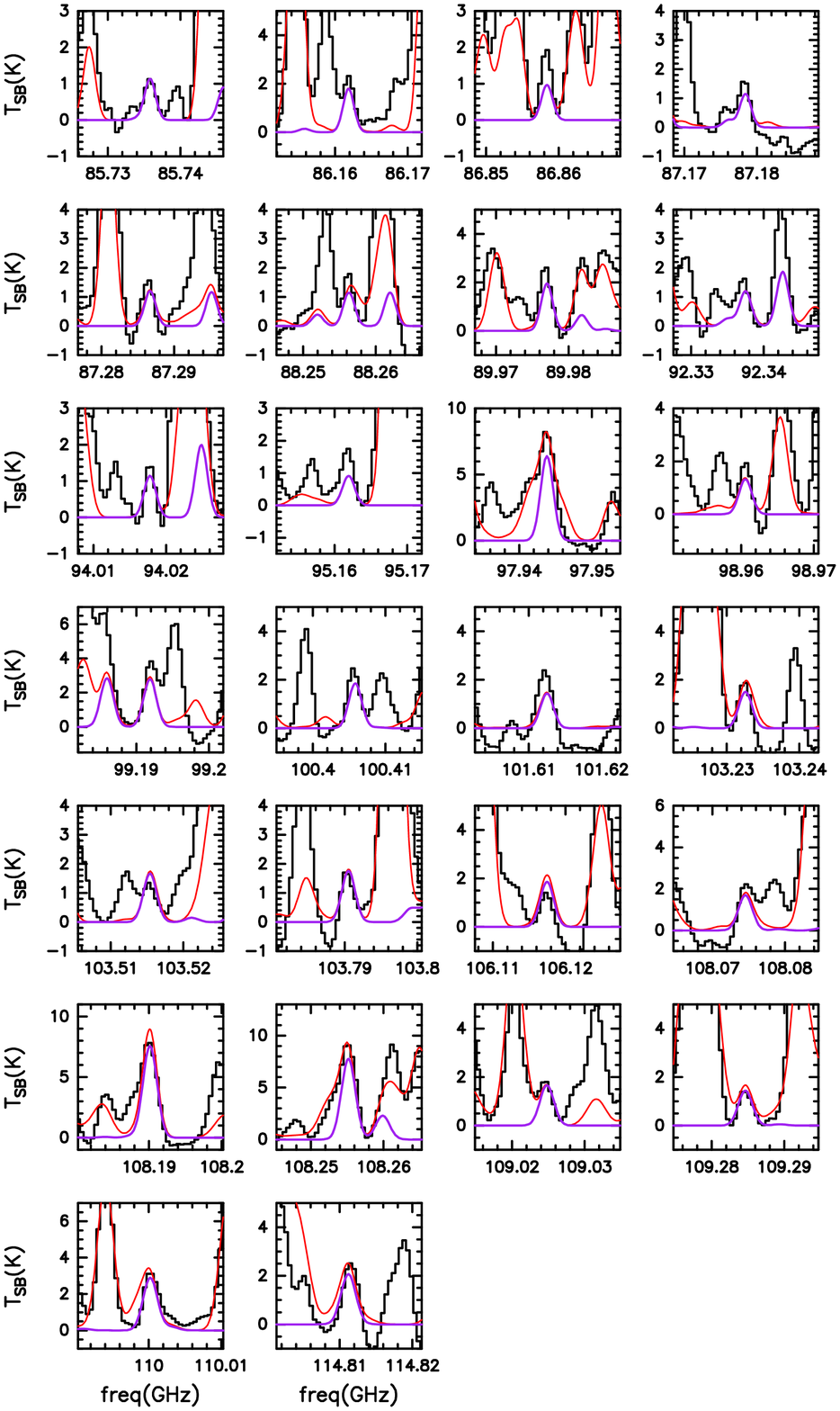}
\caption{Transitions listed in Table~\ref{transitions} and used to fit the $v$=0, and $v_{\rm t}$=1, 2 states of acetamide (\am). The purple curve represents the best LTE fit obtained with MADCUBA (Table \ref{table-fitresult-v0}). The red curve shows the simulated spectrum taking into account all the species identified so far in the region.}
\label{fig-res-am1}
\end{figure*}

Figure \ref{fig-res-am1} shows the non-contaminated or slightly contaminated transitions used to fit \am. This is the first time that \am\;is detected towards this source. In this case, both the ground vibrational state and the excited ones ($v_{\rm t}$=1, 2) are optically thin ($\tau <$0.01), and the range of upper energies of the levels is similar for the three vibrational states ($E_{\rm up}$ from $\sim$50 up to 250 K, see Table \ref{transitions}). Moreover, since all the vibrational levels are optically thin, we have been able to fit them simultaneously with a single LTE fit. 
The best-fit parameters obtained with MADCUBA are $T_{\rm ex}$ = 285$\pm$50 K, FWHM = 6.2$\pm$0.4 km s$^{-1}$, $N$ = (8$\pm$4)$\times$10$^{16}$ cm$^{-2}$, and $X$ = (8$\pm$4)$\times$10$^{-9}$.

\subsubsection{N-methylformamide (\nm)}
\label{res-am}
\begin{figure*}
\centering
\includegraphics[width=35pc]{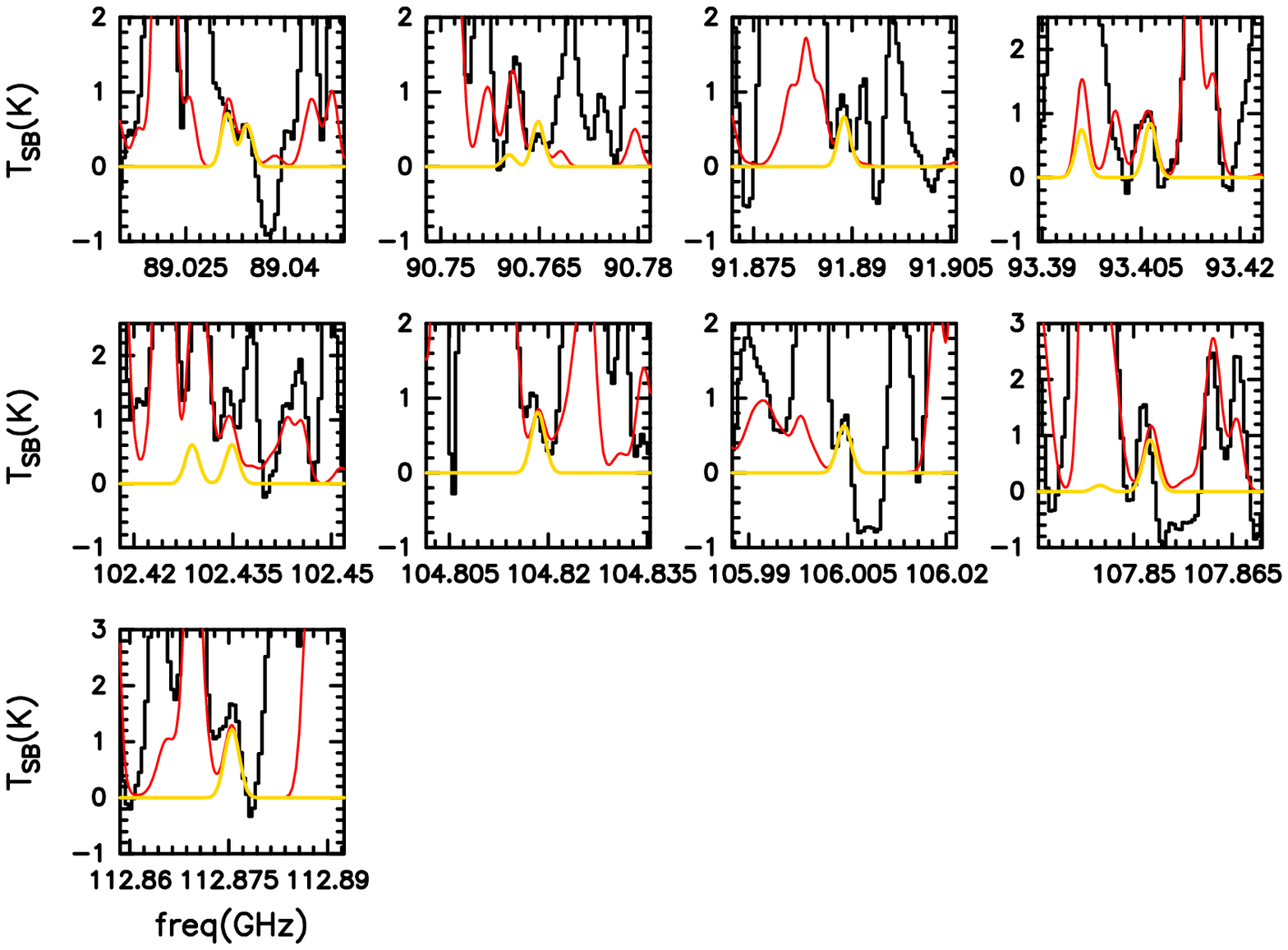}
\caption{Transitions listed in Table~\ref{transitions} and used to fit the $v$=0 state of N-methylformamide (\nm). The gold curve represents the best LTE fit obtained with MADCUBA (Table \ref{table-fitresult-v0}). The red curve shows the simulated spectrum taking into account all the species identified so far in the region.}
\label{fig-res-nm}
\end{figure*}

Figure \ref{fig-res-nm} shows the non-contaminated and slightly contaminated transitions used to fit the ground vibrational state of \nm. Only the transition at 102.434 GHz is slightly contaminated ($\sim$50\% of contamination) with ethylene glycol at the same frequency. This is the first time that \nm\;is detected towards this source.

To fit this molecule we have fixed $T_{\rm ex}$ to 285 K (same value as for its isomer \am), $\varv_{\rm LSR}$ to 96.5 km s$^{-1}$, and FWHM to 7 km s$^{-1}$, because the fit did not converge leaving them free. The best fit of the column density obtained with MADCUBA gives $N$ = (3.7$\pm$1.6)$\times$10$^{16}$ cm$^{-2}$, and $X$ = (3.7$\pm$1.7)$\times$10$^{-9}$.
In this case, the vibrationally excited states were not detected and upper limits to their column densities and abundances are provided (Table \ref{table-fitresult-v0}). In particular, the column density upper limits have been derived like for HNCO $v_{6}$=1, assuming the same $T_{\rm ex}$, $\varv_{\rm LSR}$, and FWHM of \am, and increasing $N$ until the observed spectrum could be reproduced. The derived upper limits are consistent with the column density found for the ground state, within the errors.

\subsubsection{Non detections}
\label{res-ei}

Unlike for the other species, no unblended transitions of \ei, \ur, \cfm, and \gly\;were found. Thus, we have derived upper limits for the column densities of the ground vibrational state.  
All of the transitions of these molecules are contaminated with those of other species, making it difficult to derive upper limits for the column densities. Thus, for \ei\;we have assumed $T_{\rm ex}$, $\varv_{\rm LSR}$, and FWHM derived for \mi\;$v_{\rm b}$=1 (since the ground state is optically thick, see Sect.~\ref{res-mi}), and have increased $N$ to the maximum value compatible with the observed spectrum. We have obtained $N\le$5$\times$10$^{15}$ cm$^{-2}$, and a molecular abundance $X\le$5$\times$10$^{-10}$.
This gives a \mi/\ei\;ratio $>$24, which is consistent with the \mi/\ei\;ratio $>$10 found towards the HMCs Orion KL and Sgr B2 by \citet{kolesnikova2018}. Moreover, for \ur, \cfm, and \gly\;we have assumed $T_{\rm ex}$ = 150 K, $\varv_{\rm LSR}$=96.5 km~s$^{-1}$, and  FWHM = 7 km s$^{-1}$. Also in this case we have increased $N$ to the maximum value compatible with the observed spectrum, and found $N\le$1.6$\times$10$^{14}$ cm$^{-2}$ for \ur, $N\le$3$\times$10$^{15}$ cm$^{-2}$ for \cfm, and $N\le$7$\times$10$^{14}$ cm$^{-2}$ for \gly.

\subsection{Integrated intensity maps}
\label{integrated-maps}

\begin{figure*}
\centering
\hspace*{-3cm}
\includegraphics[width=60pc]{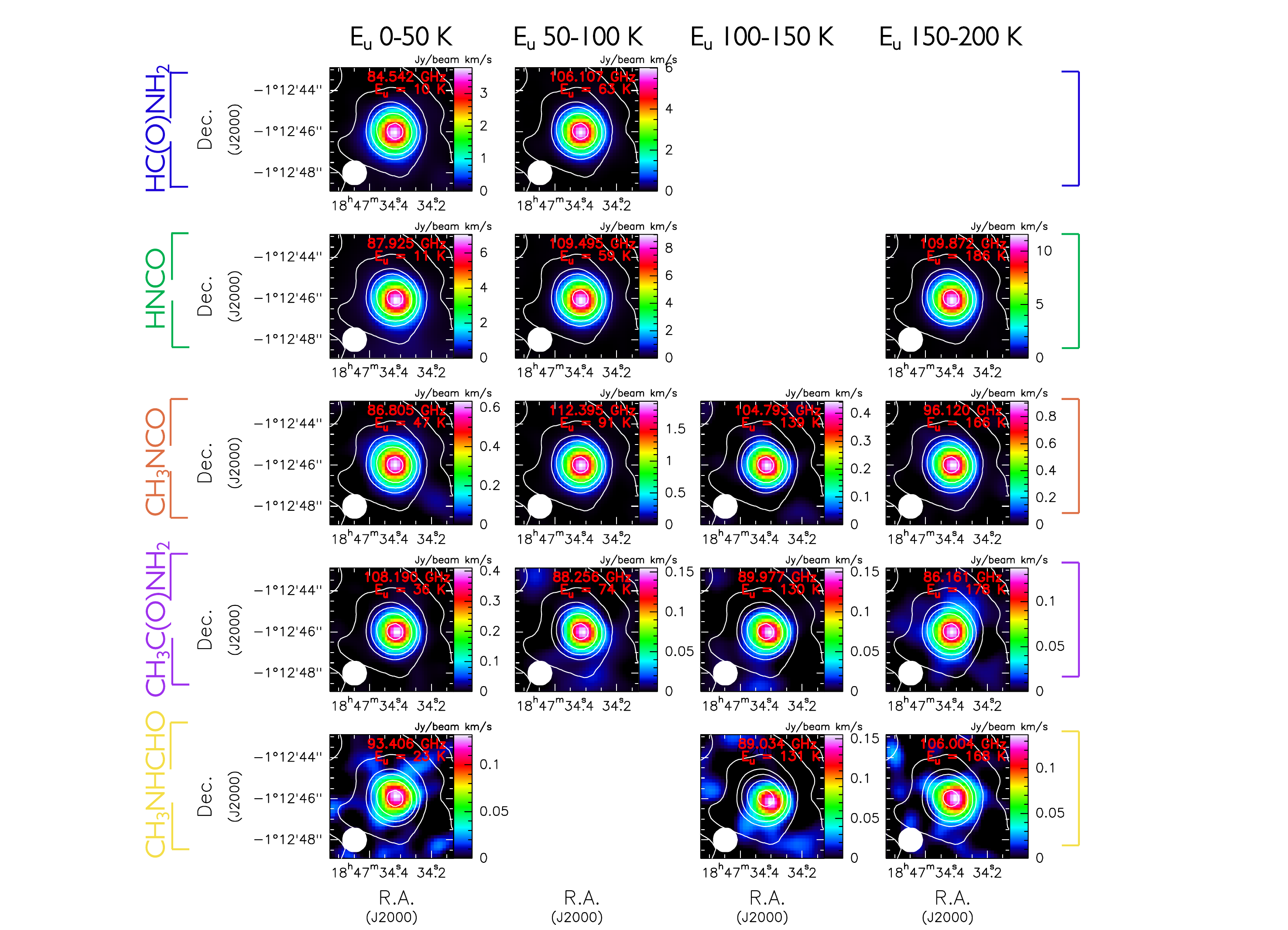}
\caption{ALMA 1\farcs2 resolution integrated emission maps at 3 mm of transitions with different $E_{\rm up}$ (from 0 up to 200 K, different columns) of \fm, \ia, \mi, \am, and \nm\;(different rows) obtained with the GUAPOS survey. In each panel, the white contours show the the continuum emission levels at 5, 10, 20, 40, 60, 100, and 200 times the rms value of 0.8 mJy/beam. The white ellipse in the lower left corner represents the synthesized beam.}
\label{fig-maps-allenergies}
\end{figure*}

\begin{figure*}
\centering
\includegraphics[width=45pc]{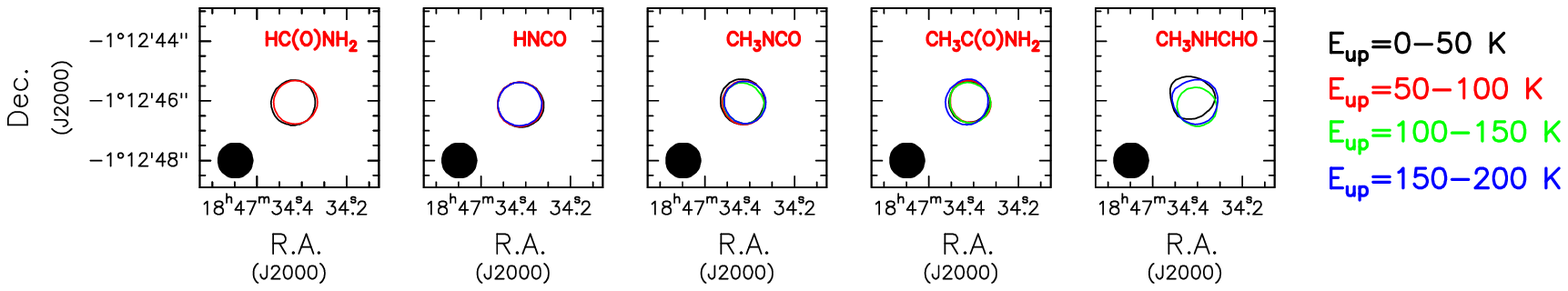}
\caption{Maps of the integrated emission for the molecules studied in this paper. Different colors within the same panel represent the range of upper energy levels of the transitions taken into account. The contours represents 0.5 times the integrated emission peak level of the corresponding map. The beam is indicated in the left-bottom corner of the maps.}
\label{fig-comparison-fluxes}
\end{figure*}

\begin{figure*}
\centering
\includegraphics[width=35pc]{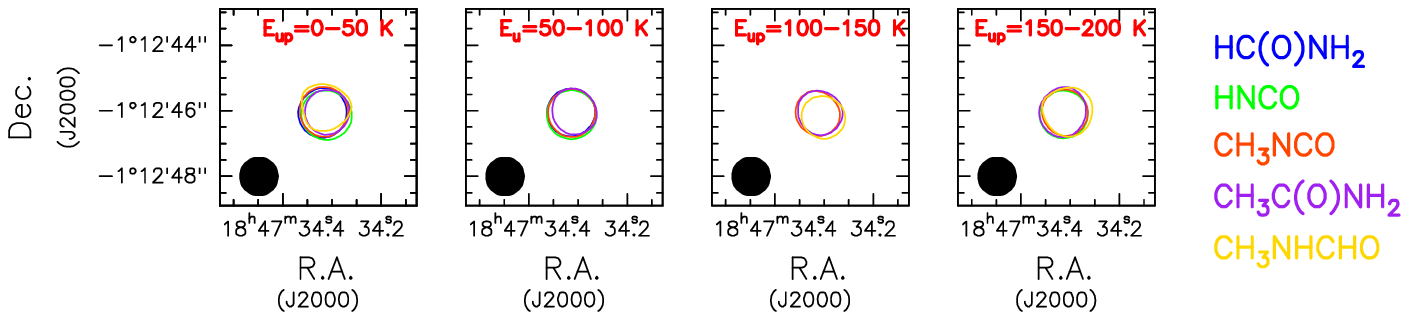}
\caption{Maps of the integrated emission for the molecules studied in this paper. Different colors within the same panel represent different molecules. The contours represents 0.5 times the integrated emission peak level of the corresponding map. The beam is indicated in the left-bottom corner of the maps.}
\label{fig-comparison-molecules}
\end{figure*}

\begin{figure*}
\centering
\includegraphics[width=30pc]{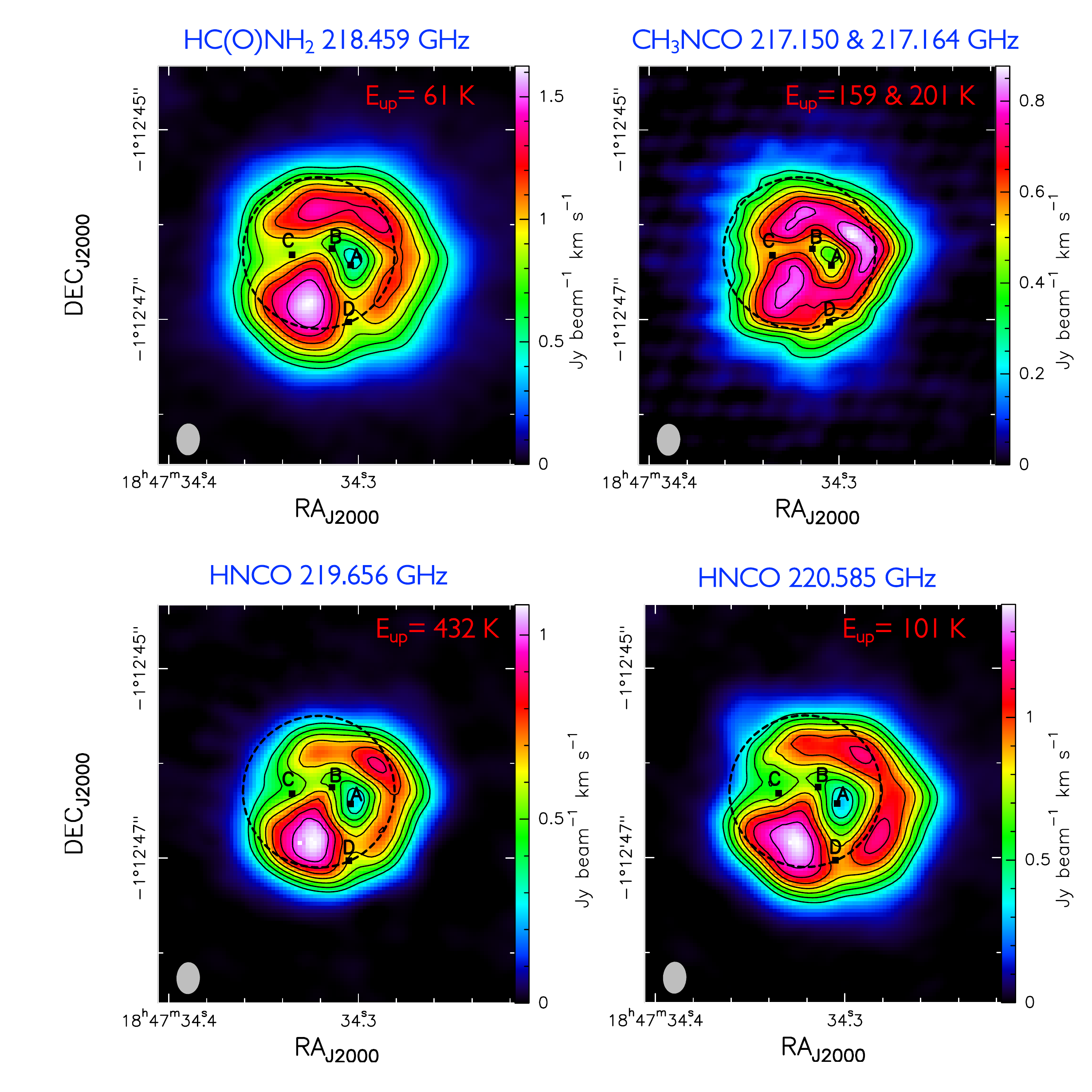}
\caption{ALMA 0\farcs2 resolution integrated emission maps at 1.4 mm of \fm\;(10$_{1,9}$--9$_{1,8}$) (\emph{top-left panel}), \mi\;(25$_{2,23,0}$--24$_{2,22,0}$ and 25$_{-3,-,1}$--24$_{-3,-,1}$) (\emph{top-right panel}), HNCO (10$_{3,8}$--9$_{3,7}$ and 10$_{3,7}$--9$_{3,6}$) (\emph{bottom-left panel}), and HNCO (10$_{1,9}$--9$_{1,8}$) (\emph{bottom-right panel}) in a velocity range between 93 and 100~km s$^{-1}$ for each transition. The contour levels are 0.3, 0.4, 0.5, 0.6, 0.7, 0.8, and 0.9 the maximum value of the maps. The maximum values of the maps are 1.625 Jy beam$^{-1}$ km s$^{-1}$ for \fm, 0.877 Jy beam$^{-1}$ km s$^{-1}$ for \mi, 1.082 Jy beam$^{-1}$ km s$^{-1}$ for HNCO at 219.656 GHz, and 1.396~Jy beam$^{-1}$ km s$^{-1}$ for HNCO at 220.585 GHz.
The black dashed circle indicates in all the panels the area in which the spectrum analysed in this work was extracted, which matches the GUAPOS beam of 1\farcs2 centered at the 3mm continuum peak. The frequency and $E_{\rm up}$ energies of the transitions are shown in blue above each panel, and in red in the top-right corner of each panel, respectively. The synthesized beam is represented by the grey ellipse in the lower left corner. The black squares indicates the position of the continuum sources A, B, C, and D resolved with 7 mm VLA observations by \citet{beltran2021}.}
\label{fig-maps-high-res}
\end{figure*}

Figure \ref{fig-maps-allenergies} shows the 1\farcs2 resolution integrated intensity maps of the 3mm GUAPOS survey of the most unblended transitions of \ia, \fm, \mi, \am, and \nm, with different values of the upper energy level ($E_{\rm up}$). The velocity range used for the integrated intensity maps goes from 93 to 100 km s$^{-1}$. \ei, \ur, \cfm, and \gly\;are not included in the figure because they were not detected, as explained in Sect.~\ref{res-ei}. The transitions of HNCO overlap with those of HN$^{13}$CO, and therefore both species contribute to the integrated emission maps.

These integrated intensity maps have been obtained from the final cubes after the calibration carried out with the CASA\footnote{https://casa.nrao.edu} (Common Astronomy Software Applications) package (\citealt{mcmullin2007}). In particular, for each transitions we have cropped the cube to frequencies $\pm$20 MHz around the rest frequency. Then, in the GUAPOS spectrum we have identified the channels in which the line intensity is zero and we have used these spectral windows to subtract the continuum pixel by pixel. Finally, we obtained the integrated intensity maps from the four channels around the rest frequencies. In fact, taking into account the spectral resolution of $\sim$0.48 MHz, the width of four channels at $\sim$90 GHz corresponds to 6.5 km s$^{-1}$, comparable to the FWHM of the molecular lines in this source. 
All these operations were made with the MADCUBA software.

The emission of the molecular species studied here arises entirely from the HMC, and comes from a region of $\sim$2\asec\;($\sim$ 7500 au, see Fig.~\ref{fig-maps-allenergies}). 
Moreover, the derivation of column densities from transitions at different energies is reasonable when their emission comes from the same region. Thus, we have compared the emission of the different molecules and of different range of energies for the transitions studied in this work. 
Figure~\ref{fig-comparison-fluxes} shows the comparison between the emitting region of different upper energy levels of the same molecule taken from maps of Fig.~\ref{fig-maps-allenergies}. Moreover, Fig.~\ref{fig-comparison-molecules} shows the comparison among transitions with similar upper energy levels of different molecular species. From these figures it is clear that the emission of different transitions arises from the same region regardless of the molecule and of the upper energy level. Furthermore, we found a similar result when comparing different molecules. Only N-methylformamide (\nm) presents some shifts that are not significant since they are smaller than the beam size of 1\farcs2.
Thus, we can conclude that the column densities derived from transitions at different upper energy levels and the ratios derived between molecules are not affected by a different spatial distribution of different transitions or molecules.

Note, however, that this similar spatial distribution could be in part due to the fact that the emission is only partially resolved at an angular resolution of 1\farcs2. Therefore, to test this and eventually unveil spatial differences unresolved at 1\farcs2 resolution, we have also analysed higher angular resolution (0\farcs2) data at 1.4 mm obtained with ALMA. Figure~\ref{fig-maps-high-res} shows the integrated emission maps of the ground vibrational state $J_{K_{\rm a},K_{\rm c}}$ = 10$_{1,9}$--9$_{1,8}$ rotational transition of \fm, of the ground vibrational state $J_{K_{\rm a},K_{\rm c}}$ = 10$_{3,8}$--9$_{3,7}$, 10$_{3,7}$--9$_{3,6}$\footnote{It is a doublet transition that is not resolved in frequency because the line width ($\sim$8 km s$^{-1}$) is higher than the spectral separation between the two lines (1.8$\times$10$^{-3}$ km s$^{-1}$ at 219.656 GHz).}, and 10$_{1,9}$--9$_{1,8}$ rotational transitions of HNCO, and of the ground vibrational state $J_{K_{\rm a},K_{\rm c},m}$ = 25$_{2,23,0}$--24$_{2,22,0}$ and the $J_{K,m}$ = 25$_{-3,1}$--24$_{-3,1}$ rotational transitions of \mi. The map of the latter molecule was obtained by combining two transitions because they emit at similar frequencies and have a similar upper energy level of about 160-200 K. Note that the $E_{\rm up}$ chosen for each of these transitions was comparable to that used for the analysis and the results of the 1\farcs2 resolution data. Moreover, we have also selected a higher $E_{\rm up}$ transition of HNCO (432 K) to show possible spatial differences. In this case, despite the different $E_{\rm up}$ of the two HNCO levels (101 K and 432 K), both transitions fill the beam of the GUAPOS observations (bottom panels of Fig.~\ref{fig-maps-high-res}).

We did not find strong and unblended transitions of \am\;and \nm\;in the narrow spectral bands of the high-angular resolution observations at 1.4 mm. Thus, these two molecules will not be discussed in the following of this section.

The 0\farcs2 resolution integrated intensity maps show that the overall behaviour at 1.4 mm of these molecules (\ia, \fm, and \mi) is quite similar. One of the main differences is that the maps at higher $E_{\rm up}$ (159-201 K, and 432 K, of \mi\;and \ia, respectively) present a more compact structure than the lower energy ones (61 and 101 K, of \fm\;and \ia, respectively). This is expected because of the temperature gradient present towards the HMC, with higher temperatures to the center with respect to the outer part as already discussed in Sect.~\ref{analysis}. A striking feature of the integrated emission at high angular resolution is its ring-like morphology. This spatial distribution is similar to that traced by other COMs, such as CH$_{3}$CN and CH$_{3}$OCHO, observed with the same angular resolution of 0\farcs2 (\citealt{beltran2018}). 
The explanation for this morphology is that most of the emission comes from a  rotating and infalling toroid surrounding a small protocluster of 4 massive protostars (indicated as black squares and named A, B, C, and D in Fig.~\ref{fig-maps-high-res}, see also \citealt{beltran2021}). Since the material is flowing inwards and locally the continuum temperature is higher than the $T_{\rm ex}$ of the molecules, the gas is seen in red-shifted absorption towards the center (see \citealt{beltran2018} for more details), and the integrated emission shows this characteristic ring structure.

The high-angular resolution maps at 1.4 mm show that the molecular emission arise from the whole HMC, filling the 1\farcs2 beam of GUAPOS (black dashed line in Fig.~\ref{fig-maps-high-res}). 
The column densities obtained from a region of 1\farcs2 in the high-angular resolution maps are consistent to those derived from the GUAPOS data. In fact, if we fix $T_{\rm ex}$, $\varv_{\rm LSR}$ and FWHM to the values derived at 3 mm (Table \ref{table-fitresult-v0}), the column densities
of the three molecules obtained at 1.4 mm are consistent
within a factor of 2 with those obtained at 3 mm. In particular, from 1.4 mm observations, $N$(\fm)=(2.5$\pm$1.9)$\times$10$^{16}$ cm$^{-2}$, $N$(\ia)=(1.12$\pm$0.04)$\times$10$^{17}$ cm$^{-2}$, $N$(\ia)=(9.1$\pm$1.8)$\times$10$^{16}$ cm$^{-2}$, from the transitions at 219.656 and 219.656 GHz, respectively, and $N$(\mi)=(6.9$\pm$1.8)$\times$10$^{16}$ cm$^{-2}$. This mean that both observations (at 1.4 and 3 mm), despite of the different angular resolution, are sensitive to approximately the same gas.

\section{Discussion}
\label{discussion}

To obtain a complete overview of the chemical processes that lead to the formation of peptide-like bond molecules in the ISM, we have compared the results obtained in G31 with those in other interstellar sources. Comparisons have been made with works containing at least the detection of \fm\;together with that of \mi\;and/or \am. These sources are the HMCs in the GC Sgr B2(N) (\citealt{belloche2013}), Sgr B2(N2) (\citealt{belloche2017}), Sgr B2(N1S) (\citealt{belloche2019}), Sgr B2(N3) and Sgr B2(N5) (\citealt{bonfand2019}), the HMCs in the Galactic disk G10.47+0.03 (hereafter G10.47), Orion BN/KL A and B, and NGC 6334I (\citealt{gorai2020}; \citealt{cernicharo2016}; \citealt{ligterink2020}, respectively), the hot core precursor G328.2551-0.5321 (hereafter G328.2551) A, B (related to accretion shock positions), and envelope (hereafter env) positions (\citealt{csengeri2019}), the hot corinos IRAS 16293-2422 (hereafter IRAS 16293) A and B (\citealt{martin-domenech2017}; \citealt{ligterink2017}; \citealt{ligterink2018}; \citealt{manigand2020}), and the GC molecular cloud G+0.693-0.027 (hereafter G+0.693, \citealt{zeng2018}).
The Sgr B2(N) and G+0.693 observations were taken with different single-dish telescope (IRAM 30m and Green Bank Telescope) while the rest were observed using ALMA. Moreover, we have also compared the column densities with those of \ia\;and \fm\;and the upper limits of \mi\;and \am\;estimated in the comet 67P/Churyumov-Gerasimenko (hereafter 67P) by the ROSINA experiment on ESA’s Rosetta mission reported by \citet{altwegg2017}.

\subsection{Molecular abundances}
\label{disc-ab}

\begin{figure*}
\centering
\includegraphics[width=45pc]{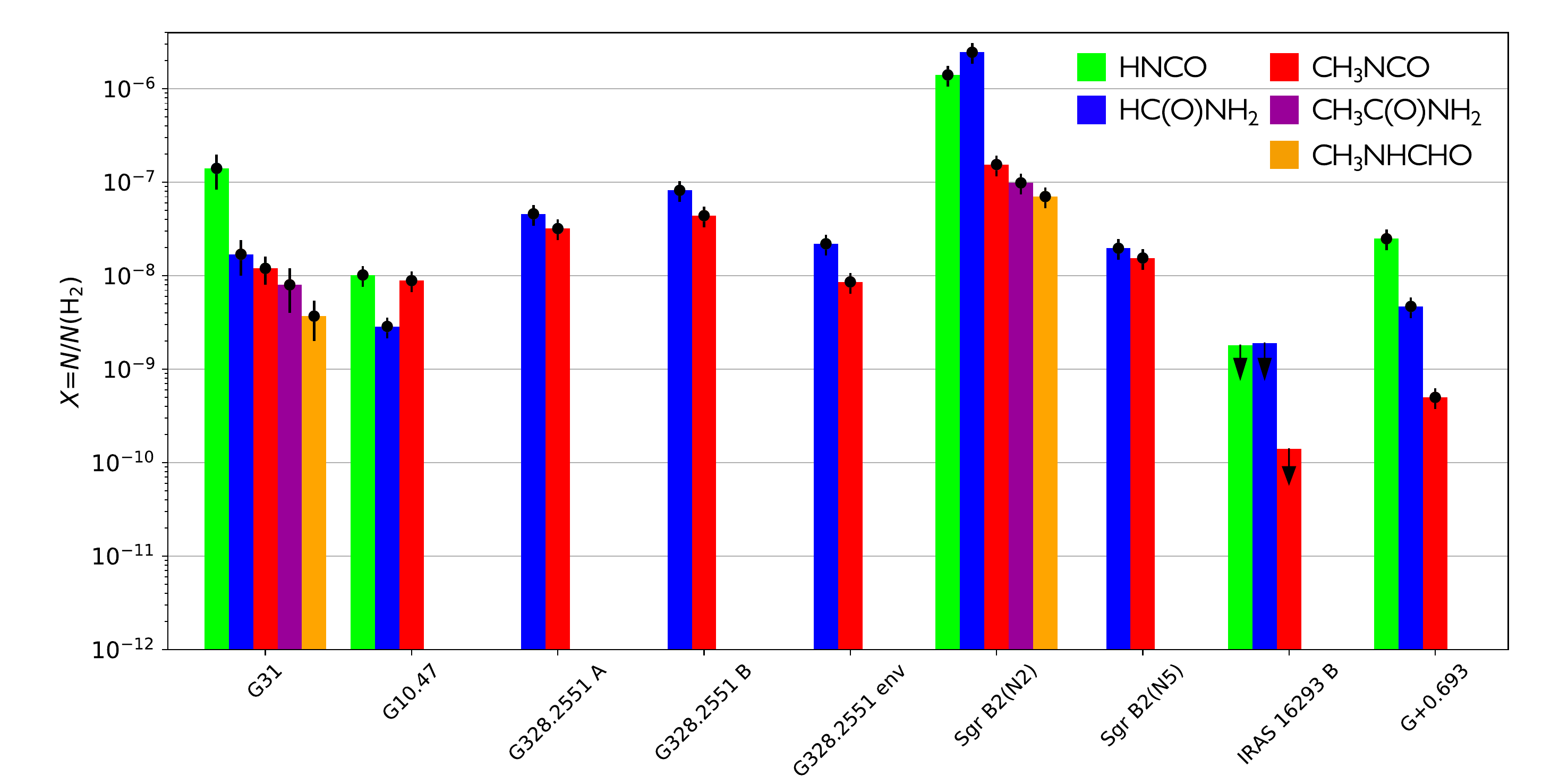}
\caption{Molecular abundances ($X$) with respect to H$_{2}$ towards G31, G10.47, G328.2551 A, B, and env, Sgr B2(N2), Sgr B2(N5), IRAS 16293 B (upper limits), and G+0.693. Different colors represent the different molecules for which the abundances are shown, as indicated in the legend in the upper-right corner. 
Data are taken from: G31 (this work), G10.47 (\citealt{gorai2020}), G328.2551 (\citealt{csengeri2019}), Sgr B2(N2) (\citealt{belloche2017}), Sgr B2(N5) (\citealt{bonfand2019}), IRAS 16293 B (\citealt{martin-domenech2017}), and G+0.693 (\citealt{zeng2018}) (from the left to the right).}
\label{fig-comp-ab}
\end{figure*}

\begin{figure*}
\centering
\includegraphics[width=40pc]{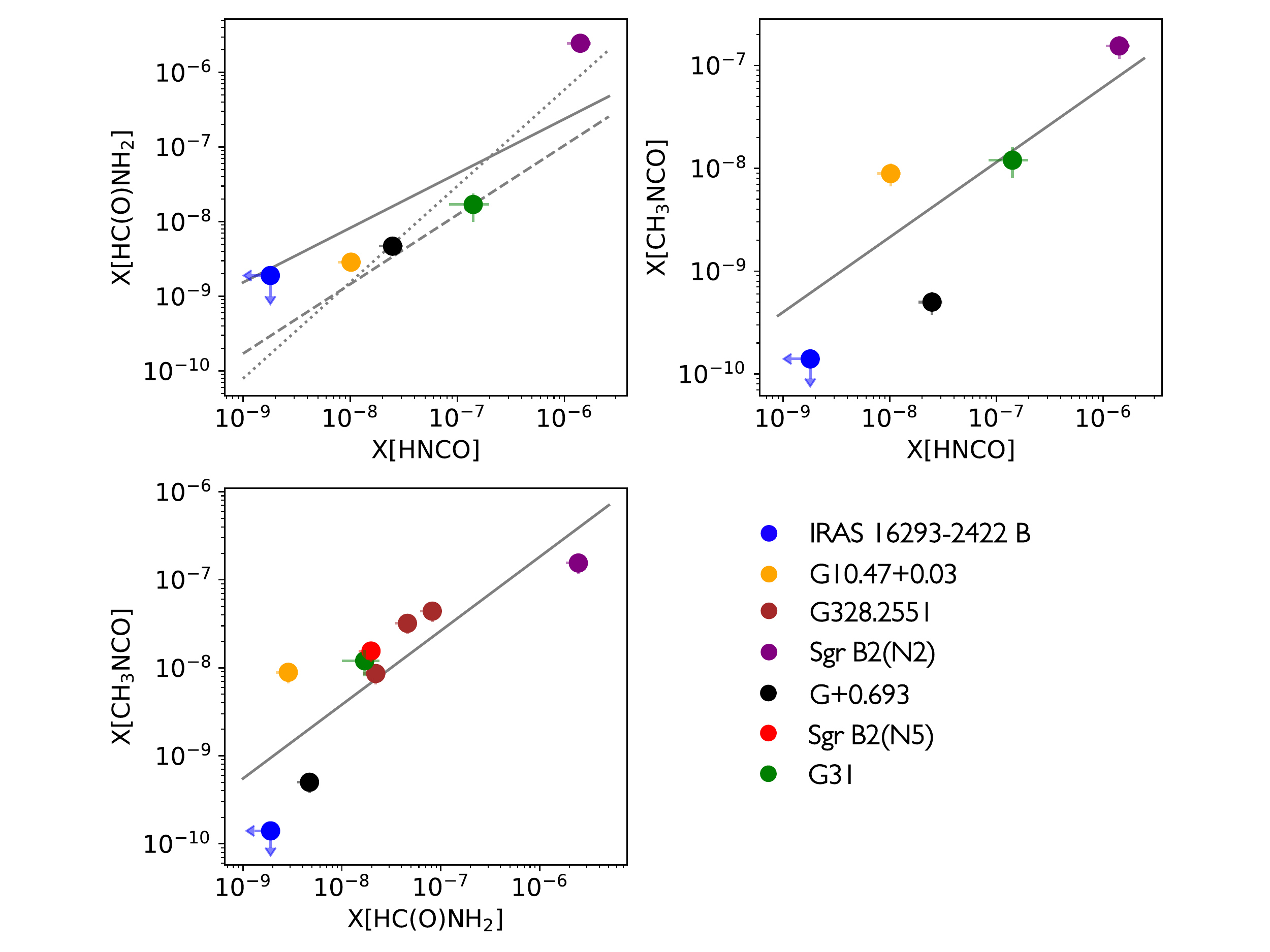}
\caption{\emph{Upper-left panel}: $X$[\fm] as a function of $X$[\ia]. The solid line is the power-law fit obtained in this work, while the dashed and pointed lines are those obtained by \citet{lopez-sepulcre2015} and \citet{quenard2018}, respectively. \emph{Upper-right panel}: $X$[\mi] as a function of $X$[\ia]. The solid line is the power-law fit obtained in this work. \emph{Lower panel}: $X$[\mi] as a function of $X$[\fm]. The solid line is the power-law fit obtained in this work.
In all the panels, the different colors represent the different sources, as indicated in the bottom-right legend. Data are taken from the same works indicated in Fig.~\ref{fig-comp-ab}.}
\label{fig-corr-ab}
\end{figure*}

Six of the works listed above provide an estimate of $N$(H$_{2}$) and make it possible to compare with the molecular abundances obtained in this work. Figure \ref{fig-comp-ab} shows the molecular abundances derived in G31 and in the other regions. For abundances whose errors were not provided we assumed a value of 25\%. We note however that the $N$(H$_{2}$) used for IRAS 16293 B should be considered as a lower limit since dust might be optically thick (e.g. \citealt{jorgensen2016}, and thus the abundances for this source should be taken as upper limits.
At a glance, we can see that G31, in addition to Sgr B2(N2), is the only other source in which all of the five molecules (\ia, \fm, \mi, \am, and \nm) have been detected. The behaviour of these two HMCs looks similar, except for \fm, which shows an over-abundance towards Sgr B2(N2). 
In most of the sources \ia\;is the most abundant species, followed by \fm, \mi, \am, and \nm. The only exceptions are IRAS 16293 B and Sgr B2(N2), whose HNCO and \fm\;abundances are similar between them, and G10.47, for which \mi\;is more abundant than \fm. Moreover, both \ia\;and \fm\;abundances towards G31 are comparable with those of the GC molecular cloud G+0.693 (\citealt{zeng2018}). \fm\;and \mi\;abundances together are similar to those derived towards the HMCs G10.47 and Sgr B2(N5), and the HMC precursor G328.2551 (\citealt{gorai2020}; \citealt{bonfand2019}; \citealt{csengeri2019}), while only the \fm\;abundance is similar to that of the low-mass protostar IRAS 16293 B (\citealt{martin-domenech2017}). All these similarities are within a factor of 4, and should be taken with caution since possible opacity effects of HNCO and \fm\;might have affected the derived abundances towards some of these sources (see Sect.~\ref{results}). For G31 this is not the case since column densities, and thus abundances, have been derived from the $^{13}$C-isotopologues or the vibrationally excited states, which are optically thin (see Table \ref{transitions}).

The high molecular abundances found in HMCs, hot corinos, and G+0.693 (10$^{-10}$--10$^{-6}$) are consistent with their formation through pathways on the surface of dust grains at earlier phases, and subsequent desorption, induced by thermal heating (for HMCs and hot corinos) or by grain sputtering produced by shocks (for G+0.693 and the accretion shocks G328.2551 A and B, \citealt{zeng2020}; \citealt{csengeri2019}). In fact, in absence of efficient desorption mechanisms, the gas-phase abundances are expected to be lower, as occurs in low-mass pre-stellar cores such as L1544 where the upper limits for the abundances of \fm\;and \mi\;that have been reported are very low ($X$(\fm)$<$8.7$\times$10$^{-13}$ and $X$(\mi)$<$4.2$\times$10$^{-11}$, \citealt{jimenez-serra2016}).

Figure~\ref{fig-corr-ab} shows the comparison between pairs of molecular abundances, \fm\;and \ia, \mi\;and \ia, and \mi\;and \fm. It is already known from previous observations that there is a correlation between HNCO and \fm\;(\citealt{lopez-sepulcre2015,lopez-sepulcre2019}; \citealt{allen2020}). In the top-left panel of Fig.~\ref{fig-corr-ab} the best power-law fits derived by \citet{lopez-sepulcre2015}, $X$[\fm]=0.04$\times$ $X$[\ia]$^{0.93}$, and by \citet{quenard2018}, $X$[\fm]=32.14$\times$ $X$[\ia]$^{1.29}$, are compared with the one derived here, $X$[\fm]=0.006$\times$ $X$[\ia]$^{0.73}$ (with a Pearson coefficient of 0.99 and a P-value $<$0.05, indicating a strong positive correlation). Thus, the sample of sources discussed in this work, which also includes HMCs and a shock-dominated molecular cloud, is in agreement with the correlation found previously for low- and intermediate-mass pre-stellar and protostellar objects, which holds across several orders of magnitude in abundance. Based on this tight correlation, it has been proposed that the two species are chemically related and that the formation of \fm\;might occur through H-addition to solid-phase HNCO (e.g. \citealt{tielens1982}; \citealt{charnley2004}). Experimental works first suggested that this process is not efficient (\citealt{noble2015}; \citealt{fedoseev2015}), while recent works revised this possibility and found that a correlation between these two molecular species can be understood by H-abstraction and addition reactions (e.g. \citealt{nguyen2011}; \citealt{haupa2019}; \citealt{suhasaria2020}). Moreover, hydrogenation of NO combined with UV-photon exposure and radical-radical reactions on grains has been suggested as the main formation pathways for both \ia\;and \fm\;(e.g. \citealt{jones2011}; \citealt{fedoseev2016}; \citealt{ligterink2018}; \citealt{dulieu2019}). \citet{coutens2016} found that the deuteration  (D/H ratio) of \fm\;in IRAS 16293 B is similar to that of HNCO, in agreement with the hypothesis that both species are chemically related via grain-surface reactions. Gas-phase formation routes have also been proposed (see e.g. NH$_{2}$ + H$_{2}$CO, \citealt{barone2015}; \citealt{skouteris2017}).
Laboratory experiments by \citet{martin-domenech2020} show that both \ia\;and\;\fm\;could form upon UV photoprocessing or electron irradiation of ice samples, indicating that energetic processing (like UV photons and cosmic rays) of ISM CO-rich ices could form both species, without the need of a chemical link and/or a similar precursor between the two.
This was predicted by the chemical modelling of \citet{quenard2018}, who showed that the formation of \fm\;at different temperature regimes is governed by different chemical processes. While at low temperatures the formation of \fm\;is driven by gas-phase formation via the reaction NH$_{2}$ + H$_{2}$CO $\rightarrow$ \fm\; + H, at high temperature its formation occurs on the surface of dust grains via radical-radical addition reactions. Moreover, they showed that for HNCO grain-surface and gas-phase reactions are equally efficient at low temperature, while at high temperatures the gas-phase formation predominates and the small fraction formed on grains is released into the gas phase via thermal desorption. \citet{rimola2018} also showed via theoretical quantum chemical computations that \fm\;can form on grain surfaces starting from CN, which can quickly react with water-rich amorphous ices. 
Thus, the correlation between \ia\;and \fm\;is mainly due to a similar response to the temperature of the two molecules, and not to a direct chemical link. In fact, the increase of the temperature triggers processes on the ice-mantle of grains, such as thermal evaporation. Moreover, as discussed above, other processes, like UV photons, cosmic rays and shocks, could help both on the formation of these molecules on grain surfaces and on their desorption in the gas.

Similarly to the HNCO vs. \fm\;relation, we have found similar correlations (Fig.~\ref{fig-corr-ab}) between \mi\;and \ia, $X$[\mi]~=~0.0015$\times$ $X$[\ia]$^{0.73}$, and between \mi\;and\;\fm, $X$[\mi]=0.02$\times$ $X$[\fm]$^{0.84}$ (both with a Pearson coefficient of 0.99 and a P-value $<$0.05, indicating strong positive correlations), suggesting links also between \mi\;and \ia, and \fm. A correlation between \mi\;and \ia\;was already suggested by \citet{ligterink2021}, who found that the \mi\;and \ia\;column density ratio is almost constant in different regions. It was proposed by \citet{halfen2015} that CH$_{3}$NCO could form in gas-phase through HNCO from the reaction:
\begin{equation}
\label{HNCO-CH3}
{\rm HNCO} + {\rm CH}_{3} \rightarrow \mieq + {\rm H},
\end{equation}
or from
\begin{align}
&{\rm HNCO} + {\rm CH}_{5}^{+} \rightarrow {\rm CH}_{3}{\rm NCOH}^{+} + {\rm H}_{2} \\
&{\rm CH}_{3}{\rm NCOH}^{+} + {\rm e}^{-} \rightarrow \mieq + {\rm H}.
\end{align}
\citet{cernicharo2016} found a similar spatial distribution for \ia, \fm\;and \mi\;towards Orion BN/KL, as observed for G31 in this work (Fig.~\ref{fig-maps-high-res}), and suggested reaction \eqref{HNCO-CH3} as a possible grain-surface reaction to form \mi. Moreover, the HNCO/\mi\;abundance ratio found for G31 of 12$\pm$6 is consistent with the range of values predicted by the HMC model of \citet{belloche2017}, who proposed \mi\;grain-surface formation and ice sublimation during the warm-up phase (e.g. through thermal desorption as suggested for \ia\;and \fm). Formation of \mi\;through \ia\;and methane has also been considered on ices, where favourable thermodynamic conditions could be created (e.g. reduction of the energy barrier, \citealt{cassone2021}). More recently \citet{majumdar2018} found that reaction \eqref{HNCO-CH3} is endothermic, and they proposed alternative routes for the formation of methyl-isocyanate on grains:
\begin{align}
\label{}
&{\rm CH}_{3} + {\rm OCN} \rightarrow \mieq,\\
&{\rm N} + {\rm CH}_{3}{\rm CO} \rightarrow {\rm CH}_{3}{\rm C(N)O} \rightarrow \mieq,
\end{align}
or through the HCN$\cdots$CO van der Waals complex
\begin{align}
\label{}
&{\rm H} + {\rm HCN \cdots CO} \rightarrow {\rm H}_{2}{\rm CN \cdots CO} \rightarrow {\rm H}_{2}{\rm CNCO} \\
&{\rm H} + {\rm H}_{2}{\rm CNCO} \rightarrow \mieq.
\end{align}
Interestingly a similar process has also been found to be important for the formation of HNCO:
\begin{equation}
\label{}
{\rm N \cdots CO} + {\rm H} \rightarrow {\rm HNCO},
\end{equation}
indicating that a possible link between the two species could be the van der Waals complexes involving CO.
Finally, the correlation between \mi\;and \fm\;(bottom panel of Fig.~\ref{fig-corr-ab}) is probably due to the fact that both molecules form on grains and are desorbed on gas-phase because of similar physical effects. For example, both molecules are expected to be already efficiently thermally desorbed at the high temperatures of HMCs ($>$100 K). In fact, temperature-programmed desorption experiments show that the \fm\;peak desorption temperature is around 200 K (\citealt{ligterink2018}), while that of \mi\;is around 150 K (\citealt{ligterink2017}). Thus, we expect most of both molecules to have been already released back to the gas-phase in hot cores. 

Thus, a strong correlation between two molecules in a sample of sources does not directly imply that these molecules are chemically related (e.g. \citealt{belloche2020}). Whether these results are a consequence of a direct chemical link or an effect caused by similar chemical responses to physical conditions cannot be firmly concluded yet, and more dedicated physico-chemical models are needed to disentangle all of the possible effects.

\subsection{Abundances with respect to CH$_{3}$OH}
\label{ab-meth}

\begin{figure*}
\centering
\includegraphics[width=43pc]{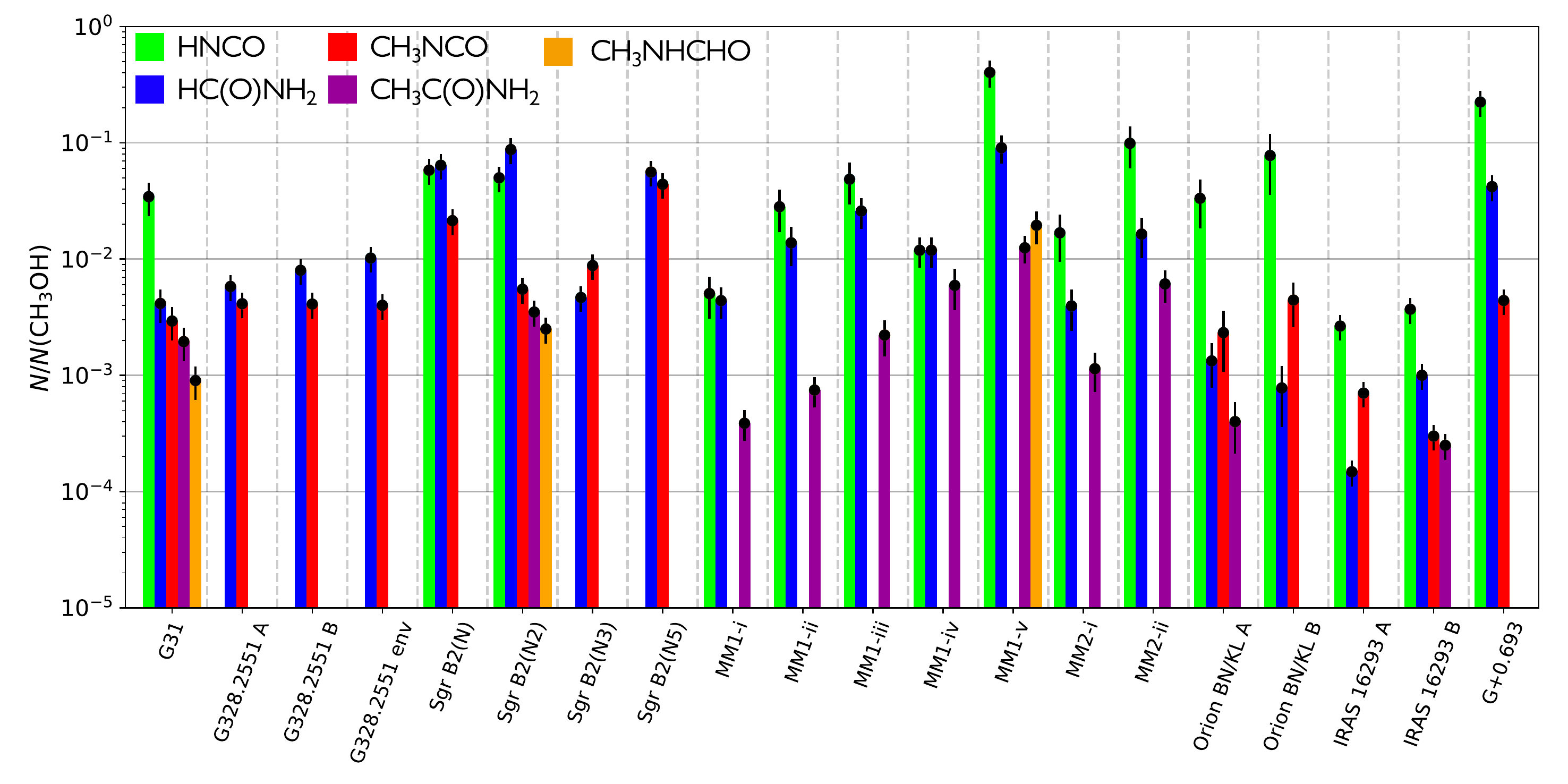}
\caption{Molecular abundances with respect to CH$_{3}$OH towards G31, G328.2551 A, B, and env, Sgr B2(N), Sgr B2(N2), Sgr B2(N3), Sgr B2(N5), NGC 6334I MM1-i--v and MM2-i--ii, Orion BN/KL A and B, IRAS 16293 A and B, and G+0.693. Different colors represent the different molecules for which the abundances are shown, as indicated in the legend in the upper-left corner. 
Data are taken from: G31 (this work, Mininni et al. in prep.), G328.2551 (\citealt{csengeri2019}), Sgr B2(N) (\citealt{belloche2013}), Sgr B2(N2) (\citealt{belloche2016}; \citealt{belloche2017}), Sgr B2(N3) and Sgr B2(N5) (\citealt{bonfand2019}), NGC 6334I (\citealt{bogelund2018}; \citealt{ligterink2020}), Orion BN/KL A and B (\citealt{cernicharo2016}), IRAS 16293 A (\citealt{ligterink2017}; \citealt{manigand2020}), IRAS 16293 B (\citealt{ligterink2017,ligterink2018}; \citealt{jorgensen2018}), and G+0.693 (\citealt{zeng2018}; \citealt{rodriguez-almeida2021}) (from the left to the right).}
\label{fig-comp-ab-meth}
\end{figure*}

\begin{figure*}
\centering
\includegraphics[width=42pc]{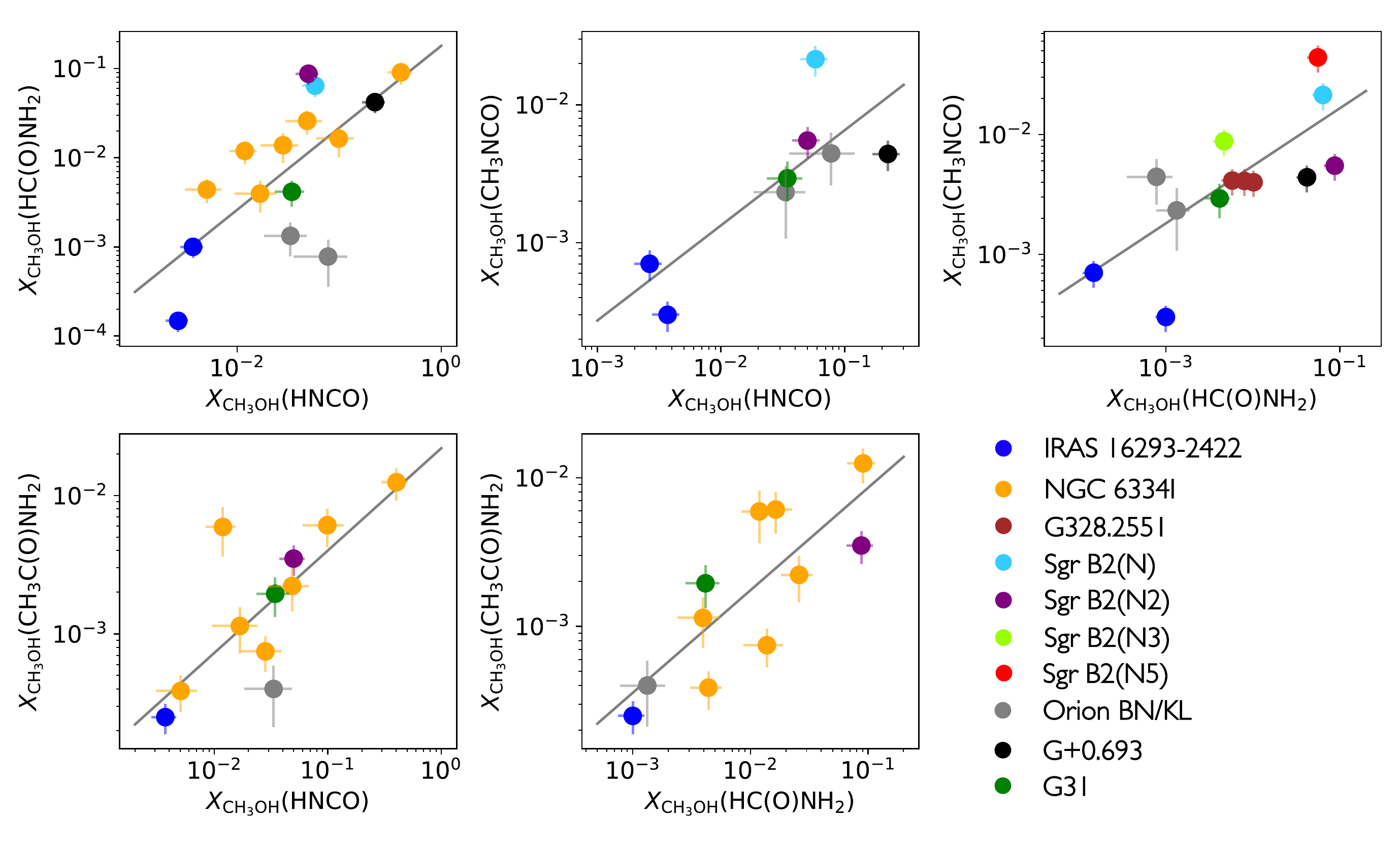}
\caption{\emph{Upper-left panel}: $X_{{\rm CH}_{3}{\rm OH}}$(\fm) as a function of $X_{{\rm CH}_{3}{\rm OH}}$(\ia). \emph{Upper-central panel}: $X_{{\rm CH}_{3}{\rm OH}}$(\mi) as a function of $X_{{\rm CH}_{3}{\rm OH}}$(\ia). \emph{Upper-right panel}: $X_{{\rm CH}_{3}{\rm OH}}$(\mi) as a function of $X_{{\rm CH}_{3}{\rm OH}}$(\fm). \emph{Bottom-left panel}: $X_{{\rm CH}_{3}{\rm OH}}$(\am) as a function of $X_{{\rm CH}_{3}{\rm OH}}$(\ia). \emph{Bottom-central panel}: $X_{{\rm CH}_{3}{\rm OH}}$(\am) as a function of $X_{{\rm CH}_{3}{\rm OH}}$(\fm).
In all the panels, the solid lines are the power-law fit obtained in this work, and the different colors represent the different sources, as indicated in the bottom-right legend. Data are taken from the same works indicated in Fig.~\ref{fig-comp-ab-meth}.}
\label{fig-corr-ab-meth}
\end{figure*}

In this section we have performed an analysis similar to that shown in Sect.~\ref{disc-ab} comparing the abundances derived with respect to methanol, CH$_{3}$OH. All the sources are included, except Sgr B2 (N1S) for which the CH$_{3}$OH column density is not found in the literature, and G10.47 for which whose CH$_{3}$OH column density has been derived from Submillimeter Array observations (\citealt{rolffs2011}) and not from ALMA observations as for the rest of the sources. 
Thus, these two sources are excluded from the discussion in this section. Moreover, for G31, G+0.693, Sgr B2(N), NGC 6334I, and IRAS 16293 A and B, $N$(CH$_{3}$OH) has been derived from the optically thin isotopologues $^{13}$CH$_{3}$OH (for G31, Mininni et al. in prep) and CH$_{3}^{18}$OH (for the other sources, \citealt{rodriguez-almeida2021}; \citealt{belloche2013}; \citealt{bogelund2018}; \citealt{manigand2020}; \citealt{jorgensen2018}), after taking into account the \cratio\;and $^{16}$O/$^{18}$O ratios corrections as a function of the galactocentric distance (\citealt{yan2019} for G31; \citealt{wilson1994} for G+0.693, Sgr B2(N), and IRAS 16293 B; \citealt{wilson1999} for NGC 6334I and IRAS 16293 A). Conversely, $N$(CH$_{3}$OH) for Sgr B2(N2), Sgr B2(N3), Sgr B2(N5), Orion BN/KL and G328.2551 has been obtained from the main isotopologue and could be affected by line opacity effects (\citealt{belloche2016}; \citealt{bonfand2019}; \citealt{cernicharo2016}; \citealt{csengeri2019}).

Figure \ref{fig-comp-ab-meth} shows the abundances with respect to CH$_{3}$OH ($N$/$N$(CH$_{3}$OH)). First of all, G328.2551 and Sgr B2(N3) present similar $N$/$N$(CH$_{3}$OH) for \fm\; and \mi\;with respect to G31, and Sgr B2(N2) presents similar $N$/$N$(CH$_{3}$OH) for all the molecules except for \fm. Except for \ia, Sgr B2(N) and Sgr B2(N5) present higher column density ratios than G31 with respect to CH$_{3}$OH. Moreover, NGC 6334I has similar $N$/$N$(CH$_{3}$OH) with respect to G31 towards all the positions, except for MM1-v for which higher ratios have been found for \ia, \fm, and \nm. Orion BN/KL A and B have ratios similar to those of G31 for all the peptide-like bond molecular species, while IRAS 16293 A and B have lower $N$/$N$(CH$_{3}$OH) ratios. Finally, with respect to G31, G+0.693 presents higher ratios for \ia\;and \fm, and a similar one for \mi.
It should be noted that all the similarities are within a factor of 4.

The general trend is that most of the sources show similar $N$/$N$(CH$_{3}$OH) ratios, except for IRAS 16293 A and B, which present lower values, and Sgr B2(N), Sgr B2(N5), and G+0.693 that show higher values with respect to G31 and the other sources. Moreover, on average the $N$/$N$(CH$_{3}$OH) of \ia\;is the highest, followed by that of \fm, \mi, \am, and \nm, similar to what was found for the abundances derived with respect to $N$(H$_{2}$).

In Fig.~\ref{fig-corr-ab-meth} we show the comparison between pairs of $N$/$N$(CH$_{3}$OH) ($X_{{\rm CH}_{3}{\rm OH}}$). In particular the best power-law fits are:
\begin{align}
\label{}
& X_{{\rm CH}_{3}{\rm OH}}({\rm HC(O)NH}_{2})=0.18\times X_{{\rm CH}_{3}{\rm OH}}({\rm HNCO})^{0.92}, \\
& X_{{\rm CH}_{3}{\rm OH}}({\rm CH}_{3}{\rm NCO})=0.032\times X_{{\rm CH}_{3}{\rm OH}}({\rm HNCO})^{0.69}, \\
& X_{{\rm CH}_{3}{\rm OH}}({\rm CH}_{3}{\rm NCO})=0.05\times X_{{\rm CH}_{3}{\rm OH}}({\rm HC(O)NH}_{2})^{0.48}, \\
& X_{{\rm CH}_{3}{\rm OH}}({\rm CH}_{3}{\rm C(O)NH}_{2})=0.022\times X_{{\rm CH}_{3}{\rm OH}}({\rm HNCO})^{0.74}, \\
& X_{{\rm CH}_{3}{\rm OH}}({\rm CH}_{3}{\rm C(O)NH}_{2})= 0.042\times X_{{\rm CH}_{3}{\rm OH}}({\rm HC(O)NH}_{2})^{0.69}. 
\end{align}
Thus, also in this case we have found positive correlations between \fm\;and \ia, \mi\;and \ia, and \mi\;and \ia, as already discussed in Sect.~\ref{disc-ab}. Moreover, thanks to the available data we have also found for the first time correlations between \am\;and \ia, and \am\;and \fm.

\subsection{Column density ratios}
\begin{figure*}
\centering
\includegraphics[width=45pc]{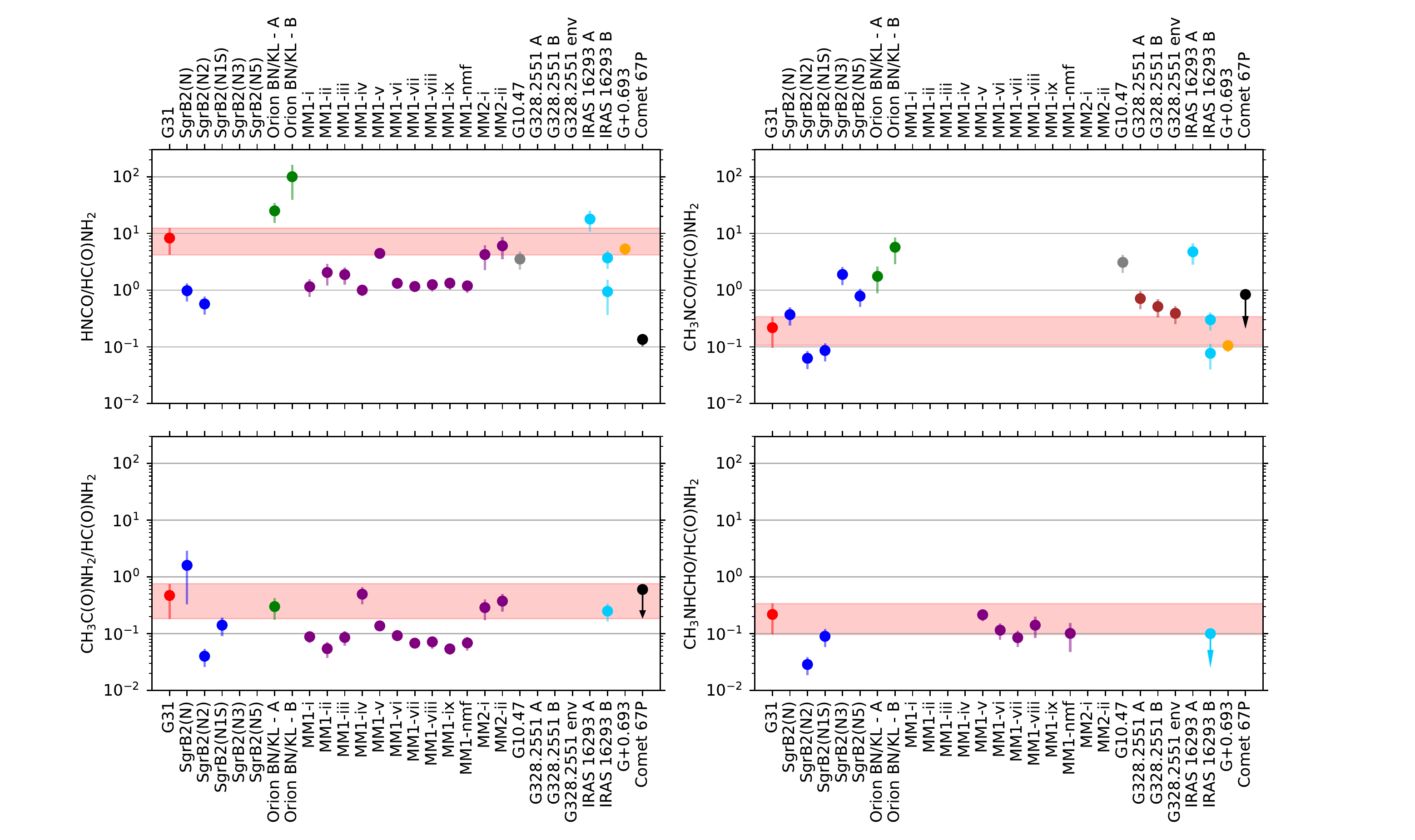}
\caption{HNCO, \mi, \am, and \nm\;column density ratios with respect to \fm (\emph{upper-left}, \emph{upper-right}, \emph{bottom-left}, and \emph{bottom-right} panels, respectively) towards G31 (red points and shaded areas) and high- and low-mass star-forming regions, a GC molecular cloud, and the 67P comet (from the left to the right in all panels). Different colors represent: G31 in red, Sgr B2 in blue, Orion BN/KL in green, NGC 6334I in purple, G10.47 in grey, G328.2551 in brown, IRAS 16293 in light blue, G+0.693 in orange, and the comet 67P in black. Data are taken from: G31 (this work), Sgr B2(N) (\citealt{belloche2013}), Sgr B2(N2) (\citealt{belloche2017}), Sgr B2(N1S) (\citealt{belloche2019}), Sgr B2(N3) and Sgr B2(N5) (\citealt{bonfand2019}), Orion BN/KL - A and B (\citealt{cernicharo2016}), NGC 6334I MM1-i--MM2ii (\citealt{ligterink2020}), G10.47 (\citealt{gorai2020}), G328.2551 A, B and env (\citealt{csengeri2019}), IRAS 16293 A and B  (\citealt{ligterink2017}; \citealt{manigand2020}; \citealt{ligterink2018} and \citealt{martin-domenech2017}); G+0.693 (\citealt{zeng2018}), and comet 67P (\citealt{altwegg2017}) (from the left to the right). }
\label{fig-colden}
\end{figure*}

Figure \ref{fig-colden} shows the column density ratios of HNCO, \mi, \am, and \nm\;with respect to \fm\;in the different astronomical sources. For column densities whose errors were not provided we assumed a value of 25\%. For IRAS 16293 B two values are shown, corresponding to those derived by \citet{martin-domenech2017}, \citet{ligterink2017}, and \citet{ligterink2018}. From the left to the right, Fig.~\ref{fig-colden} shows massive and low-mass star-forming regions, the GC G+0.693 molecular cloud, and the comet 67P values.

We note that overall there are similarities between molecular abundance ratios towards different regions, typically within $\sim$1 order of magnitude in the 80\% of the sources. The HNCO/\fm\;ratio found in G31 is similar, within the errors, to those derived towards Orion BN/KL-A, NGC 6334I MM1-v, MM2-i, and MM2-ii, G10.47, IRAS 16293 B, and G+0.693 (top-left panel of Fig.~\ref{fig-colden}). The \mi/\fm\;ratio is consistent, within the errors, to those of Sgr B2(N), Sgr B2(N1S), IRAS 16293 B, G328.2551 B and env, and G+0.693 (top-right panel of Fig.~\ref{fig-colden}). The \am/\fm\;ratio is consistent, within the errors, to those of Sgr B2(N1S), Orion BN/KL-A, NGC 6334I MM1-iv, MM2-i, and MM2-ii, and IRAS 16293 B (bottom-left panel of Fig.~\ref{fig-colden}). The \nm/\fm\;ratio is similar, within the errors, to those of Sgr B2(N1S), NGC 6334I MM1-v, MM1-vi, MM1-vii, MM1-viii, and MM1-nmf, which are all the sources in which \nm\;has been detected, except Sgr B2(N2) (bottom-right panel of Fig.~\ref{fig-colden}). The \nm/\am\;ratio is overall similar in all the sources, as observed in the bottom panels of Fig.~\ref{fig-colden} where their column density ratios with respect to \fm\;are shown.

The similarities of the different molecular ratios in interstellar regions with very different physical properties (e.g. masses from $\sim$0.5 M$_{\odot}$ up to $\sim$100 M$_{\odot}$ and luminosities from $\sim$1 L$_{\odot}$ up to $\sim$10$^{7}$ L$_{\odot}$) and location in the Galaxy, such as high-mass and low-mass star-forming regions (HMCs and hot corinos, respectively) in the GC and in the galactic disk, and a GC molecular cloud with no signs of star formation yet (\citealt{zeng2020}), suggest that these molecules were formed during very early phases of evolution. In fact, as discussed in Sect.~\ref{disc-ab}, these species could have been formed mostly on grains, and, in a later stage, released back to gas-phase through thermal desorption, in HMCs and hot corinos, or through shock-induced grain sputtering, in the case of G+0.693 and G328.2551 accretion shock positions. This has also been suggested by \citet{coletta2020}, who found constant abundance ratios for H$_{3}$OCHO, CH$_{3}$OCH$_{3}$, C$_{2}$H$_{5}$CN and \fm\;towards low- and high-mass star-forming regions in different evolutionary stages. For the GC a similar conclusion has been proposed by \citet{requena-torres2006} who compared the abundances of O-bearing COMs derived towards giant molecular clouds in the GC with those measured in hot corinos and hot cores. They found that all of these abundances are consistent within a factor of 10 and suggested that COMs are ejected from grain mantles by shocks.

Regarding the formation of \am, \citet{quan2007} have proposed that it could be formed in gas-phase via radiative association reaction, like:
\begin{align}
\label{}
&{\rm HC(O)NH}_{2} + {\rm CH}_{3}^{+} \rightarrow {\rm CH}_{3}{\rm CHONH}_{2}^{+} + h\nu \label{am1}\\
&{\rm CH}_{3}{\rm CHONH}_{2}^{+} + {\rm e}^{-} \rightarrow {\rm CH}_{3}{\rm C(O)NH}_{2} + {\rm  H}, \label{am2}
\end{align}
\citet{halfen2011} have suggested that ion-molecule processes might lead to the formation of both \am\;and \nm, and more recently \citet{redondo2014} have studied the viability of ion-molecule gas-phase reactions, such as CH$_{5}^{+}$ + \fm, to form \am. However, only gas-phase reactions are not enough to reproduce the observed abundances and the \am/\fm\;ratios in the ISM. \citet{frigge2018} show that \nm\;could form in a mixture of methylamine (CH$_{3}$NH$_{2}$) and CO ices, upon irradiation with energetic electrons. In particular, they proposed the following reactions on grain surfaces:
\begin{align}
&{\rm CH}_{3}{\rm NH}_{2} + {\rm CR} \rightarrow {\rm CH}_{3}{\rm NH} + {\rm H} \\
&{\rm CH}_{3}{\rm NH} + {\rm HCO} \rightarrow {\rm CH}_{3}{\rm NHCHO},
\end{align}
where CR are the cosmic rays simulated by energetic electrons. Moreover, \citet{garrod2008} have proposed CH$_{3}$ + HNCO as a possible formation route on grains, while \citet{belloche2017} suggest that \am\;is predominantly formed by H-abstraction from \fm, followed by methyl-group (CH$_{3}$) addition, and by the reaction NH$_{2}$ + CH$_{3}$CO. Moreover, these authors found that \nm\;could be formed on grains either through the direct addition of functional-group radicals (e.g. CH$_{3}$ + HNCHO) or through the hydrogenation of \mi. \am\;has also been identified in carbonaceous chondrites (\citealt{cooper1995}), and was found to form in experiments with irradiated ices (e.g. \citealt{berger1961}; \citealt{ligterink2018}), favouring also the grain-surface formation.

The lower left panel of Fig.~\ref{fig-colden} shows that \am\;is well correlated with \fm, as we already found from the correlation of their abundances with respect to CH$_{3}$OH (see Sect.~\ref{ab-meth}). The \am/\fm\;ratios are all within $\sim$1 order of magnitude.  This might indicate a direct chemical link, as suggested by reactions \eqref{am1} and \eqref{am2}, or that both molecules are mainly formed on grain surface and desorb under similar physical
conditions, as proposed by \citet{quenard2018} for HNCO and \fm. Indeed, this hypothesis is supported by the temperature programmed desorption experiment of \citet{ligterink2018}, who have showed that the peak desorption temperatures of \am\;and \fm\;are very similar (219 and 210 K, respectively, see also \citealt{corazzi2020}). 

Finally, a comparison with the values found in the comet 67P shows that the \am/\fm\;and \nm/\fm\;upper limits (bottom-right panel of Fig.~\ref{fig-colden}) are consistent with the values observed in the ISM, while the HNCO/\fm\;ratio is smaller than the ISM values. This could be due to an over-abundance of \fm\;in the 67P comet with respect to the ISM that could indicate a chemical reprocessing during later stages, e.g. the protoplanetary disk phase. However, the measurements with Rosina (Rosetta Orbiter Spectrometer for Ion and Neutral Analysis) cannot distinguish between the different structural isomers, so, if the other isomers were also formed, the abundance of \fm\;would be lower.

\section{Conclusions}
\label{conclusions}
In this work we have studied the peptide-like bond molecules \ia, \fm, \mi, \am, \nm, \ei, \ur, \cfm, and \gly\;in the context of the GUAPOS spectral survey, obtained with the ALMA interferometer towards the HMC G31. This is the first time that all of these molecules have been studied together towards G31 and outside the Galactic center. The main results and conclusions of our study are summarised below:
\begin{enumerate}
\item The column densities of \ia, \fm, and \mi\;have been derived from their optically thin $^{13}$C-isotopologues, or from vibrationally excited states in case the $^{13}$C-species were only tentatively detected. \am\;is found to be optically thin in all the vibrational states allowing the derivation of the column density taking into account all the states together, while for \nm\;only transitions from the optical thin ground vibrational state have been detected. \ei\;has not been detected and the upper limit derived provides a \mi/\ei\;ratio $>$24, consistent with what previously measured towards Orion KL and Sgr B2. On the other hand, also \ur, \cfm, and \gly\;have not been detected and we have derived their upper limits. Our findings in G31 show that the molecules follow the subsequent order of abundances compared to H$_{2}$ (from 10$^{-7}$ down to a few 10$^{-9}$): $X$(HNCO) $>$ $X$(\fm) $\ge$ $X$(\mi) $\ge$ $X$(\am) $\ge$ $X$(\nm). Moreover, we have found abundances with respect to CH$_{3}$OH that range from 10$^{-3}$ to $\sim$4$\times$10$^{-2}$.
\item The emission of all the species towards the HMC is compact ($\sim$2\asec, i.e. $\sim$7500 au), and this has also been confirmed with higher angular resolution observations for \ia, \fm, and \mi. The five molecular species trace hot molecular gas (temperature higher than 100 K), without significant spatial emission differences among them.
\item The comparison with other sources in the ISM (HMCs in the GC and in the Galactic disk, hot corinos, and a shock-dominated GC molecular cloud) shows tight correlations between the abundances of \ia\;and \fm, \mi\;and \ia, and for the first time between \mi\;and \fm, \am\;and \ia, and \am\;and \fm\;abundances. This suggests either a possible chemical link between these species, a common precursor, or a similar response to the physical conditions of the molecular clouds. 
\item The column density abundance ratios are quite similar in all the sources investigated, regardless of their physical conditions (e.g. mass and luminosity), and Galactic environment (GC or Galactic disk). Moreover, HMC, hot corinos and the shock-induced G+0.693 molecular cloud show abundances several orders of magnitude higher than low-mass pre-stellar cores. This suggests that most of the observed molecular abundances come from surface chemistry formation at early evolutionary stages. These molecules are subsequently released back to the gas phase, either by thermal (HMCs, hot corinos) or shock-induced desorption (G+0.693 and G328.2551 A and B).  
\end{enumerate}

\begin{acknowledgements}
We thank the anonymous referee for the careful reading of the article and the useful comments.
L.~C. acknowledges financial support from the Spanish State Research Agency (AEI) through the project No. ESP2017-86582-C4-1-R. L.~C. and V.~M.R. acknowledge support from the Comunidad de Madrid through the Atracci\'on de Talento Investigador Senior Grant (COOL: Cosmic Origins Of Life; 2019-T1/TIC-15379). I.~J.-S. has received partial support from the Spanish State Research Agency (AEI; project number PID2019-105552RB-C41). C.~M. acknowledges support from the Italian Ministero dell’Istruzione, Università e Ricerca through the grant Progetti Premiali 2012 - iALMA (CUP C52I13000140001). C.~.M akso acknowledges funding from the European Research Council (ERC) under the European Union's  Horizon 2020 program, through the ECOGAL Synergy grant (grant ID 855130). A.~S.-M. carried out this research within the Collaborative Research Centre 956 (subproject A6), funded by the Deutsche Forschungsgemeinschaft (DFG) - project ID 184018867. 
This paper makes use of the following ALMA data: ADS/JAO.ALMA\#2013.1.00489.S and ADS/JAO.ALMA\#2017.1.00501.S. ALMA is a partnership of ESO (representing its member states), NSF (USA) and NINS (Japan), together with NRC (Canada), MOST and ASIAA (Taiwan), and KASI (Republic of Korea), in cooperation with the Republic of Chile. The Joint ALMA Observatory is operated by ESO, AUI/NRAO and NAOJ. We thank Arnaud Belloche for providing us the \am, $v$=0, $v_{\rm t}$=1, 2 spectroscopic entries. Most of the figures were generated with the PYTHON-based package MATPLOTLIB (\citealt{hunter2007}).

\end{acknowledgements}

\bibliographystyle{aa} 
 \bibliography{bibliography} 

\begin{thebibliography}{154}
\expandafter\ifx\csname natexlab\endcsname\relax\def\natexlab#1{#1}\fi

\bibitem[{{Adams}(2010)}]{adams2010}
{Adams}, F.~C. 2010, \araa, 48, 47

\bibitem[{{Agarwal} {et~al.}(1985){Agarwal}, {Schutte}, {Greenberg}, {Ferris},
  {Briggs}, {Connor}, {van de Bult}, \& {Baas}}]{agarwal1985}
{Agarwal}, V.~K., {Schutte}, W., {Greenberg}, J.~M., {et~al.} 1985, Origins of
  Life, 16, 21

\bibitem[{{Allen} {et~al.}(2020){Allen}, {van der Tak}, {L{\'o}pez-Sepulcre},
  {S{\'a}nchez-Monge}, {Rivilla}, \& {Cesaroni}}]{allen2020}
{Allen}, V., {van der Tak}, F.~F.~S., {L{\'o}pez-Sepulcre}, A., {et~al.} 2020,
  \aap, 636, A67

\bibitem[{{Altwegg} {et~al.}(2017){Altwegg}, {Balsiger}, {Berthelier},
  {Bieler}, {Calmonte}, {Fuselier}, {Goesmann}, {Gasc}, {Gombosi}, {Le Roy},
  {de Keyser}, {Morse}, {Rubin}, {Schuhmann}, {Taylor}, {Tzou}, \&
  {Wright}}]{altwegg2017}
{Altwegg}, K., {Balsiger}, H., {Berthelier}, J.~J., {et~al.} 2017, \mnras, 469,
  S130

\bibitem[{{Arce} {et~al.}(2008){Arce}, {Santiago-Garc{\'\i}a}, {J{\o}rgensen},
  {Tafalla}, \& {Bachiller}}]{arce2008}
{Arce}, H.~G., {Santiago-Garc{\'\i}a}, J., {J{\o}rgensen}, J.~K., {Tafalla},
  M., \& {Bachiller}, R. 2008, \apjl, 681, L21

\bibitem[{{Bacmann} {et~al.}(2012){Bacmann}, {Taquet}, {Faure}, {Kahane}, \&
  {Ceccarelli}}]{bacmann2012}
{Bacmann}, A., {Taquet}, V., {Faure}, A., {Kahane}, C., \& {Ceccarelli}, C.
  2012, \aap, 541, L12

\bibitem[{{Barone} {et~al.}(2015){Barone}, {Latouche}, {Skouteris}, {Vazart},
  {Balucani}, {Ceccarelli}, \& {Lefloch}}]{barone2015}
{Barone}, V., {Latouche}, C., {Skouteris}, D., {et~al.} 2015, \mnras, 453, L31

\bibitem[{{Belloche} {et~al.}(2019){Belloche}, {Garrod}, {M{\"u}ller},
  {Menten}, {Medvedev}, {Thomas}, \& {Kisiel}}]{belloche2019}
{Belloche}, A., {Garrod}, R.~T., {M{\"u}ller}, H.~S.~P., {et~al.} 2019, \aap,
  628, A10

\bibitem[{{Belloche} {et~al.}(2020){Belloche}, {Maury}, {Maret}, {Anderl},
  {Bacmann}, {Andr{\'e}}, {Bontemps}, {Cabrit}, {Codella}, {Gaudel}, {Gueth},
  {Lef{\`e}vre}, {Lefloch}, {Podio}, \& {Testi}}]{belloche2020}
{Belloche}, A., {Maury}, A.~J., {Maret}, S., {et~al.} 2020, \aap, 635, A198

\bibitem[{{Belloche} {et~al.}(2017){Belloche}, {Meshcheryakov}, {Garrod},
  {Ilyushin}, {Alekseev}, {Motiyenko}, {Margul{\`e}s}, {M{\"u}ller}, \&
  {Menten}}]{belloche2017}
{Belloche}, A., {Meshcheryakov}, A.~A., {Garrod}, R.~T., {et~al.} 2017, \aap,
  601, A49

\bibitem[{{Belloche} {et~al.}(2016){Belloche}, {M{\"u}ller}, {Garrod}, \&
  {Menten}}]{belloche2016}
{Belloche}, A., {M{\"u}ller}, H.~S.~P., {Garrod}, R.~T., \& {Menten}, K.~M.
  2016, \aap, 587, A91

\bibitem[{{Belloche} {et~al.}(2013){Belloche}, {M{\"u}ller}, {Menten},
  {Schilke}, \& {Comito}}]{belloche2013}
{Belloche}, A., {M{\"u}ller}, H.~S.~P., {Menten}, K.~M., {Schilke}, P., \&
  {Comito}, C. 2013, \aap, 559, A47

\bibitem[{{Beltr{\'a}n} {et~al.}(2005){Beltr{\'a}n}, {Cesaroni}, {Neri},
  {Codella}, {Furuya}, {Testi}, \& {Olmi}}]{beltran2005}
{Beltr{\'a}n}, M.~T., {Cesaroni}, R., {Neri}, R., {et~al.} 2005, \aap, 435, 901

\bibitem[{{Beltr{\'a}n} {et~al.}(2018){Beltr{\'a}n}, {Cesaroni}, {Rivilla},
  {S{\'a}nchez-Monge}, {Moscadelli}, {Ahmadi}, {Allen}, {Beuther}, {Etoka},
  {Galli}, {Galv{\'a}n-Madrid}, {Goddi}, {Johnston}, {Klaassen},
  {K{\"o}lligan}, {Kuiper}, {Kumar}, {Maud}, {Mottram}, {Peters}, {Schilke},
  {Testi}, {van der Tak}, \& {Walmsley}}]{beltran2018}
{Beltr{\'a}n}, M.~T., {Cesaroni}, R., {Rivilla}, V.~M., {et~al.} 2018, \aap,
  615, A141

\bibitem[{{Beltr{\'a}n} {et~al.}(2009){Beltr{\'a}n}, {Codella}, {Viti}, {Neri},
  \& {Cesaroni}}]{beltran2009}
{Beltr{\'a}n}, M.~T., {Codella}, C., {Viti}, S., {Neri}, R., \& {Cesaroni}, R.
  2009, \apjl, 690, L93

\bibitem[{{Beltr{\'a}n} {et~al.}(2013){Beltr{\'a}n}, {Olmi}, {Cesaroni},
  {Schisano}, {Elia}, {Molinari}, {Di Giorgio}, {Kirk}, {Mottram},
  {Pestalozzi}, {Testi}, \& {Thompson}}]{beltran2013}
{Beltr{\'a}n}, M.~T., {Olmi}, L., {Cesaroni}, R., {et~al.} 2013, \aap, 552,
  A123

\bibitem[{{Beltr{\'a}n} {et~al.}(2019){Beltr{\'a}n}, {Padovani}, {Girart},
  {Galli}, {Cesaroni}, {Paladino}, {Anglada}, {Estalella}, {Osorio}, {Rao},
  {S{\'a}nchez-Monge}, \& {Zhang}}]{beltran2019}
{Beltr{\'a}n}, M.~T., {Padovani}, M., {Girart}, J.~M., {et~al.} 2019, \aap,
  630, A54

\bibitem[{{Beltr{\'a}n} {et~al.}(2021){Beltr{\'a}n}, {Rivilla}, {Cesaroni},
  {Maud}, {Galli}, {Moscadelli}, {Lorenzani}, {Ahmadi}, {Beuther}, {Csengeri},
  {Etoka}, {Goddi}, {Klaassen}, {Kuiper}, {Kumar}, {Peters},
  {S{\'a}nchez-Monge}, {Schilke}, {van der Tak}, {Vig}, \&
  {Zinnecker}}]{beltran2021}
{Beltr{\'a}n}, M.~T., {Rivilla}, V.~M., {Cesaroni}, R., {et~al.} 2021, \aap,
  648, A100

\bibitem[{{Berger}(1961)}]{berger1961}
{Berger}, R. 1961, Proc. Natl. Acad. Sci., 47(9), 1434

\bibitem[{{Bisschop} {et~al.}(2007){Bisschop}, {J{\o}rgensen}, {van Dishoeck},
  \& {de Wachter}}]{bisschop2007}
{Bisschop}, S.~E., {J{\o}rgensen}, J.~K., {van Dishoeck}, E.~F., \& {de
  Wachter}, E.~B.~M. 2007, \aap, 465, 913

\bibitem[{Blanco {et~al.}(2006)Blanco, López, Lesarri, \& Alonso}]{blanco2006}
Blanco, S., López, J.~C., Lesarri, A., \& Alonso, J.~L. 2006, Journal of the
  American Chemical Society, 128, 12111

\bibitem[{{B{\o}gelund} {et~al.}(2018){B{\o}gelund}, {McGuire}, {Ligterink},
  {Taquet}, {Brogan}, {Hunter}, {Pearson}, {Hogerheijde}, \& {van
  Dishoeck}}]{bogelund2018}
{B{\o}gelund}, E.~G., {McGuire}, B.~A., {Ligterink}, N. F.~W., {et~al.} 2018,
  \aap, 615, A88

\bibitem[{{Bonal} {et~al.}(2009){Bonal}, {Huss}, {Nagashima}, \&
  {Krot}}]{bonal2009}
{Bonal}, L., {Huss}, G.~R., {Nagashima}, K., \& {Krot}, A.~N. 2009, Meteoritics
  and Planetary Science Supplement, 72, 5178

\bibitem[{{Bonfand} {et~al.}(2019){Bonfand}, {Belloche}, {Garrod}, {Menten},
  {Willis}, {St{\'e}phan}, \& {M{\"u}ller}}]{bonfand2019}
{Bonfand}, M., {Belloche}, A., {Garrod}, R.~T., {et~al.} 2019, \aap, 628, A27

\bibitem[{{Brown} {et~al.}(1975){Brown}, {Godfrey}, \& {Storey}}]{brown1975}
{Brown}, R.~D., {Godfrey}, P.~D., \& {Storey}, J. 1975, Journal of Molecular
  Spectroscopy, 58, 445

\bibitem[{Brown {et~al.}(1997)Brown, Berghout, \& Crim}]{brown1997raman}
Brown, S.~S., Berghout, H.~L., \& Crim, F.~F. 1997, J. Chem. Phys., 107, 9764

\bibitem[{{Canelo} {et~al.}(2021){Canelo}, {Bronfman}, {Mendoza}, {Duronea},
  {Merello}, {Carvajal}, {Fria{\c{c}}a}, \& {Lepine}}]{canelo2021}
{Canelo}, C.~M., {Bronfman}, L., {Mendoza}, E., {et~al.} 2021, \mnras, 504,
  4428

\bibitem[{{Cassone} {et~al.}(2021){Cassone}, {Saija}, {Sponer}, {Sponer},
  {Jim{\'e}nez-Escobar}, {Ciaravella}, \& {Cecchi-Pestellini}}]{cassone2021}
{Cassone}, G., {Saija}, F., {Sponer}, J., {et~al.} 2021, \mnras, 504, 1565

\bibitem[{{Cernicharo} {et~al.}(2016){Cernicharo}, {Kisiel}, {Tercero},
  {Kolesnikov{\'a}}, {Medvedev}, {L{\'o}pez}, {Fortman}, {Winnewisser}, {de
  Lucia}, {Alonso}, \& {Guillemin}}]{cernicharo2016}
{Cernicharo}, J., {Kisiel}, Z., {Tercero}, B., {et~al.} 2016, \aap, 587, L4

\bibitem[{{Cesaroni}(2019)}]{cesaroni2019}
{Cesaroni}, R. 2019, \aap, 631, A65

\bibitem[{{Cesaroni} {et~al.}(2010){Cesaroni}, {Hofner}, {Araya}, \&
  {Kurtz}}]{cesaroni2010}
{Cesaroni}, R., {Hofner}, P., {Araya}, E., \& {Kurtz}, S. 2010, \aap, 509, A50

\bibitem[{{Cesaroni} {et~al.}(2017){Cesaroni}, {S{\'a}nchez-Monge},
  {Beltr{\'a}n}, {Johnston}, {Maud}, {Moscadelli}, {Mottram}, {Ahmadi},
  {Allen}, {Beuther}, {Csengeri}, {Etoka}, {Fuller}, {Galli},
  {Galv{\'a}n-Madrid}, {Goddi}, {Henning}, {Hoare}, {Klaassen}, {Kuiper},
  {Kumar}, {Lumsden}, {Peters}, {Rivilla}, {Schilke}, {Testi}, {van der Tak},
  {Vig}, {Walmsley}, \& {Zinnecker}}]{cesaroni2017}
{Cesaroni}, R., {S{\'a}nchez-Monge}, {\'A}., {Beltr{\'a}n}, M.~T., {et~al.}
  2017, \aap, 602, A59

\bibitem[{{Charnley} {et~al.}(2004){Charnley}, {Ehrenfreund}, {Millar},
  {Boogert}, {Markwick}, {Butner}, {Ruiterkamp}, \& {Rodgers}}]{charnley2004}
{Charnley}, S.~B., {Ehrenfreund}, P., {Millar}, T.~J., {et~al.} 2004, \mnras,
  347, 157

\bibitem[{{Christiansen}(2005)}]{christiansen2005}
{Christiansen}, J.~J. 2005, Journal of Molecular Spectroscopy, 231, 131

\bibitem[{{Codella} {et~al.}(2020){Codella}, {Ceccarelli}, {Bianchi},
  {Balucani}, {Podio}, {Caselli}, {Feng}, {Lefloch}, {L{\'o}pez-Sepulcre},
  {Neri}, {Spezzano}, \& {De Simone}}]{codella2020}
{Codella}, C., {Ceccarelli}, C., {Bianchi}, E., {et~al.} 2020, \aap, 635, A17

\bibitem[{{Coletta} {et~al.}(2020){Coletta}, {Fontani}, {Rivilla}, {Mininni},
  {Colzi}, {S{\'a}nchez-Monge}, \& {Beltr{\'a}n}}]{coletta2020}
{Coletta}, A., {Fontani}, F., {Rivilla}, V.~M., {et~al.} 2020, \aap, 641, A54

\bibitem[{{Colzi} {et~al.}(2018b){Colzi}, {Fontani}, {Rivilla},
  {S{\'a}nchez-Monge}, {Testi}, {Beltr{\'a}n}, \& {Caselli}}]{colzi18b}
{Colzi}, L., {Fontani}, F., {Rivilla}, V.~M., {et~al.} 2018b, \mnras, 478, 3693

\bibitem[{{Cooper} \& {Cronin}(1995)}]{cooper1995}
{Cooper}, G.~W. \& {Cronin}, J.~R. 1995, \gca, 59, 1003

\bibitem[{{Corazzi} {et~al.}(2020){Corazzi}, {Fedele}, {Poggiali}, \&
  {Brucato}}]{corazzi2020}
{Corazzi}, M.~A., {Fedele}, D., {Poggiali}, G., \& {Brucato}, J.~R. 2020, \aap,
  636, A63

\bibitem[{{Coutens} {et~al.}(2016){Coutens}, {J{\o}rgensen}, {van der Wiel},
  {M{\"u}ller}, {Lykke}, {Bjerkeli}, {Bourke}, {Calcutt}, {Drozdovskaya},
  {Favre}, {Fayolle}, {Garrod}, {Jacobsen}, {Ligterink}, {{\"O}berg},
  {Persson}, {van Dishoeck}, \& {Wampfler}}]{coutens2016}
{Coutens}, A., {J{\o}rgensen}, J.~K., {van der Wiel}, M.~H.~D., {et~al.} 2016,
  \aap, 590, L6

\bibitem[{{Csengeri} {et~al.}(2019){Csengeri}, {Belloche}, {Bontemps},
  {Wyrowski}, {Menten}, \& {Bouscasse}}]{csengeri2019}
{Csengeri}, T., {Belloche}, A., {Bontemps}, S., {et~al.} 2019, \aap, 632, A57

\bibitem[{{Cuadrado} {et~al.}(2017){Cuadrado}, {Goicoechea}, {Cernicharo},
  {Fuente}, {Pety}, \& {Tercero}}]{cuadrado2017}
{Cuadrado}, S., {Goicoechea}, J.~R., {Cernicharo}, J., {et~al.} 2017, \aap,
  603, A124

\bibitem[{{Dulieu} {et~al.}(2019){Dulieu}, {Nguyen}, {Congiu}, {Baouche}, \&
  {Taquet}}]{dulieu2019}
{Dulieu}, F., {Nguyen}, T., {Congiu}, E., {Baouche}, S., \& {Taquet}, V. 2019,
  \mnras, 484, L119

\bibitem[{Durig {et~al.}(2010)Durig, Zhou, Zheng, \&
  Durig}]{durig2010conformational}
Durig, J.~R., Zhou, S.~X., Zheng, C., \& Durig, D.~T. 2010, J. Mol. Struct.,
  971, 23

\bibitem[{{Endres} {et~al.}(2016){Endres}, {Schlemmer}, {Schilke}, {Stutzki},
  \& {M{\"u}ller}}]{endres2016}
{Endres}, C.~P., {Schlemmer}, S., {Schilke}, P., {Stutzki}, J., \&
  {M{\"u}ller}, H. S.~P. 2016, Journal of Molecular Spectroscopy, 327, 95

\bibitem[{{Fedoseev} {et~al.}(2016){Fedoseev}, {Chuang}, {van Dishoeck},
  {Ioppolo}, \& {Linnartz}}]{fedoseev2016}
{Fedoseev}, G., {Chuang}, K.~J., {van Dishoeck}, E.~F., {Ioppolo}, S., \&
  {Linnartz}, H. 2016, \mnras, 460, 4297

\bibitem[{{Fedoseev} {et~al.}(2015){Fedoseev}, {Ioppolo}, {Zhao}, {Lamberts},
  \& {Linnartz}}]{fedoseev2015}
{Fedoseev}, G., {Ioppolo}, S., {Zhao}, D., {Lamberts}, T., \& {Linnartz}, H.
  2015, \mnras, 446, 439

\bibitem[{{Frigge} {et~al.}(2018){Frigge}, {Zhu}, {Turner}, {Abplanalp},
  {Bergantini}, {Sun}, {Chen}, {Chang}, \& {Kaiser}}]{frigge2018}
{Frigge}, R., {Zhu}, C., {Turner}, A.~M., {et~al.} 2018, \apj, 862, 84

\bibitem[{{Gardner} {et~al.}(1980){Gardner}, {Godfrey}, \&
  {Williams}}]{gardner1980}
{Gardner}, F.~F., {Godfrey}, P.~D., \& {Williams}, D.~R. 1980, \mnras, 193, 713

\bibitem[{{Garrod} {et~al.}(2008){Garrod}, {Widicus Weaver}, \&
  {Herbst}}]{garrod2008}
{Garrod}, R.~T., {Widicus Weaver}, S.~L., \& {Herbst}, E. 2008, \apj, 682, 283

\bibitem[{{Girart} {et~al.}(2009){Girart}, {Beltr{\'a}n}, {Zhang}, {Rao}, \&
  {Estalella}}]{girart2009}
{Girart}, J.~M., {Beltr{\'a}n}, M.~T., {Zhang}, Q., {Rao}, R., \& {Estalella},
  R. 2009, Science, 324, 1408

\bibitem[{{Gorai} {et~al.}(2020){Gorai}, {Bhat}, {Sil}, {Mondal}, {Ghosh},
  {Chakrabarti}, \& {Das}}]{gorai2020}
{Gorai}, P., {Bhat}, B., {Sil}, M., {et~al.} 2020, \apj, 895, 86

\bibitem[{{Gorai} {et~al.}(2021){Gorai}, {Das}, {Shimonishi}, {Sahu}, {Mondal},
  {Bhat}, \& {Chakrabarti}}]{gorai2021}
{Gorai}, P., {Das}, A., {Shimonishi}, T., {et~al.} 2021, \apj, 907, 108

\bibitem[{Gordy {et~al.}(1984)Gordy, Cook, \& Weissberger}]{gordy1984microwave}
Gordy, W., Cook, R.~L., \& Weissberger, A. 1984, Microwave molecular spectra,
  Vol.~18 (Wiley New York)

\bibitem[{{Grim} {et~al.}(1989){Grim}, {Greenberg}, {de Groot}, {Baas},
  {Schutte}, \& {Schmitt}}]{grim1989}
{Grim}, R.~J.~A., {Greenberg}, J.~M., {de Groot}, M.~S., {et~al.} 1989, \aaps,
  78, 161

\bibitem[{{Guzm{\'a}n} {et~al.}(2013){Guzm{\'a}n}, {Goicoechea}, {Pety},
  {Gratier}, {Gerin}, {Roueff}, {Le Petit}, {Le Bourlot}, \&
  {Faure}}]{guzman2013}
{Guzm{\'a}n}, V.~V., {Goicoechea}, J.~R., {Pety}, J., {et~al.} 2013, \aap, 560,
  A73

\bibitem[{{Halfen} {et~al.}(2011){Halfen}, {Ilyushin}, \&
  {Ziurys}}]{halfen2011}
{Halfen}, D.~T., {Ilyushin}, V., \& {Ziurys}, L.~M. 2011, \apj, 743, 60

\bibitem[{{Halfen} {et~al.}(2015){Halfen}, {Ilyushin}, \&
  {Ziurys}}]{halfen2015}
{Halfen}, D.~T., {Ilyushin}, V.~V., \& {Ziurys}, L.~M. 2015, \apjl, 812, L5

\bibitem[{{Haupa} {et~al.}(2019){Haupa}, {Tarczay}, \& {Lee}}]{haupa2019}
{Haupa}, K.~A., {Tarczay}, G., \& {Lee}, Y.-P. 2019, in 74th International
  Symposium on Molecular Spectroscopy

\bibitem[{Heineking {et~al.}(1994)Heineking, Grabow, \&
  Stahl}]{heineking1994microwave}
Heineking, N., Grabow, J.-U., \& Stahl, W. 1994, Mol. Phys., 81, 1177

\bibitem[{{Helmich} \& {van Dishoeck}(1997)}]{helmich1997}
{Helmich}, F.~P. \& {van Dishoeck}, E.~F. 1997, \aaps, 124, 205

\bibitem[{{Hirota} {et~al.}(1974){Hirota}, {Sugisaki}, {Nielsen}, \&
  {S{\o}rensen}}]{hirota1974}
{Hirota}, E., {Sugisaki}, R., {Nielsen}, C.~J., \& {S{\o}rensen}, G.~O. 1974,
  Journal of Molecular Spectroscopy, 49, 251

\bibitem[{Hocking {et~al.}(1975)Hocking, Gerry, \&
  Winnewisser}]{hocking1975microwave}
Hocking, W., Gerry, M., \& Winnewisser, G. 1975, Can. J. Phys., 53, 1869

\bibitem[{{Hocking} {et~al.}(1975){Hocking}, {Gerry}, \&
  {Winnewisser}}]{hocking1975}
{Hocking}, W.~H., {Gerry}, M.~C.~L., \& {Winnewisser}, G. 1975, Canadian
  Journal of Physics, 53, 1869

\bibitem[{{Hollis} {et~al.}(2004){Hollis}, {Jewell}, {Lovas}, {Remijan}, \&
  {M{\o}llendal}}]{hollis2004}
{Hollis}, J.~M., {Jewell}, P.~R., {Lovas}, F.~J., {Remijan}, A., \&
  {M{\o}llendal}, H. 2004, \apjl, 610, L21

\bibitem[{{Hollis} {et~al.}(2006){Hollis}, {Lovas}, {Remijan}, {Jewell},
  {Ilyushin}, \& {Kleiner}}]{hollis2006}
{Hollis}, J.~M., {Lovas}, F.~J., {Remijan}, A.~J., {et~al.} 2006, \apjl, 643,
  L25

\bibitem[{Hunter(2007)}]{hunter2007}
Hunter, J.~D. 2007, Computing in Science \& Engineering, 9, 90

\bibitem[{{Ilyushin} {et~al.}(2004){Ilyushin}, {Alekseev}, {Dyubko}, {Kleiner},
  \& {Hougen}}]{ilyushin2004}
{Ilyushin}, V.~V., {Alekseev}, E.~A., {Dyubko}, S.~F., {Kleiner}, I., \&
  {Hougen}, J.~T. 2004, Journal of Molecular Spectroscopy, 227, 115

\bibitem[{{Immer} {et~al.}(2019){Immer}, {Li}, {Quiroga-Nu{\~n}ez}, {Reid},
  {Zhang}, {Moscadelli}, \& {Rygl}}]{immer2019}
{Immer}, K., {Li}, J., {Quiroga-Nu{\~n}ez}, L.~H., {et~al.} 2019, \aap, 632,
  A123

\bibitem[{{Jim{\'e}nez-Serra} {et~al.}(2020){Jim{\'e}nez-Serra},
  {Mart{\'\i}n-Pintado}, {Rivilla}, {Rodr{\'\i}guez-Almeida}, {Alonso Alonso},
  {Zeng}, {Cocinero}, {Mart{\'\i}n}, {Requena-Torres}, {Mart{\'\i}n-Domenech},
  \& {Testi}}]{jimenez-serra2020}
{Jim{\'e}nez-Serra}, I., {Mart{\'\i}n-Pintado}, J., {Rivilla}, V.~M., {et~al.}
  2020, Astrobiology, 20, 1048

\bibitem[{{Jim{\'e}nez-Serra} {et~al.}(2016){Jim{\'e}nez-Serra}, {Vasyunin},
  {Caselli}, {Marcelino}, {Billot}, {Viti}, {Testi}, {Vastel}, {Lefloch}, \&
  {Bachiller}}]{jimenez-serra2016}
{Jim{\'e}nez-Serra}, I., {Vasyunin}, A.~I., {Caselli}, P., {et~al.} 2016,
  \apjl, 830, L6

\bibitem[{{Jones} {et~al.}(2011){Jones}, {Bennett}, \& {Kaiser}}]{jones2011}
{Jones}, B.~M., {Bennett}, C.~J., \& {Kaiser}, R.~I. 2011, \apj, 734, 78

\bibitem[{{J{\o}rgensen} {et~al.}(2012){J{\o}rgensen}, {Favre}, {Bisschop},
  {Bourke}, {van Dishoeck}, \& {Schmalzl}}]{jorgensen2012}
{J{\o}rgensen}, J.~K., {Favre}, C., {Bisschop}, S.~E., {et~al.} 2012, \apjl,
  757, L4

\bibitem[{{J{\o}rgensen} {et~al.}(2018){J{\o}rgensen}, {M{\"u}ller}, {Calcutt},
  {Coutens}, {Drozdovskaya}, {{\"O}berg}, {Persson}, {Taquet}, {van Dishoeck},
  \& {Wampfler}}]{jorgensen2018}
{J{\o}rgensen}, J.~K., {M{\"u}ller}, H.~S.~P., {Calcutt}, H., {et~al.} 2018,
  \aap, 620, A170

\bibitem[{{J{\o}rgensen} {et~al.}(2016){J{\o}rgensen}, {van der Wiel},
  {Coutens}, {Lykke}, {M{\"u}ller}, {van Dishoeck}, {Calcutt}, {Bjerkeli},
  {Bourke}, {Drozdovskaya}, {Favre}, {Fayolle}, {Garrod}, {Jacobsen},
  {{\"O}berg}, {Persson}, \& {Wampfler}}]{jorgensen2016}
{J{\o}rgensen}, J.~K., {van der Wiel}, M.~H.~D., {Coutens}, A., {et~al.} 2016,
  \aap, 595, A117

\bibitem[{{Kasten} \& {Dreizler}(1986a)}]{kasten1986a}
{Kasten}, W. \& {Dreizler}, H. 1986a, Zeitschrift Naturforschung Teil A, 41,
  637

\bibitem[{{Kasten} \& {Dreizler}(1986b)}]{kasten1986b}
{Kasten}, W. \& {Dreizler}, H. 1986b, Zeitschrift Naturforschung Teil A, 41,
  1173

\bibitem[{{Kolesnikov{\'a}} {et~al.}(2018){Kolesnikov{\'a}}, {Alonso},
  {Tercero}, {Cernicharo}, \& {Alonso}}]{kolesnikova2018}
{Kolesnikov{\'a}}, L., {Alonso}, E.~R., {Tercero}, B., {Cernicharo}, J., \&
  {Alonso}, J.~L. 2018, \aap, 616, A173

\bibitem[{{Kolesnikov{\'a}} {et~al.}(2019){Kolesnikov{\'a}}, {Kisiel},
  {Alonso}, {Guillemin}, {Alonso}, {Medvedev}, \&
  {Winnewisser}}]{kolesnikova2019}
{Kolesnikov{\'a}}, L., {Kisiel}, Z., {Alonso}, E.~R., {et~al.} 2019, \apjs,
  245, 31

\bibitem[{{Koput}(1986)}]{koput1986}
{Koput}, J. 1986, Journal of Molecular Spectroscopy, 115, 131

\bibitem[{Korschinek {et~al.}(2020)Korschinek, Faestermann, Poutivtsev, Arazi,
  Knie, Rugel, \& Wallner}]{korschinek2020}
Korschinek, G., Faestermann, T., Poutivtsev, M., {et~al.} 2020, Phys. Rev.
  Lett., 125, 031101

\bibitem[{{Kretschmer} {et~al.}(1996){Kretschmer}, {Consalvo}, {Knaack},
  {Schade}, {Stahl}, \& {Dreizler}}]{kretschmer1996}
{Kretschmer}, U., {Consalvo}, D., {Knaack}, A., {et~al.} 1996, Molecular
  Physics, 87, 1159

\bibitem[{{Kryvda} {et~al.}(2009){Kryvda}, {Gerasimov}, {Dyubko}, {Alekseev},
  \& {Motiyenko}}]{kryvda2009}
{Kryvda}, A.~V., {Gerasimov}, V.~G., {Dyubko}, S.~F., {Alekseev}, E.~A., \&
  {Motiyenko}, R.~A. 2009, Journal of Molecular Spectroscopy, 254, 28

\bibitem[{{Kukolich} \& {Nelson}(1971)}]{kukolich1971b}
{Kukolich}, S.~G. \& {Nelson}, A.~C. 1971, Chemical Physics Letters, 11, 383

\bibitem[{Kukolich {et~al.}(1971)Kukolich, Nelson, \&
  Yamanashi}]{kukolich1971a}
Kukolich, S.~G., Nelson, A.~C., \& Yamanashi, B.~S. 1971, Journal of the
  American Chemical Society, 93, 6769

\bibitem[{{Kurland} \& {Bright Wilson}(1957)}]{kurland1957}
{Kurland}, R.~J. \& {Bright Wilson}, E., J. 1957, \jcp, 27, 585

\bibitem[{{Lapinov} {et~al.}(2007){Lapinov}, {Golubiatnikov}, {Markov}, \&
  {Guarnieri}}]{lapinov2007}
{Lapinov}, A.~V., {Golubiatnikov}, G.~Y., {Markov}, V.~N., \& {Guarnieri}, A.
  2007, Astronomy Letters, 33, 121

\bibitem[{{Lattelais} {et~al.}(2010){Lattelais}, {Pauzat}, {Ellinger}, \&
  {Ceccarelli}}]{lattelais2010}
{Lattelais}, M., {Pauzat}, F., {Ellinger}, Y., \& {Ceccarelli}, C. 2010, \aap,
  519, A30

\bibitem[{{Li} \& {Draine}(2001)}]{li2001}
{Li}, A. \& {Draine}, B.~T. 2001, \apj, 554, 778

\bibitem[{{Lichtenberg} {et~al.}(2019){Lichtenberg}, {Golabek}, {Burn},
  {Meyer}, {Alibert}, {Gerya}, \& {Mordasini}}]{lichtenberg2019}
{Lichtenberg}, T., {Golabek}, G.~J., {Burn}, R., {et~al.} 2019, Nature
  Astronomy, 3, 307

\bibitem[{{Ligterink} {et~al.}(2021){Ligterink}, {Ahmadi}, {Coutens},
  {Tychoniec}, {Calcutt}, {van Dishoeck}, {Linnartz}, {J{\o}rgensen}, {Garrod},
  \& {Bouwman}}]{ligterink2021}
{Ligterink}, N.~F.~W., {Ahmadi}, A., {Coutens}, A., {et~al.} 2021, \aap, 647,
  A87

\bibitem[{{Ligterink} {et~al.}(2017){Ligterink}, {Coutens}, {Kofman},
  {M{\"u}ller}, {Garrod}, {Calcutt}, {Wampfler}, {J{\o}rgensen}, {Linnartz}, \&
  {van Dishoeck}}]{ligterink2017}
{Ligterink}, N.~F.~W., {Coutens}, A., {Kofman}, V., {et~al.} 2017, \mnras, 469,
  2219

\bibitem[{{Ligterink} {et~al.}(2020){Ligterink}, {El-Abd}, {Brogan}, {Hunter},
  {Remijan}, {Garrod}, \& {McGuire}}]{ligterink2020}
{Ligterink}, N. F.~W., {El-Abd}, S.~J., {Brogan}, C.~L., {et~al.} 2020, \apj,
  901, 37

\bibitem[{{Ligterink} {et~al.}(2018){Ligterink}, {Terwisscha van Scheltinga},
  {Taquet}, {J{\o}rgensen}, {Cazaux}, {van Dishoeck}, \&
  {Linnartz}}]{ligterink2018}
{Ligterink}, N.~F.~W., {Terwisscha van Scheltinga}, J., {Taquet}, V., {et~al.}
  2018, \mnras, 480, 3628

\bibitem[{{L{\'o}pez-Sepulcre} {et~al.}(2019){L{\'o}pez-Sepulcre}, {Balucani},
  {Ceccarelli}, {Codella}, {Dulieu}, \& {Theul{\'e}}}]{lopez-sepulcre2019}
{L{\'o}pez-Sepulcre}, A., {Balucani}, N., {Ceccarelli}, C., {et~al.} 2019, ACS
  Earth and Space Chemistry, 3, 2122

\bibitem[{{L{\'o}pez-Sepulcre} {et~al.}(2015){L{\'o}pez-Sepulcre}, {Jaber},
  {Mendoza}, {Lefloch}, {Ceccarelli}, {Vastel}, {Bachiller}, {Cernicharo},
  {Codella}, {Kahane}, {Kama}, \& {Tafalla}}]{lopez-sepulcre2015}
{L{\'o}pez-Sepulcre}, A., {Jaber}, A.~A., {Mendoza}, E., {et~al.} 2015, \mnras,
  449, 2438

\bibitem[{{Majumdar} {et~al.}(2018){Majumdar}, {Loison}, {Ruaud}, {Gratier},
  {Wakelam}, \& {Coutens}}]{majumdar2018}
{Majumdar}, L., {Loison}, J.~C., {Ruaud}, M., {et~al.} 2018, \mnras, 473, L59

\bibitem[{{Manigand} {et~al.}(2020){Manigand}, {J{\o}rgensen}, {Calcutt},
  {M{\"u}ller}, {Ligterink}, {Coutens}, {Drozdovskaya}, {van Dishoeck}, \&
  {Wampfler}}]{manigand2020}
{Manigand}, S., {J{\o}rgensen}, J.~K., {Calcutt}, H., {et~al.} 2020, \aap, 635,
  A48

\bibitem[{{Marcelino} {et~al.}(2007){Marcelino}, {Cernicharo}, {Ag{\'u}ndez},
  {Roueff}, {Gerin}, {Mart{\'\i}n-Pintado}, {Mauersberger}, \&
  {Thum}}]{marcelino2007}
{Marcelino}, N., {Cernicharo}, J., {Ag{\'u}ndez}, M., {et~al.} 2007, \apjl,
  665, L127

\bibitem[{{Maris}(2004)}]{maris2004}
{Maris}, A. 2004, Physical Chemistry Chemical Physics (Incorporating Faraday
  Transactions), 6, 2611

\bibitem[{{Mart{\'\i}n} {et~al.}(2019){Mart{\'\i}n}, {Mart{\'\i}n-Pintado},
  {Blanco-S{\'a}nchez}, {Rivilla}, {Rodr{\'\i}guez-Franco}, \&
  {Rico-Villas}}]{martin2019}
{Mart{\'\i}n}, S., {Mart{\'\i}n-Pintado}, J., {Blanco-S{\'a}nchez}, C.,
  {et~al.} 2019, \aap, 631, A159

\bibitem[{{Mart{\'\i}n-Dom{\'e}nech} {et~al.}(2020){Mart{\'\i}n-Dom{\'e}nech},
  {{\"O}berg}, \& {Rajappan}}]{martin-domenech2020}
{Mart{\'\i}n-Dom{\'e}nech}, R., {{\"O}berg}, K.~I., \& {Rajappan}, M. 2020,
  \apj, 894, 98

\bibitem[{{Mart{\'\i}n-Dom{\'e}nech} {et~al.}(2017){Mart{\'\i}n-Dom{\'e}nech},
  {Rivilla}, {Jim{\'e}nez-Serra}, {Qu{\'e}nard}, {Testi}, \&
  {Mart{\'\i}n-Pintado}}]{martin-domenech2017}
{Mart{\'\i}n-Dom{\'e}nech}, R., {Rivilla}, V.~M., {Jim{\'e}nez-Serra}, I.,
  {et~al.} 2017, \mnras, 469, 2230

\bibitem[{{McGuire}(2018)}]{mcguire2018}
{McGuire}, B.~A. 2018, \apjs, 239, 17

\bibitem[{{McMullin} {et~al.}(2007){McMullin}, {Waters}, {Schiebel}, {Young},
  \& {Golap}}]{mcmullin2007}
{McMullin}, J.~P., {Waters}, B., {Schiebel}, D., {Young}, W., \& {Golap}, K.
  2007, in Astronomical Society of the Pacific Conference Series, Vol. 376,
  Astronomical Data Analysis Software and Systems XVI, ed. R.~A. {Shaw},
  F.~{Hill}, \& D.~J. {Bell}, 127

\bibitem[{{Mininni} {et~al.}(2020){Mininni}, {Beltr{\'a}n}, {Rivilla},
  {S{\'a}nchez-Monge}, {Fontani}, {M{\"o}ller}, {Cesaroni}, {Schilke}, {Viti},
  {Jim{\'e}nez-Serra}, {Colzi}, {Lorenzani}, \& {Testi}}]{mininni2020}
{Mininni}, C., {Beltr{\'a}n}, M.~T., {Rivilla}, V.~M., {et~al.} 2020, \aap,
  644, A84

\bibitem[{{Miyake} \& {Nakagawa}(1993)}]{miyake1993}
{Miyake}, K. \& {Nakagawa}, Y. 1993, \icarus, 106, 20

\bibitem[{{Moskienko} \& {Dyubko}(1991)}]{moskienko1991}
{Moskienko}, E.~M. \& {Dyubko}, S.~F. 1991, Radiophysics and Quantum
  Electronics, 34, 181

\bibitem[{{Motiyenko} {et~al.}(2012){Motiyenko}, {Tercero}, {Cernicharo}, \&
  {Margul{\`e}s}}]{motiyenko2012}
{Motiyenko}, R.~A., {Tercero}, B., {Cernicharo}, J., \& {Margul{\`e}s}, L.
  2012, \aap, 548, A71

\bibitem[{{M{\"u}ller} {et~al.}(2005){M{\"u}ller}, {Schl{\"o}der}, {Stutzki},
  \& {Winnewisser}}]{muller2005}
{M{\"u}ller}, H. S.~P., {Schl{\"o}der}, F., {Stutzki}, J., \& {Winnewisser}, G.
  2005, Journal of Molecular Structure, 742, 215

\bibitem[{{M{\"u}ller} {et~al.}(2001){M{\"u}ller}, {Thorwirth}, {Roth}, \&
  {Winnewisser}}]{muller2001}
{M{\"u}ller}, H.~S.~P., {Thorwirth}, S., {Roth}, D.~A., \& {Winnewisser}, G.
  2001, \aap, 370, L49

\bibitem[{{Nazari} {et~al.}(2021){Nazari}, {van Gelder}, {van Dishoeck},
  {Tabone}, {van 't Hoff}, {Ligterink}, {Beuther}, {Boogert}, {Garatti},
  {Klaassen}, {Linnartz}, {Taquet}, \& {Tychoniec}}]{nazari2021}
{Nazari}, P., {van Gelder}, M.~L., {van Dishoeck}, E.~F., {et~al.} 2021, arXiv
  e-prints, arXiv:2104.03326

\bibitem[{{Nguyen} {et~al.}(2011){Nguyen}, {Abbott}, {Dawley}, {Orlando},
  {Leszczynski}, \& {Nguyen}}]{nguyen2011}
{Nguyen}, V.~S., {Abbott}, H.~L., {Dawley}, M.~M., {et~al.} 2011, Journal of
  Physical Chemistry A, 115, 841

\bibitem[{{Nguyen-Q-Rieu} {et~al.}(1991){Nguyen-Q-Rieu}, {Henkel}, {Jackson},
  \& {Mauersberger}}]{nguyen-q-rieu1991}
{Nguyen-Q-Rieu}, {Henkel}, C., {Jackson}, J.~M., \& {Mauersberger}, R. 1991,
  \aap, 241, L33

\bibitem[{Niedenhoff {et~al.}(1996)Niedenhoff, Yamada, \&
  Winnewisser}]{niedenhoff1996pure}
Niedenhoff, M., Yamada, K., \& Winnewisser, G. 1996, J. Mol. Spectrosc., 176,
  342

\bibitem[{{Niedenhoff} {et~al.}(1995){Niedenhoff}, {Yamada}, {Belov}, \&
  {Winnewisser}}]{niedenhoff1995}
{Niedenhoff}, M., {Yamada}, K.~M.~T., {Belov}, S.~P., \& {Winnewisser}, G.
  1995, Journal of Molecular Spectroscopy, 174, 151

\bibitem[{{Noble} {et~al.}(2015){Noble}, {Theule}, {Congiu}, {Dulieu},
  {Bonnin}, {Bassas}, {Duvernay}, {Danger}, \& {Chiavassa}}]{noble2015}
{Noble}, J.~A., {Theule}, P., {Congiu}, E., {et~al.} 2015, \aap, 576, A91

\bibitem[{{Osorio} {et~al.}(2009){Osorio}, {Anglada}, {Lizano}, \&
  {D'Alessio}}]{osorio2009}
{Osorio}, M., {Anglada}, G., {Lizano}, S., \& {D'Alessio}, P. 2009, \apj, 694,
  29

\bibitem[{Pascal {et~al.}(2005)Pascal, Boiteau, \& Commeyras}]{pascal2005}
Pascal, R., Boiteau, L., \& Commeyras, A. 2005, From the Prebiotic Synthesis of
  $\alpha$-Amino Acids Towards a PrimitiveTranslation Apparatus for the
  Synthesis of Peptides, ed. P.~Walde (Berlin, Heidelberg: Springer Berlin
  Heidelberg), 69--122

\bibitem[{{P{\'e}rez} {et~al.}(2012){P{\'e}rez}, {Carpenter}, {Chand ler},
  {Isella}, {Andrews}, {Ricci}, {Calvet}, {Corder}, {Deller}, {Dullemond},
  {Greaves}, {Harris}, {Henning}, {Kwon}, {Lazio}, {Linz}, {Mundy}, {Sargent},
  {Storm}, {Testi}, \& {Wilner}}]{perez2012}
{P{\'e}rez}, L.~M., {Carpenter}, J.~M., {Chand ler}, C.~J., {et~al.} 2012,
  \apjl, 760, L17

\bibitem[{Pickett(1991)}]{pickett1991}
Pickett, H.~M. 1991, J. Mol. Spectrosc., 148, 371

\bibitem[{{Pickett} {et~al.}(1998){Pickett}, {Poynter}, {Cohen}, {Delitsky},
  {Pearson}, \& {M{\"u}ller}}]{pickett1998}
{Pickett}, H.~M., {Poynter}, R.~L., {Cohen}, E.~A., {et~al.} 1998, \jqsrt, 60,
  883

\bibitem[{{Quan} \& {Herbst}(2007)}]{quan2007}
{Quan}, D. \& {Herbst}, E. 2007, \aap, 474, 521

\bibitem[{{Qu{\'e}nard} {et~al.}(2018){Qu{\'e}nard}, {Jim{\'e}nez-Serra},
  {Viti}, {Holdship}, \& {Coutens}}]{quenard2018}
{Qu{\'e}nard}, D., {Jim{\'e}nez-Serra}, I., {Viti}, S., {Holdship}, J., \&
  {Coutens}, A. 2018, \mnras, 474, 2796

\bibitem[{{Raunier} {et~al.}(2004){Raunier}, {Chiavassa}, {Duvernay}, {Borget},
  {Aycard}, {Dartois}, \& {d'Hendecourt}}]{raunier2004}
{Raunier}, S., {Chiavassa}, T., {Duvernay}, F., {et~al.} 2004, \aap, 416, 165

\bibitem[{{Redondo} {et~al.}(2014){Redondo}, {Barrientos}, \&
  {Largo}}]{redondo2014}
{Redondo}, P., {Barrientos}, C., \& {Largo}, A. 2014, \apj, 793, 32

\bibitem[{{Remijan} {et~al.}(2014){Remijan}, {Snyder}, {McGuire}, {Kuo},
  {Looney}, {Friedel}, {Golubiatnikov}, {Lovas}, {Ilyushin}, {Alekseev},
  {Dyubko}, {McCall}, \& {Hollis}}]{remijan2014}
{Remijan}, A.~J., {Snyder}, L.~E., {McGuire}, B.~A., {et~al.} 2014, \apj, 783,
  77

\bibitem[{{Requena-Torres} {et~al.}(2006){Requena-Torres},
  {Mart{\'\i}n-Pintado}, {Rodr{\'\i}guez-Franco}, {Mart{\'\i}n},
  {Rodr{\'\i}guez-Fern{\'a}ndez}, \& {de Vicente}}]{requena-torres2006}
{Requena-Torres}, M.~A., {Mart{\'\i}n-Pintado}, J., {Rodr{\'\i}guez-Franco},
  A., {et~al.} 2006, \aap, 455, 971

\bibitem[{{Rimola} {et~al.}(2018){Rimola}, {Skouteris}, {Balucani},
  {Ceccarelli}, {Enrique-Romero}, {Taquet}, \& {Ugliengo}}]{rimola2018}
{Rimola}, A., {Skouteris}, D., {Balucani}, N., {et~al.} 2018, ACS Earth and
  Space Chemistry, 2, 720

\bibitem[{{Rivilla} {et~al.}(2017){Rivilla}, {Beltr{\'a}n}, {Cesaroni},
  {Fontani}, {Codella}, \& {Zhang}}]{rivilla2017}
{Rivilla}, V.~M., {Beltr{\'a}n}, M.~T., {Cesaroni}, R., {et~al.} 2017, \aap,
  598, A59

\bibitem[{Rivilla {et~al.}(2021)Rivilla, Jim{\'e}nez-Serra,
  Mart{\'\i}n-Pintado, Briones, Rodr{\'\i}guez-Almeida, Rico-Villas, Tercero,
  Zeng, Colzi, de~Vicente, Mart{\'\i}n, \& Requena-Torres}]{rivilla2021}
Rivilla, V.~M., Jim{\'e}nez-Serra, I., Mart{\'\i}n-Pintado, J., {et~al.} 2021,
  Proceedings of the National Academy of Sciences, 118
  [\eprint{https://www.pnas.org/content/118/22/e2101314118.full.pdf}]

\bibitem[{{Rivilla} {et~al.}(2019){Rivilla}, {Mart{\'\i}n-Pintado},
  {Jim{\'e}nez-Serra}, {Zeng}, {Mart{\'\i}n}, {Armijos-Abenda{\~n}o},
  {Requena-Torres}, {Aladro}, \& {Riquelme}}]{rivilla2019}
{Rivilla}, V.~M., {Mart{\'\i}n-Pintado}, J., {Jim{\'e}nez-Serra}, I., {et~al.}
  2019, \mnras, 483, L114

\bibitem[{{Rodr{\'\i}guez-Almeida} {et~al.}(2021){Rodr{\'\i}guez-Almeida},
  {Jim{\'e}nez-Serra}, {Rivilla}, {Mart{\'\i}n-Pintado}, {Zeng}, {Tercero}, {de
  Vicente}, {Colzi}, {Rico-Villas}, {Mart{\'\i}n}, \&
  {Requena-Torres}}]{rodriguez-almeida2021}
{Rodr{\'\i}guez-Almeida}, L.~F., {Jim{\'e}nez-Serra}, I., {Rivilla}, V.~M.,
  {et~al.} 2021, \apjl, 912, L11

\bibitem[{{Rodr{\'\i}guez-Fern{\'a}ndez}
  {et~al.}(2010){Rodr{\'\i}guez-Fern{\'a}ndez}, {Tafalla}, {Gueth}, \&
  {Bachiller}}]{rodriguez-fernandez2010}
{Rodr{\'\i}guez-Fern{\'a}ndez}, N.~J., {Tafalla}, M., {Gueth}, F., \&
  {Bachiller}, R. 2010, \aap, 516, A98

\bibitem[{{Rolffs} {et~al.}(2011){Rolffs}, {Schilke}, {Zhang}, \&
  {Zapata}}]{rolffs2011}
{Rolffs}, R., {Schilke}, P., {Zhang}, Q., \& {Zapata}, L. 2011, \aap, 536, A33

\bibitem[{Sakaizumi {et~al.}(1976)Sakaizumi, Yamada, Ushida, Ohashi, \&
  Yamaguchi}]{sakaizumi1976microwave}
Sakaizumi, T., Yamada, O., Ushida, K., Ohashi, O., \& Yamaguchi, I. 1976, Bull.
  Chem. Soc. Jpn, 49, 2908

\bibitem[{{S{\'a}nchez-Monge} {et~al.}(2018){S{\'a}nchez-Monge}, {Schilke},
  {Ginsburg}, {Cesaroni}, \& {Schmiedeke}}]{sanchez-monge2018}
{S{\'a}nchez-Monge}, {\'A}., {Schilke}, P., {Ginsburg}, A., {Cesaroni}, R., \&
  {Schmiedeke}, A. 2018, \aap, 609, A101

\bibitem[{{Sanz-Novo} {et~al.}(2020){Sanz-Novo}, {Belloche}, {Alonso},
  {Kolesnikov{\'a}}, {Garrod}, {Mata}, {M{\"u}ller}, {Menten}, \&
  {Gong}}]{sanz-novo2020}
{Sanz-Novo}, M., {Belloche}, A., {Alonso}, J.~L., {et~al.} 2020, \aap, 639,
  A135

\bibitem[{{Skouteris} {et~al.}(2017){Skouteris}, {Vazart}, {Ceccarelli},
  {Balucani}, {Puzzarini}, \& {Barone}}]{skouteris2017}
{Skouteris}, D., {Vazart}, F., {Ceccarelli}, C., {et~al.} 2017, \mnras, 468, L1

\bibitem[{{Snyder} \& {Buhl}(1972)}]{snyder1972}
{Snyder}, L.~E. \& {Buhl}, D. 1972, \apj, 177, 619

\bibitem[{{Stubgaard}(1978)}]{stubgaard1978}
{Stubgaard}, M. 1978, PhD thesis, Københavns Universitet

\bibitem[{{Suhasaria} \& {Mennella}(2020)}]{suhasaria2020}
{Suhasaria}, T. \& {Mennella}, V. 2020, \aap, 641, A88

\bibitem[{{Tielens} \& {Hagen}(1982)}]{tielens1982}
{Tielens}, A.~G.~G.~M. \& {Hagen}, W. 1982, \aap, 114, 245

\bibitem[{{Turner} {et~al.}(1999){Turner}, {Terzieva}, \&
  {Herbst}}]{turner1999}
{Turner}, B.~E., {Terzieva}, R., \& {Herbst}, E. 1999, \apj, 518, 699

\bibitem[{{Vorob'eva} \& {Dyubko}(1994)}]{vorobeva1994}
{Vorob'eva}, E.~M. \& {Dyubko}, S.~F. 1994, Radiophysics and Quantum
  Electronics, 37, 155

\bibitem[{Wallner {et~al.}(2020)Wallner, Feige, Fifield, Froehlich, Golser,
  Hotchkis, Koll, Leckenby, Martschini, Merchel, Panjkov, Pavetich, Rugel, \&
  Tims}]{wallner2020}
Wallner, A., Feige, J., Fifield, L.~K., {et~al.} 2020, Proceedings of the
  National Academy of Sciences, 117, 21873

\bibitem[{{Wilson}(1999)}]{wilson1999}
{Wilson}, T.~L. 1999, Reports on Progress in Physics, 62, 143

\bibitem[{{Wilson} \& {Rood}(1994)}]{wilson1994}
{Wilson}, T.~L. \& {Rood}, R. 1994, \araa, 32, 191

\bibitem[{{Winnewisser} {et~al.}(2005){Winnewisser}, {Medvedev}, {De Lucia},
  {Herbst}, {Koput}, {Sastry}, \& {Butler}}]{winnewisser2005}
{Winnewisser}, M., {Medvedev}, I.~R., {De Lucia}, F.~C., {et~al.} 2005, \apjs,
  159, 189

\bibitem[{Yamada(1977)}]{yamada1977type}
Yamada, K. 1977, J. Mol. Spectrosc., 68, 423

\bibitem[{Yamada \& Winnewisser(1977)}]{yamada1977b}
Yamada, K. \& Winnewisser, M. 1977, J. Mol. Spectrosc., 68, 307

\bibitem[{{Yan} {et~al.}(2019){Yan}, {Zhang}, {Henkel}, {Mufakharov}, {Jia},
  {Tang}, {Wu}, {Li}, {Zeng}, {Wang}, {Li}, {Huang}, \& {Jian}}]{yan2019}
{Yan}, Y.~T., {Zhang}, J.~S., {Henkel}, C., {et~al.} 2019, \apj, 877, 154

\bibitem[{{Zeng} {et~al.}(2018){Zeng}, {Jim{\'e}nez-Serra}, {Rivilla},
  {Mart{\'\i}n}, {Mart{\'\i}n-Pintado}, {Requena-Torres},
  {Armijos-Abenda{\~n}o}, {Riquelme}, \& {Aladro}}]{zeng2018}
{Zeng}, S., {Jim{\'e}nez-Serra}, I., {Rivilla}, V.~M., {et~al.} 2018, \mnras,
  478, 2962

\bibitem[{{Zeng} {et~al.}(2020){Zeng}, {Zhang}, {Jim{\'e}nez-Serra}, {Tercero},
  {Lu}, {Mart{\'\i}n-Pintado}, {de Vicente}, {Rivilla}, \& {Li}}]{zeng2020}
{Zeng}, S., {Zhang}, Q., {Jim{\'e}nez-Serra}, I., {et~al.} 2020, \mnras, 497,
  4896

\end{thebibliography}
 
\begin{appendix}
 \clearpage
 
\section{Continuum determination} 
\label{cont-sub}
\begin{figure*}
\centering
\includegraphics[width=35pc]{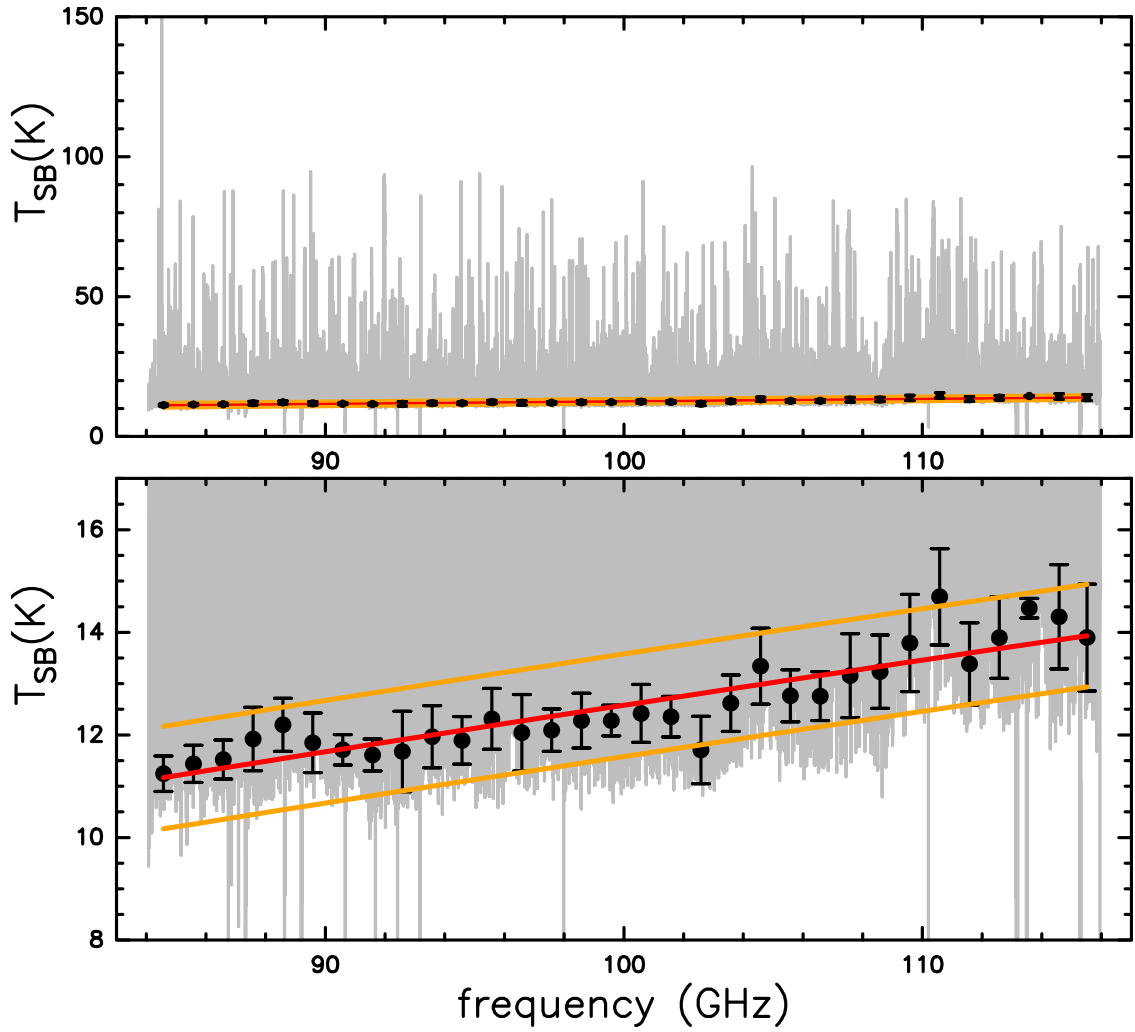}
\caption{\emph{Top panel}: Continuum level derived with STATCONT overimposed to the GUAPOS spectrum. \emph{Bottom panel}: y-axis zoom of the top panel. The red solid line represents the result from the linear regression fit, as explained in Appendix \ref{cont-sub}, and the orange lines represent the erorr of $\pm$1 K on the fit.}
\label{fig-contsub}
\end{figure*}

In this appendix we present the results from the continuum determination and subtraction procedure.
First, we have divided the final spectrum obtained by \citet{mininni2020} in 32 spectral windows of 1 GHz each. Then, we have applied the corrected sigma clipping method (c-SMC) of STATCONT (\citealt{sanchez-monge2018}) to each of them, which produces continuum-subtracted spectra. Secondly, we have averaged them together with the MADCUBA software. The final spectrum is shown in Fig.~\ref{fig-total-spectra-1}.

STATCONT gives also the continuum emission levels, with its uncertainty, for each of the 32 spectral windows, which are shown in Fig.~\ref{fig-contsub} as black dots. The synthesized beam brightness temperature can be described with the function:
\begin{equation}
\label{func-T}
T_{\rm SB}(\nu)=T_{\rm SB}(\nu_{0}) \biggl(\frac{\nu}{\nu_{0}}\biggr)^{\beta},
\end{equation}
where $\nu_{0}$=84.579 GHz, and $\beta$ is the spectral index of the dust opacity $\kappa_{\rm \nu}$ ($\kappa_{\rm \nu} \propto \nu^{\rm \beta}$), which is related to the slope $\alpha$ of the spectral energy distribution ($S_{\rm \nu} \propto \nu^{\rm \alpha}$), by
\begin{equation}
\alpha = 2 + \beta,
\end{equation}
for optically thin dust emission (e.g. \citealt{miyake1993}). 
Equation~\ref{func-T} can be written as:
\begin{equation}
\log(T_{\rm SB}) = \log (T_{\rm SB}(\nu_{0})) + \beta \log\biggl(\frac{\nu}{\nu_{0}}\biggr),
\end{equation}
where log indicates the logarithm to base 10.
We applied the linear regression fit to the latter equation, obtaining $\beta$=0.71$\pm$0.06 and $\log(T_{\rm SB}(\nu_{0})$)=1.048$\pm$0.005, which corresponds to $T_{\rm SB}(\nu_{0})$=11$\pm$1. 

The millimeter dust opacity slope for small ISM dust grains is $\beta \sim$1.7 (\citealt{li2001}). The $\beta$ of 0.7 obtained towards G31 is lower, and could indicate a larger sized grain (centimeter-size) distribution (e.g. \citealt{perez2012}). This value is also consistent with what found by \citet{beltran2013} ($\beta$=0.8) from the SED fitting towards the G29.96-0.02 hot molecular core.

The red solid line of Fig.~\ref{fig-contsub} represents Eq.~\eqref{func-T}, with $\beta$=0.71. The orange solid lines correspond to the error of $\pm$1 found from the fit to $T_{\rm SB}(\nu_{0})$. The error of 1 K obtained with this procedure is consistent with the uncertainty of 1.2 K obtained from the combination of the spectra to derive the final spectra performed by \citet{mininni2020}.
Thus, being conservative, we adopted an uncertainty value of $\pm$1.2 K for the continuum determination, which corresponds to the 11\% of $T_{\rm SB}(\nu_{0})$, as an additional error in the derivation of the molecular parameters from the fit procedure (see Sects.~\ref{analysis} and \ref{results}).

\clearpage
\section{Spectroscopy}
\label{spectroscopy}

\begin{table*}
\begin{center}
\caption{Spectroscopy references for the molecules studied in this work.}
\begin{tabular}{cccccc}
  \hline
  Molecule & Catalogue\tablefootmark{a} & Entry & Date & Line list & Dipole moment \\ 
  & & & & reference & reference \\
  \hline 
   \multicolumn{6}{c}{Isocyanic acid}  \\
   \hline
 HNCO, $v$=0 & CDMS & 43511 & May 2009& (1), (2), (3), (4) & (3)\\
 HNCO, $v_{4}$=1 & MADCUBA & -- & December 2020 & (5), (6), (7), Appendix \ref{spectroscopy-HNCO} & (3) \\
HNCO, $v_{5}$=1 & MADCUBA & -- & December 2020 & Appendix \ref{spectroscopy-HNCO} & (3)  \\
 HNCO, $v_{6}$=1 & MADCUBA & -- & December 2020 & Appendix \ref{spectroscopy-HNCO} & (3) \\
 HN$^{13}$CO, $v$=0 & JPL & 44008 & July 1987 & (3) & (3)\\
 H$^{15}$NCO, $v$=0 & JPL & 44007 & July 1987 & (3) & (3)\\
 \hline
  \multicolumn{6}{c}{Formamide}  \\
  \hline
 \fm, $v$=0 & CDMS & 455512 & April 2013 & (8), (9), (10), (11), & (16)\\
 &&& & (12), (13), (14), (15)  & \\
 \fm, $v_{12}$=1 & CDMS & 45516 & April 2013 & (9), (11), (12), (14), (15) & (16)\\
 H$^{13}$C(O)NH$_{2}$, $v$=0 & CDMS & 46512 & April 2013 & (10), (13), (14), (15), (17) & (16) \\
 \hline
  \multicolumn{6}{c}{Methyl isocyanate}  \\
  \hline
 \mi, $v_{\rm b}$=0& CDMS & 57505 & March 2016 &  (18), (19), (20) & (21) \\
 \mi, $v_{\rm b}$=1 & CDMS & 57506 & March 2016 & (18), (19), (20) & (21) \\
 $^{13}$CH$_{3}$NCO, $v_{\rm b}$=0,1 & MADCUBA & --& December 2020 & (20) & (21) \\
 CH$_{3}$N$^{13}$CO, $v_{\rm b}$=0,1 & MADCUBA & --& December 2020 & (20)& (21) \\
 \hline
  \multicolumn{6}{c}{Acetamide}  \\
  \hline
 \am, $v$=0, $v_{\rm t}$=1, 2 & MADCUBA & -- &  December 2020 &
  (22), (23)& (22)\\
 \hline
  \multicolumn{6}{c}{N-methylformamide}  \\
  \hline
 \nm,$v$=0, $v_{\rm t}$=1, 2 & MADCUBA & -- &  December 2020 &(23)  &  (23)\\
 \hline
  \multicolumn{6}{c}{Ethyl isocyanate}  \\
  \hline
CH$_{3}$CH$_{2}$NCO, $v$=0 & MADCUBA & -- & December 2020 & 
(24), Appendix \ref{spectroscopy-CH3NCO}& (25) \\
\hline
  \ur  $v$=0 & CDMS &60517 & October 2017 & (26), (27), (28), (29)& (27) \\
    \hline
 \cfm,  $v$=0 & CDMS & 70504 & June 2006 & (30), (31)& (30) \\
    \hline
\gly,  $v$=0 & CDMS & 75517 & December 2020 & (32)& (33) \\
 \hline
  \normalsize
  \label{tab-spectroscopy}
   \end{tabular}
   \end{center}
  \tablebib{
(1)~\citet{lapinov2007}; (2) \citet{niedenhoff1995}; (3) \citet{hocking1975}; (4) \citet{kukolich1971a}; (5) \citet{yamada1977type}; (6) \citet{yamada1977b} ; (7) \citet{niedenhoff1996pure}; (8) \citet{kukolich1971b}; (9) \citet{hirota1974}; (10) \citet{gardner1980}; (11) \citet{moskienko1991}; (12) \citet{vorobeva1994}; (13) \citet{blanco2006}; (14) \citet{kryvda2009}; (15) \citet{motiyenko2012}; (16) \citet{kurland1957}; (17) \citet{stubgaard1978}; (18) \citet{cernicharo2016}; (19) \citet{koput1986}; (20) \citet{kolesnikova2019}; (21) \citet{kasten1986a}; (22) \citet{ilyushin2004}; (23) \citet{belloche2017}; (24) \citet{kolesnikova2018}; (25)\citet{sakaizumi1976microwave}; (26) \citet{remijan2014}; (27) \citet{brown1975}; (28) \citet{kasten1986b}; (29)\citet{kretschmer1996};  (30) \citet{christiansen2005}; (31) \citet{winnewisser2005};  (32) \citet{sanz-novo2020}; (33) \citet{maris2004}.
}
\tablefoot{\tablefoottext{a}{The species labelled with MADCUBA were imported into MADCUBA-SLIM, using the spectroscopic works indicated in the table.}
}
 \end{table*}

The transitions of the molecules studied in this work (\ia, \fm, \mi, \am, \nm, and \ei) were taken from the catalogues and spectroscopic works listed in Table \ref{tab-spectroscopy}. We have taken the entries from the Cologne Database for Molecular Spectroscopy\footnote{\url{http://cdms.astro.uni-koeln.de/classic/}.}(CDMS, \citealt{muller2001, muller2005}; \citealt{endres2016}) and the Jet Propulsion Laboratory\footnote{\url{https://spec.jpl.nasa.gov/ftp/pub/catalog/catdir.html}.} (JPL, \citealt{pickett1998}) catalogues. Moreover, for species whose spectroscopy was not present in the catalogues, we have added to MADCUBA-SLIM entries using available spectroscopic works. A detailed explanation about how we evaluate the partition function of \ia\;and \ei\;is given in Sects.~\ref{spectroscopy-HNCO} and \ref{spectroscopy-CH3NCO}, respectively.

\subsection{\ia\;data and partition function evaluation}
\label{spectroscopy-HNCO}

The spectral predictions of HNCO in the three lowest vibrationally excited states (\citealt{brown1997raman}), namely $v_4=1$ ($E=776.6$~cm$^{-1}$), $v_5=1$ ($E=577.4$~cm$^{-1}$), and $v_6=1$ ($E=656.3$~cm$^{-1}$), have been obtained using the spectroscopic data reported in \citet{yamada1977type,yamada1977b,niedenhoff1996pure}.
The dipole moment components ($\mu_a = 1.58$~D and $\mu_b = 1.35$~D) were assumed to be identical to those determined for the ground vibrational state \citep{hocking1975microwave}.

The ro-vibrational partition function of HNCO has been computed at temperatures, $T$, between 2.725 and 300~K using the rotational partition function values from CDMS (see Col.~2 of Table~\ref{tab:qrot1}) and calculating the vibrational correction (\citealt{gordy1984microwave}) as:

\begin{equation}\label{eq:vibfact}
    Q_\mathrm{vib} = \left[ \prod_{k=1}^{3N-6} \left( 1 - e^{-\frac{h \omega_{\rm k}}{k_{\rm B} T}} \right) \right]^{-1} \,,
\end{equation}
where $\omega_{\rm k}$ is the vibrational harmonic frequency of the $k$-th vibrational mode, $h$ is the Planck constant, and $k_{\rm B}$ is the Boltzmann constant.
The sum runs over the $k = 6$ normal modes of HNCO, with the harmonic frequencies ($\omega_k$) taken from \citet{brown1997raman}.
The values of $Q_\mathrm{vibrot}$ are listed in Table~\ref{tab:qrot1}.

\begin{table}[htb!]
	\centering
	\caption{Partition function values of HNCO computed at different temperatures.}
	\label{tab:qrot1}
	\bigskip
	\begin{tabular}{. .c.}
		\hline\hline \\[-1ex]
		\mcl{1}{c}{$T$ (K)} & \mcl{1}{c}{$Q_\mathrm{rot}$\tablefootmark{a}} & \mcl{1}{c}{$Q_\mathrm{vib}$\tablefootmark{b}} & \mcl{1}{c}{$Q_\mathrm{vibrot}$\tablefootmark{c}} \\[0.5ex]
		\hline \\[-1.5ex]
		300.000 & 2695.3359 & 1.14437 & 3084.4546 \\[0.5ex]
		225.000 & 1742.4282 & 1.04874 & 1827.3538 \\[0.5ex]
		150.000 &  943.7057 & 1.00639 &  949.7374 \\[0.5ex]
		75.000  &  331.9879 & 1.00002 &  331.9943 \\[0.5ex]
		37.500  &  117.3039 & 1.00000 &  117.3039 \\[0.5ex]
		18.750  &   42.8291 & 1.00000 &   42.8291 \\[0.5ex]
		9.375   &   18.4492 & 1.00000 &   18.4492 \\[0.5ex]
		5.000   &    9.8228 & 1.00000 &    9.8228 \\[0.5ex]
		2.725   &    5.5129 & 1.00000 &    5.5129 \\[0.5ex]
		\hline\hline
	\end{tabular}
	\tablefoot{\tablefoottext{a}{Taken from CDMS.} \tablefoottext{b}{Computed using Eq.~\eqref{eq:vibfact}.} \tablefoottext{c}{ Obtained as the product of $Q_\mathrm{rot}$ and $Q_\mathrm{vib}$.}}
\end{table}

\subsection{\ei\;data and partition function evaluation}
\label{spectroscopy-CH3NCO}

The spectral predictions of C$_{2}$H$_{5}$NCO in the ground and the first three torsional excited states have been obtained using the recent spectroscopic data of \citet{kolesnikova2018}, which also contain transitions from \citet{heineking1994microwave}. The values of the dipole moment components, $\mu_a = 2.81(2)$~D and $\mu_b = 0.03$~D, were derived in \citet{sakaizumi1976microwave}.

In order to evaluate the ro-vibrational partition function of C$_{2}$H$_{5}$NCO, an approach similar to that used in \citet{cernicharo2016} has been adopted. First, the rotational partition function was computed analytically using the following equation:

\begin{equation}\label{eq:qrot-analyt}
    Q_\mathrm{rot}(T) = 5.3311 \times 10^6 \times\sqrt{\frac{T^3}{(A \times B \times C)}},
\end{equation}
where the factor 5.3311$\times$10$^6$ is derived from the units used for the temperature ($T$), K, and for the rotational constants $A$, $B$ and $C$, MHz.
The values thus obtained were used to check whether the numerical computation of \texttt{SPCAT} \citep{pickett1991} was close to convergence or not. The two methods gave comparable results, with differences smaller than 1\%. Therefore, we decided to use the values computed numerically.

As far as the vibrational partition function is concerned, we decided to account for only the vibrational states below 400~cm$^{-1}$, as done by \citet{cernicharo2016}. This choice allowed a fair determination of the [CH$_{3}$NCO]/[C$_{2}$H$_{5}$NCO] abundance ratio.

\ei\;possesses three vibrational modes below 400 cm$^{-1}$ ($\sim$600~K), namely the C-N torsion ($E = 40$~cm$^{-1}$ \citealt{kolesnikova2018}), the CNC bending ($E = 122$~cm$^{-1}$ \citealt{durig2010conformational}), and the CH$_{3}$ torsion ($E = 265$~cm$^{-1}$ \citealt{durig2010conformational}). We considered all the multiple and combined excitations of these three modes within our threshold energy of 400~cm$^{-1}$ and computed the value of $Q_\mathrm{vib}$ at different temperatures using the following equation (\citealt{gordy1984microwave}):

\begin{equation}\label{eq:qvibfact2}
    Q_\mathrm{vib} = \sum_{E_{\rm i}=0~\mathrm{cm}^{-1}}^{E_i \le 400~\mathrm{cm}^{-1}} e^{-\frac{E_{\rm i}}{k_{\rm B} T}} \,,
\end{equation}
where $E_{\rm i}$ is the vibrational energy of the $i$-th level.
Finally, $Q_\mathrm{vibrot}$ was obtained as the product of $Q_\mathrm{rot}$ and $Q_\mathrm{vib}$. All these values, computed at temperatures between 2.725 and 300~K, are listed in Table~\ref{tab:qrot2}.

\begin{table}[htb!]
	\centering
	\caption{Partition function values of C$_{2}$H$_{5}$NCO computed at different temperatures.}
	\label{tab:qrot2}
	\bigskip
	\begin{tabular}{. .c.}
		\hline\hline \\[-1ex]
		\mcl{1}{c}{$T$ (K)} & \mcl{1}{c}{$Q_\mathrm{rot}$\tablefootmark{a}} & \mcl{1}{c}{$Q_\mathrm{vib}$\tablefootmark{b}} & \mcl{1}{c}{$Q_\mathrm{vibrot}$\tablefootmark{c}} \\[0.5ex]
		\hline \\[-1.5ex]
		300.000 & 83423.9162 & 9.96669 & 831460.1637 \\[0.5ex]
		225.000 & 54116.8299 & 7.38116 & 399445.2080 \\[0.5ex]
		150.000 & 29365.9661 & 4.54674 & 133519.4629 \\[0.5ex]
		75.000  & 10348.8360 & 2.07522 &  21476.0934 \\[0.5ex]
		37.500  &  3655.0733 & 1.28668 &   4702.9099 \\[0.5ex]
		18.750  &  1292.7145 & 1.04879 &   1355.7917 \\[0.5ex]
		9.375   &   457.8892 & 1.00216 &    458.8790 \\[0.5ex]
		5.000   &   179.0092 & 1.00001 &    179.0110 \\[0.5ex]
		2.725   &    72.5214 & 1.00000 &     72.5214 \\[0.5ex]
		\hline\hline
	\end{tabular}
	\tablefoot{\tablefoottext{a}{Computed with \texttt{SPCAT}.} \tablefoottext{b}{Computed using Eq.~\eqref{eq:qvibfact2}.} \tablefoottext{c}{Obtained as the product of $Q_\mathrm{rot}$ and $Q_\mathrm{vib}$.}}
\end{table}

\clearpage
\section{Identified Spectral Lines}
\begin{figure*}
\centering
\includegraphics[width=35pc]{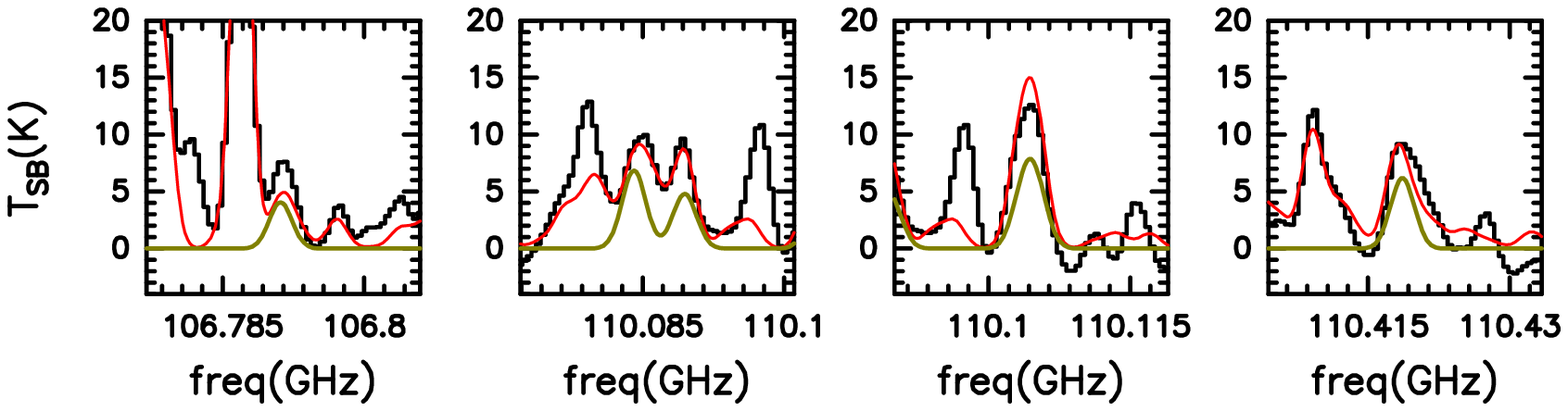}
\caption{Transitions listed in Table~\ref{transitions} and used to fit the $v_{4}$=1 state of isocyanic acid (\ia). The olive curve represents the best LTE fit obtained with MADCUBA (Table \ref{table-fitresult-v0}). The red curve shows the simulated spectrum taking into account all the species identified so far in the region.}
\label{fig-res-ia-va}
\end{figure*}

\begin{figure*}
\centering
\includegraphics[width=25pc]{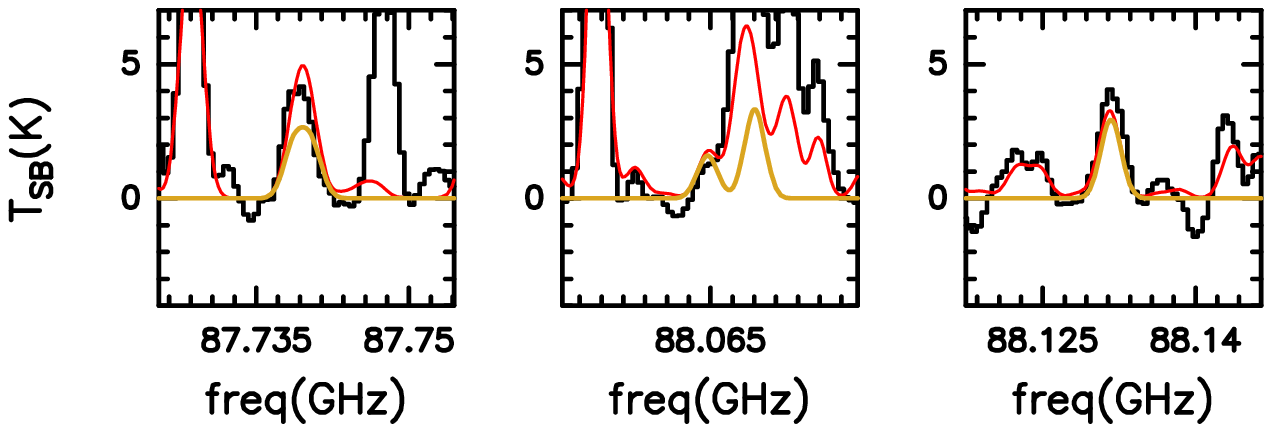}
\caption{Transitions listed in Table~\ref{transitions} and used to fit the $v_{5}$=1 state of isocyanic acid (\ia). The dark golden curve represents the best LTE fit obtained with MADCUBA (Table \ref{table-fitresult-v0}). The red curve shows the simulated spectrum taking into account all the species identified so far in the region.}
\label{fig-res-ia-vb}
\end{figure*}

\begin{figure*}
\centering
\includegraphics[width=25pc]{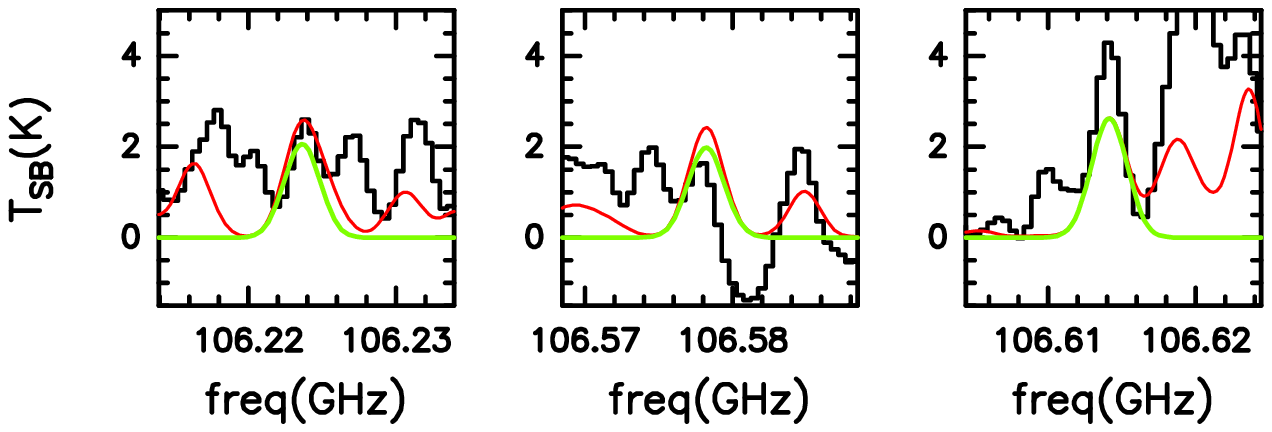}
\caption{Transitions listed in Table~\ref{transitions} and used to fit H$^{15}$NCO. The light green curve represents the best LTE fit obtained with MADCUBA (Table \ref{table-fitresult-v0}). The red curve shows the simulated spectrum taking into account all the species identified so far in the region.}
\label{fig-ia-15N}
\end{figure*}

\begin{figure*}
\centering
\includegraphics[width=35pc]{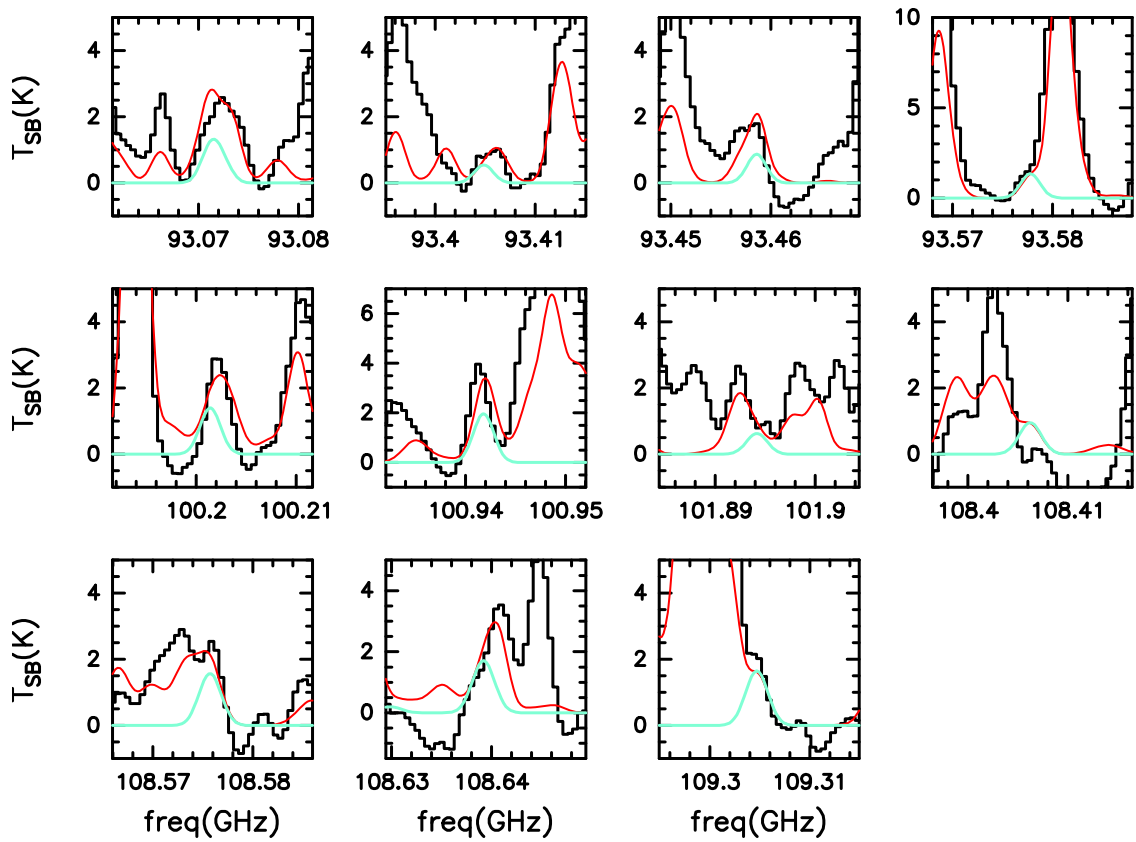}
\caption{Transitions listed in Table~\ref{transitions} and used to fit $^{13}$CH$_{3}$NCO. The turquoise curve represents the best LTE fit obtained with MADCUBA (Table \ref{table-fitresult-v0}). The red curve shows the simulated spectrum taking into account all the species identified so far in the region. Note that at 108.406 GHz the baseline derived from STATCONT is slightly high, 
 and the simulated spectra do not match exactly the observed one.}
\label{fig-mi-13C}
\end{figure*}

In this Appendix we list the unblended or slightly blended transitions of the molecules studied in this work. These transitions are also shown in Figs.~\ref{fig-res-ia}--\ref{fig-res-nm}, \ref{fig-res-ia-va}, \ref{fig-res-ia-vb}, \ref{fig-ia-15N}, and \ref{fig-mi-13C}.


\onecolumn

\begin{longtable}{l c c  c c c}
\caption{\label{transitions}Unblended or slightly blended transitions of the molecules studied in this work. The first column indicates the molecule and vibrational state for which the transitions are listed. The second column shows the frequency of the rotational transition given in the third column. log$I$ is the base 10 logarithm of the integrated intensity at 300 K, $E_{\rm up}$ is the energy of the upper level and $\tau$ is the optical depth of the transition derived from the LTE fit, as described in Sect.~\ref{analysis}.}
\tabularnewline \hline 
Molecule & Frequency	& Transition\tablefootmark{a}	& log$I$	& $E_{\rm up}$		&$\tau$	\\
& (MHz) &   & (nm$^{2}$ MHz) &  (K) & \\
\hline
\endfirsthead
\caption{Continued.}\\ 
\hline 
Molecule & Frequency	& Transition\tablefootmark{a}	& log$I$	 & $E_{\rm up}$		&$\tau$	\\
& (MHz) & & (nm$^{2}$ MHz) &  (K) & \\
\hline
 \endhead
\hline
HNCO, $v$=0 & 87925.2334  &	4,0,4--3,0,3 & -3.8060	& 10.55 &  0.18$\pm$0.04 \\
HNCO, $v$=0 & 109495.9991 &	5,1,5--4,1,4	 & -3.6057	& 59.04 &  0.21$\pm$0.04 \\
HNCO, $v$=0 & 109872.3731 &	5,2,4--4,2,3 & -3.8446	& 186.11&  0.10$\pm$0.02 \\
HNCO, $v$=0 & 109872.7789 &	5,2,3--4,2,2 & -3.8446	& 186.11&  0.10$\pm$0.02 \\
HNCO, $v$=0 & 109905.7485 &	5,0,5--4,0,4 & -3.5221	& 15.83 &  0.27$\pm$0.06 \\
HNCO, $v$=0 & 110298.0874 &	5,1,4--4,1,3 & -3.5995	& 59.16&  0.21$\pm$0.04 \\
\hline
HNCO, $v_{4}$=1 & 106791.4186 & 17,0,17--16,1,16  & -4.8853  & 992.59  &   0.0192$\pm$0.0012 \\
HNCO, $v_{4}$=1 & 110084.3531 & 5,0,5--4,0,4  & -4.7236  &  846.79  &  0.032$\pm$0.002 \\
HNCO, $v_{4}$=1 & 110089.7202 & 5,3,3--4,3,2  & -5.1069  &  977.73 &  0.0114$\pm$0.0007 \\
HNCO, $v_{4}$=1 & 110089.7207 & 5,3,2--4,3,1  & -5.1069  &  977.73  &  0.0114$\pm$0.0007 \\
HNCO, $v_{4}$=1 & 110104.0506 & 5,2,4--4,2,3  & -4.8776  &  900.99  &  0.0213$\pm$0.0014 \\
HNCO, $v_{4}$=1 & 110105.2994 & 5,2,3--4,2,2  & -4.8776  &  900.99  &  0.0213$\pm$0.0014 \\
HNCO, $v_{4}$=1 & 110418.9327 & 5,1,4--4,1,3  & -4.7579  &  860.07  &  0.029$\pm$0.002 \\
\hline
HNCO, $v_{5}$=1 & 87738.9007 & 4,2,3--3,2,2 &  -5.3855 &  1014.08 & 0.0091$\pm$0.0016 \\
HNCO, $v_{5}$=1 & 87740.6295 & 4,2,2--3,2,1 &  -5.3854 &  1014.08 & 0.0091$\pm$0.0016 \\
HNCO, $v_{5}$=1 & 88064.919 & 4,3,2--3,3,1 &  -5.7274 &  1090.86 & 0.0037$\pm$0.0007 \\
HNCO, $v_{5}$=1 & 88064.9193 & 4,3,1--3,3,0 &  -5.7274 &  1090.86 & 0.0037$\pm$0.0007 \\
HNCO, $v_{5}$=1 & 88131.9182 & 4,1,3--3,1,2 &  -5.2254 &  973.16 & 0.014$\pm$0.003 \\
\hline
HN$^{13}$CO, $v$=0 & 87926.8683  &  4,0,4,3--3,0,3,3  &  -5.3655  &  10.55 &  0.0014$\pm$0.0001 \\
HN$^{13}$CO, $v$=0 & 87927.64  &  4,0,4,3--3,0,3,2  &  -4.3075  &  10.55 &  0.0160$\pm$0.0011 \\
HN$^{13}$CO, $v$=0 & 87927.64  &  4,0,4,4--3,0,3,3  &  -4.1894  &  10.55 &  0.0210$\pm$0.0014 \\
HN$^{13}$CO, $v$=0 & 87927.64  &  4,0,4,5--3,0,3,4  &  -4.0743  &  10.55 &  0.027$\pm$0.002 \\
HN$^{13}$CO, $v$=0 & 109497.4203  &  5,1,5,4--4,1,4,4  &  -5.3589  &  58.94 &  0.00107$\pm$0.00007\\
HN$^{13}$CO, $v$=0 & 109498.34  &  5,1,5,5--4,1,4,4  &  -3.9787  &  58.94 &  0.026$\pm$0.002 \\
HN$^{13}$CO, $v$=0 & 109498.34  &  5,1,5,4--4,1,4,3  &  -4.0701  &  58.94 &  0.021$\pm$0.002 \\
HN$^{13}$CO, $v$=0 & 109498.34  &  5,1,5,6--4,1,4,5  &  -3.8884  &  58.94 &  0.032$\pm$0.002 \\
HN$^{13}$CO, $v$=0 & 109499.1736  &  5,1,5,5--4,1,4,5  &  -5.3589  &  58.94 &  0.00107$\pm$0.00007 \\
HN$^{13}$CO, $v$=0 & 109875.3019  &  5,2,4,4--4,2,3,4  &  -5.6014  &  188.51 &  0.00052$\pm$0.00004 \\
HN$^{13}$CO, $v$=0 & 109875.7012  &  5,2,3,4--4,2,2,4  &  -5.6014  &  188.51 &  0.00052$\pm$0.00004 \\
HN$^{13}$CO, $v$=0 & 109875.73  &  5,2,4,5--4,2,3,4  &  -4.2212  &  188.51 &  0.0124$\pm$0.0009 \\
HN$^{13}$CO, $v$=0 & 109875.73  &  5,2,4,6--4,2,3,5  &  -4.131  &  188.51 &  0.0153$\pm$0.0010 \\
HN$^{13}$CO, $v$=0 & 109875.73  &  5,2,4,4--4,2,3,3  &  -4.3126  &  188.51 &  0.0100$\pm$0.0007 \\
HN$^{13}$CO, $v$=0 & 109876.08  &  5,2,3,5--4,2,2,4  &  -4.2212  &  188.51 &  0.0124$\pm$0.0009 \\
HN$^{13}$CO, $v$=0 & 109876.08  &  5,2,3,6--4,2,2,5  &  -4.1309  &  188.51 &  0.0153$\pm$0.0011 \\
HN$^{13}$CO, $v$=0 & 109876.08  &  5,2,3,4--4,2,2,3  &  -4.3126  &  188.51 &  0.0101$\pm$0.0007 \\
HN$^{13}$CO, $v$=0 & 109876.0808  &  5,2,4,5--4,2,3,5  &  -5.6014  &  188.51 &  0.00052$\pm$0.00004 \\
HN$^{13}$CO, $v$=0 & 109876.4797  &  5,2,3,5--4,2,2,5  &  -5.6014  &  188.51 &  0.00052$\pm$0.00004 \\
HN$^{13}$CO, $v$=0 & 109908.0277  &  5,0,5,4--4,0,4,4  &  -5.2755  &  15.83 &  0.00137$\pm$0.00009 \\
HN$^{13}$CO, $v$=0 & 109908.95  &  5,0,5,4--4,0,4,3  &  -3.9867  &  15.83 &  0.027$\pm$0.002 \\
HN$^{13}$CO, $v$=0 & 109908.95  &  5,0,5,5--4,0,4,4  &  -3.8953  &  15.83 &  0.033$\pm$0.002 \\
HN$^{13}$CO, $v$=0 & 109908.95  &  5,0,5,6--4,0,4,5  &  -3.805  &  15.83 &  0.040$\pm$0.003 \\
HN$^{13}$CO, $v$=0 & 109909.5554  &  5,0,5,4--4,0,4,4  &  -5.2755  &  15.83 &  0.0014$\pm$0.0001 \\
HN$^{13}$CO, $v$=0 & 110301.5491  &  5,1,4,4--4,1,3,4  &  -5.3527  &  59.05 &  0.00108$\pm$0.00008 \\
HN$^{13}$CO, $v$=0 & 110301.98  &  5,1,4,5--4,1,3,4 &  -3.9725  &  59.05 &  0.026$\pm$0.002 \\
HN$^{13}$CO, $v$=0 & 110301.98  &  5,1,4,4--4,1,3,3  &  -4.0639  &  59.05 &  0.021$\pm$0.002 \\
HN$^{13}$CO, $v$=0 & 110301.98  &  5,1,4,6--4,1,3,5  &  -3.8822  &  59.058 &  0.032$\pm$0.002 \\
HN$^{13}$CO, $v$=0 & 110302.4762  &  5,1,4,5--4,1,3,5  &  -5.3527  &  59.05 &  0.00108$\pm$0.00007 \\
\hline
H$^{15}$NCO, $v$=0 & 106223.92 & 5,1,5--4,1,4 & -3.5257 & 58.13 & 0.0097$\pm$0.0015 \\
H$^{15}$NCO, $v$=0 & 106578.32 & 5,2,4--4,2,3 & -3.7669 & 186.69 & 0.0047$\pm$0.0007 \\
H$^{15}$NCO, $v$=0 & 106578.6409 & 5,2,3--4,2,2 & -3.7669 & 186.69 & 0.0047$\pm$0.0007 \\
H$^{15}$NCO, $v$=0 & 106614.42 & 5,0,5--4,0,4 & -3.4429 & 15.35 & 0.012$\pm$0.002 \\
\hline
\fm, $v$=0 & 84542.3304 & 4,0,4--3,0,3 & -3.6043 & 10.16 & 0.30$\pm$0.09 \\
\fm, $v$=0 & 84807.7953 & 4,2,3--3,2,2 & -3.7436 & 22.10 & 0.20$\pm$0.06 \\
\fm, $v$=0 & 85093.2721 & 4,2,2--3,2,1 & -3.7407 & 22.12 & 0.20$\pm$0.06 \\
\fm, $v$=0 & 87848.8735 & 4,1,3--3,1,2 & -3.6036 & 13.52 & 0.28$\pm$0.09 \\
\fm, $v$=0 & 105972.5989 & 5,2,4--4,2,3 & -3.4106 & 27.19 & 0.34$\pm$0.10 \\
\fm, $v$=0 & 106107.8695 & 5,4,2--4,4,1 & -3.8292 & 62.95 & 0.11$\pm$0.03 \\
\fm, $v$=0 & 106107.8951 & 5,4,1--4,4,0 & -3.8292 & 62.95 & 0.11$\pm$0.03 \\
\fm, $v$=0 & 106134.4272 & 5,3,3--4,3,2 & -3.5489 & 42.10 & 0.23$\pm$0.07 \\
\fm, $v$=0 & 106141.3997 & 5,3,2--4,3,1 & -3.5489 & 42.10 & 0.23$\pm$0.07\\
\fm, $v$=0 & 106541.6799 & 5,2,3--4,2,2 & -3.4060 & 27.24 & 0.34$\pm$0.11 \\
\hline
\fm, $v_{12}$=1 & 84481.506 & 4,0,4--3,0,3 & -4.2069 & 425.96 & 0.044$\pm$0.017 \\
\fm, $v_{12}$=1 & 85021.4563 & 4,2,2--3,2,1 & -4.3431 & 437.74 & 0.031$\pm$0.012 \\
\fm, $v_{12}$=1 & 87728.805 & 4,1,3--3,1,2 & -4.2066 & 429.27 & 0.042$\pm$0.016 \\
\fm, $v_{12}$=1 & 102051.892 & 5,1,5--4,1,4 & -3.9736 & 433.45 & 0.06$\pm$0.03 \\
\fm, $v_{12}$=1 & 105392.39 & 5,0,5--4,0,4 & -3.9246 & 431.02 & 0.07$\pm$0.03 \\
\fm, $v_{12}$=1 & 105890.6281 & 5,2,4--4,2,3 & -4.0129 & 442.80 & 0.05$\pm$0.02 \\
\fm, $v_{12}$=1 & 106023.1054 & 5,4,2--4,4,1 & -4.4307 & 478.01 & 0.0201$\pm$0.008 \\
\fm, $v_{12}$=1 & 106023.1303 & 5,4,1--4,4,0 & -4.4307 & 478.012 & 0.0201$\pm$0.008 \\
\fm, $v_{12}$=1 & 106049.2003 & 5,3,3--4,3,2 & -4.1509 & 457.48 & 0.039$\pm$0.015 \\
\fm, $v_{12}$=1 & 106056.0192 & 5,3,2--4,3,1 & -4.1509 & 457.48 & 0.039$\pm$0.015 \\
\fm, $v_{12}$=1 & 106448.387 & 5,2,3--4,2,2 & -4.0084 & 442.85 & 0.05$\pm$0.02 \\
\fm, $v_{12}$=1 & 109604.533 & 5,1,4--4,1,3 & -3.9129 & 434.54 & 0.07$\pm$0.03 \\
\hline
H$^{13}$C(O)NH$_{2}$, $v$=0 &  84390.6787 & 4,0,4--3,0,3 &  -3.6117  & 10.14 & 0.025$\pm$0.002 \\
H$^{13}$C(O)NH$_{2}$, $v$=0 &  105260.2704 & 5,0,5--4,0,4 &  -3.3296  & 15.19 & 0.037$\pm$0.003 \\
H$^{13}$C(O)NH$_{2}$, $v$=0 &  105803.157 & 5,2,4--4,2,3 &  -3.4173  & 26.85 & 0.029$\pm$0.003 \\
H$^{13}$C(O)NH$_{2}$, $v$=0 &  105947.1307 & 5,4,2--4,4,1 &  -3.8344  & 61.67 & 0.010$\pm$0.001 \\
H$^{13}$C(O)NH$_{2}$, $v$=0 &  105947.1608 & 5,4,1--4,4,0 &  -3.8344  & 61.67 & 0.010$\pm$0.001 \\
H$^{13}$C(O)NH$_{2}$, $v$=0 &  106411.1376 & 5,2,3--4,2,2 &  -3.4123  & 26.90 & 0.029$\pm$0.003\\
H$^{13}$C(O)NH$_{2}$, $v$=0 &  109656.8289 & 5,1,4--4,1,3 &  -3.3163  & 18.70 & 0.036$\pm$0.003 \\
\hline
\mi, $v_{\rm b}$=0 & 85938.962 & 10,1,10,0--9,1,9,0 & -4.5727 & 28.64 & 0.083$\pm$0.011 \\ 
\mi, $v_{\rm b}$=0 & 86680.19 & 10,0,10,0--9,0,9,0 & -4.5525 & 22.88 & 0.088$\pm$0.012 \\
\mi, $v_{\rm b}$=0 & 86686.556 & 10,0,0,1--9,0,0,1 & -4.5697 & 34.97 & 0.080$\pm$0.011 \\
\mi, $v_{\rm b}$=0 & 86780.778 & 10,2,9,0--9,2,8,0 & -4.6038 & 46.73& 0.070$\pm$0.009 \\
\mi, $v_{\rm b}$=0 & 86805.027 & 10,2,8,0--9,2,7,0 & -4.6035 & 46.74 & 0.070$\pm$0.009 \\
\mi, $v_{\rm b}$=0 & 87016.078 & 10,2,0,1--9,2,0,1 & -4.6219 & 58.81 & 0.063$\pm$0.008 \\
\mi, $v_{\rm b}$=0 & 87114.738 & 10,0,0,2--9,0,0,2 & -4.6289 & 75.86 & 0.057$\pm$0.007 \\
\mi, $v_{\rm b}$=0 & 87116.22 & 9,3,0,1--8,3,0,1 & -4.83 & 84.85 & 0.034$\pm$0.004 \\
\mi, $v_{\rm b}$=0 & 87153.379 & 10,1,0,2--9,1,0,2 & -4.6418 & 81.82 & 0.054$\pm$0.007 \\
\mi, $v_{\rm b}$=0 & 87281.104 & 10,-2,--,1--9,-2,--,1 & -4.6219 & 58.83 & 0.063$\pm$0.008 \\  
\mi, $v_{\rm b}$=0 & 87592.56 & 10,-1,--,0--9,-1,--,0 & -4.7237 & 143.73 & 0.033$\pm$0.004 \\ 
\mi, $v_{\rm b}$=0 & 94529.761 & 11,1,11,0--10,1,10,0 & -4.4541 & 33.18 & 0.097$\pm$0.013 \\ 
\mi, $v_{\rm b}$=0 & 95115.214 & 11,3,8,0--10,3,7,0 & -4.5445 & 81.09 & 0.062$\pm$0.008 \\
\mi, $v_{\rm b}$=0 & 95115.214 & 11,3,9,0--10,3,8,0 & -4.5445 & 81.09 & 0.062$\pm$0.008 \\
\mi, $v_{\rm b}$=0 & 95455.814 & 11,2,10,0--10,2,9,0 & -4.4828 & 51.31 & 0.082$\pm$0.011 \\
\mi, $v_{\rm b}$=0 & 95488.131 & 11,2,9,0--10,2,8,0 & -4.4825 & 51.32 & 0.0822$\pm$0.011 \\
\mi, $v_{\rm b}$=0 & 95765.763 & 11,-1,0,2--10,-1,0,2 & -4.5232 & 86.39 & 0.063$\pm$0.008 \\
\mi, $v_{\rm b}$=0 & 95826.264 & 10,3,0,1--9,3,0,1 & -4.6883 & 89.02 & 0.042$\pm$0.005 \\
\mi, $v_{\rm b}$=0 & 95958.329 & 11,1,--,-3--10,1,--,-3 & -4.6064 & 149.20 & 0.038$\pm$0.004 \\
\mi, $v_{\rm b}$=0 & 96120.05 & 11,2,0,3--10,2,0,3 & -4.6492 & 166.01 & 0.032$\pm$0.003 \\
\mi, $v_{\rm b}$=0 & 96120.05 & 11,2,--,-3--10,2,--,-3 & -4.6504 & 166.87 & 0.032$\pm$0.003 \\
\mi, $v_{\rm b}$=0 & 96253.318 & 11,1,10,0--10,1,9,0 & -4.439 & 33.68 & 0.098$\pm$0.013 \\
\mi, $v_{\rm b}$=0 & 96328.605 & 11,2,0,4--10,2,0,4 & -4.7763 & 253.64 & 0.0155$\pm$0.0015 \\ 
\mi, $v_{\rm b}$=0 & 96328.605 & 11,2,0,4--10,2,0,4 & -4.7763 & 253.64 & 0.0155$\pm$0.0015 \\ 
\mi, $v_{\rm b}$=0 & 102819.142 & 12,-1,--,1--11,-1,--,1 & -4.3571 & 50.45 & 0.102$\pm$0.013 \\
\mi, $v_{\rm b}$=0 & 103022.546 & 12,1,0,1--11,1,0,1 & -4.3571 & 50.45 & 0.102$\pm$0.013 \\
\mi, $v_{\rm b}$=0 & 103119.703 & 12,1,12,0--11,1,11,0 & -4.347 & 38.13 & 0.111$\pm$0.015 \\
\mi, $v_{\rm b}$=0 & 104130.01 & 12,2,11,0--11,2,10,0 & -4.3741 & 56.31 & 0.095$\pm$0.012 \\
\mi, $v_{\rm b}$=0 & 104171.962 & 12,2,10,0--11,2,9,0 & -4.3737 & 56.32 & 0.095$\pm$0.012 \\
\mi, $v_{\rm b}$=0 & 104424.13 & 12,2,0,1--11,2,0,1 & -4.3921 & 68.39 & 0.085$\pm$0.011 \\ 
\mi, $v_{\rm b}$=0 & 104470.566 & 12,-1,--,2--11,-1,--,2 & -4.4162 & 91.39 & 0.072$\pm$0.009 \\
\mi, $v_{\rm b}$=0 & 104532.714 & 12,0,0,2--11,0,0,2 & -4.4046 & 85.43 & 0.076$\pm$0.009 \\
\mi, $v_{\rm b}$=0 & 104535.872 & 11,3,0,1--10,3,0,1 & -4.563 & 93.60 & 0.051$\pm$0.006 \\
\mi, $v_{\rm b}$=0 & 104671.232 & 12,0,0,3--11,0,0,3 & -4.4943 & 147.16 & 0.046$\pm$0.005 \\
\mi, $v_{\rm b}$=0 & 104679.971 & 12,1,--,-3--11,1,--,-3 & -4.4994 & 154.24 & 0.044$\pm$0.005 \\
\mi, $v_{\rm b}$=0 & 104793.013 & 12,-3,--,2--11,-3,--,2 & -4.5102 & 139.05 & 0.046$\pm$0.005 \\
\mi, $v_{\rm b}$=0 & 104842.449 & 12,-2,--,1--11,-2,--,1 & -4.3921 & 68.41 & 0.085$\pm$0.011 \\ 
\mi, $v_{\rm b}$=0 & 104998.886 & 12,1,11,0--11,1,10,0 & -4.3321 & 38.72 & 0.113$\pm$0.015 \\
\mi, $v_{\rm b}$=0 & 105132.553 & 12,-1,--,3--11,-1,--,3 & -4.4982 & 153.39 & 0.044$\pm$0.005 \\ 
\mi, $v_{\rm b}$=0 & 111634.362 & 13,1,0,1--12,1,0,1 & -4.2599 & 55.86 & 0.115$\pm$0.015 \\
\mi, $v_{\rm b}$=0 & 111708.707 & 13,1,13,0--12,1,12,0 & -4.2498 & 43.49 & 0.12$\pm$0.02 \\
\mi, $v_{\rm b}$=0 & 112395.756 & 13,3,10,0--12,3,9,0 & -4.3315 & 91.50 & 0.08$\pm$0.01 \\
\mi, $v_{\rm b}$=0 & 112395.756 & 13,3,11,0--12,3,10,0 & -4.3315 & 91.50 & 0.08$\pm$0.01 \\
\mi, $v_{\rm b}$=0 & 112803.258 & 13,2,12,0--12,2,11,0 & -4.2755 & 61.73 & 0.107$\pm$0.014 \\
\mi, $v_{\rm b}$=0 & 113129.051 & 13,2,0,1--12,2,0,1 & -4.2935 & 73.80 & 0.096$\pm$0.012 \\
\mi, $v_{\rm b}$=0 & 113591.533 & 13,2,0,3--12,2,0,3 & -4.4417 & 176.42 & 0.041$\pm$0.004 \\
\mi, $v_{\rm b}$=0 & 113591.533 & 13,2,--,-3--12,2,--,-3 & -4.443 & 177.28 & 0.041$\pm$0.004 \\
\mi, $v_{\rm b}$=0 & 113631.453 & 13,-2,--,1--12,-2,--,1 & -4.2935 & 73.83 & 0.096$\pm$0.012 \\ 
\hline
\mi, $v_{\rm b}$=1 & 85948.138 & 10,-1,--,1--9,-1,--,1 &  -4.9638 & 304.20 & 0.028$\pm$0.014 \\  
\mi, $v_{\rm b}$=1 & 86624.236 & 10,0,0,1--9,0,0,1 &  -4.9509 &   298.28 & 0.030$\pm$0.015 \\ 
\mi, $v_{\rm b}$=1 & 86674.929 & 10,0,10,0--9,0,9,0 &  -4.932 &   285.04 & 0.034$\pm$0.017 \\
\mi, $v_{\rm b}$=1 & 87105.592 & 10,0,--,-2--9,0,--,-2 &  -5.0152 &   342.76 & 0.018$\pm$0.011 \\ 
\mi, $v_{\rm b}$=1 & 87139.46 & 10,-1,--,-2--9,-1,--,-2 &  -5.0282 &  348.72 & 0.017$\pm$0.011 \\ 
\mi, $v_{\rm b}$=1 & 87535.51 & 10,1,9,0--9,1,8,0 &  -4.9372 & 291.21 & 0.032$\pm$0.016 \\  
\mi, $v_{\rm b}$=1 & 95335.299 & 11,0,11,0--10,0,10,0 &  -4.8142 &  289.62 & 0.04	$\pm$.02 \\ 
\mi, $v_{\rm b}$=1 & 95539.843 & 11,2,9,0--10,2,8,0 &  -4.8627 & 313.46 & 0.029$\pm$0.016 \\  
\mi, $v_{\rm b}$=1 & 102699.98 & 12,3,10,0--11,3,9,0 &  -4.813 & 348.16 & 0.024$\pm$0.015 \\  
\mi, $v_{\rm b}$=1 & 102699.98 & 12,3,9,0--11,3,8,0 &  -4.813 & 348.16 & 0.024$\pm$0.015 \\  
\mi, $v_{\rm b}$=1 & 103108.301 & 12,1,12,0--11,1,11,0 &  -4.7267 &  300.28 & 0.04$\pm$0.02 \\ 
\mi, $v_{\rm b}$=1 & 103765.305 & 12,0,0,1--11,0,0,1 &  -4.7266 & 307.84 & 0.04$\pm$0.02 \\  
\mi, $v_{\rm b}$=1 & 103993.608 & 12,0,12,0--11,0,11,0 &  -4.7078 &  294.61 & 0.04$\pm$0.02 \\ 
\mi, $v_{\rm b}$=1 & 111696.508 & 13,1,13,0--12,1,12,0 &  -4.6294 &  305.64 & 0.05$\pm$0.02 \\
\hline
$^{13}$CH$_{3}$NCO, $v_{\rm b}$=0 & 93071.391 & 11,2,0,1--10,2,0,1 & -4.5009 & 63.27 & 0.0086	0.0013 \\
$^{13}$CH$_{3}$NCO, $v_{\rm b}$=0 & 93405.042 & 11,-3,--,-2--10,-3,--,-2 & -4.6222 & 133.93 & 0.0046$\pm$0.0007 \\
$^{13}$CH$_{3}$NCO, $v_{\rm b}$=0 & 93458.719 & 11,2,0,3--10,2,0,3 & -4.6492 & 165.88 & 0.0037$\pm$0.0006 \\
$^{13}$CH$_{3}$NCO, $v_{\rm b}$=0 & 93458.822 & 11,2,0,3--10,2,0,3 & -4.6504 & 166.74 & 0.0037$\pm$0.0006 \\
$^{13}$CH$_{3}$NCO, $v_{\rm b}$=0 & 93577.977 & 11,1,10,0--10,1,9,0 & -4.439 & 33.55 & 0.0114$\pm$0.0018 \\
$^{13}$CH$_{3}$NCO, $v_{\rm b}$=0 & 100201.619 & 12,1,0,1--11,1,0,1 & -4.3571 & 50.32 & 0.0119$\pm$0.0018 \\
$^{13}$CH$_{3}$NCO, $v_{\rm b}$=0 & 100942.013 & 12,3,9,0--11,3,8,0 & -4.4325 & 85.95 & 0.0083$\pm$0.0013 \\
$^{13}$CH$_{3}$NCO, $v_{\rm b}$=0 & 100942.013 & 12,3,10,0--11,3,9,0 & -4.4325 & 85.95 & 0.0083$\pm$0.0013 \\
$^{13}$CH$_{3}$NCO, $v_{\rm b}$=0 & 101894.392 & 12,-3,--,-2--11,-3,--,-2 & -4.5102 & 138.91 & 0.0053 $\pm$0.0008 \\
$^{13}$CH$_{3}$NCO, $v_{\rm b}$=0 & 108406.444 & 13,2,--,-2--12,2,--,-2 & -4.3527 & 114.44 & 0.0081$\pm$0.0013 \\
$^{13}$CH$_{3}$NCO, $v_{\rm b}$=0 & 108575.945 & 13,1,0,1--12,1,0,1 & -4.2599 & 55.72 & 0.013$\pm$0.002 \\
$^{13}$CH$_{3}$NCO, $v_{\rm b}$=0 & 108639.42 & 13,1,13,0--12,1,12,0 & -4.2498 & 43.34 & 0.015$\pm$0.002 \\
$^{13}$CH$_{3}$NCO, $v_{\rm b}$=0 & 109304.924 & 13,0,0,1--12,0,0,1 & -4.2487 & 49.79 & 0.014$\pm$0.002 \\
\hline
\am, $v$=0 & 86161.9164 & 19,14,6--19,13,7 & -4.9115 & 177.62 & 0.006$\pm$0.004 \\
\am, $v$=0 & 88256.3587 & 13,4,9--13,3,10 & -5.1006 & 74.48 & 0.004$\pm$0.003 \\
\am, $v$=0 & 89977.0936 & 17,8,9--17,7,10 & -4.8582 & 130.04 & 0.007$\pm$0.004 \\
\am, $v$=0 & 94017.835 & 16,14,3--16,13,4 & -5.0692 & 136.42 & 0.004$\pm$0.003 \\
\am, $v$=0 & 97943.8728 & 9,0,9--8,1,8 & -4.6386 & 24.49 & 0.011$\pm$0.007 \\
\am, $v$=0 & 97943.8739 & 9,1,9--8,1,8 & -5.8455 & 24.49 & 0.0007$\pm$0.0005 \\
\am, $v$=0 & 97943.8816 & 9,0,9--8,0,8 & -5.8455 & 24.49 & 0.0007$\pm$0.0005 \\
\am, $v$=0 & 97943.8828 & 9,1,9--8,0,8 & -4.6386 & 24.49 & 0.011$\pm$0.007 \\
\am, $v$=0 & 100405.9333 & 23,13,10--23,12,11 & -4.8242 & 243.83 & 0.007$\pm$0.004 \\
\am, $v$=0 & 103515.4785 & 23,17,6--23,16,7 & -4.8556 & 258.25 & 0.006$\pm$0.003 \\
\am, $v$=0 & 108190.1973 & 10,0,10--9,1,9 & -4.5115 & 35.74 & 0.013$\pm$0.009 \\
\am, $v$=0 & 108190.1975 & 10,1,10--9,1,9 & -5.9118 & 35.74 & 0.0005$\pm$0.0003 \\
\am, $v$=0 & 108190.1993 & 10,0,10--9,0,9 & -5.9118 & 35.74 & 0.0005$\pm$0.0003 \\
\am, $v$=0 & 108190.1996 & 10,1,10--9,0,9 & -4.5115 & 35.74 & 0.013$\pm$0.009 \\
\am, $v$=0 & 108255.2324 & 10,0,10--9,1,9 & -4.5097 & 29.68 & 0.013$\pm$0.009 \\
\am, $v$=0 & 108255.2325 & 10,1,10--9,1,9 & -5.7188 & 29.68 & 0.0008$\pm$0.0005 \\
\am, $v$=0 & 108255.2335 & 10,0,10--9,0,9 & -5.7188 & 29.68 & 0.0008$\pm$0.0005 \\
\am, $v$=0 & 108255.2336 & 10,1,10--9,0,9 & -4.5097 & 29.68 & 0.013$\pm$0.009 \\
\am, $v$=0 & 109024.8757 & 22,13,10--22,12,11 & -4.8380 & 222.25 & 0.006$\pm$0.003 \\
\hline
\am, $v_{\rm t}$=1 & 86858.4501 & 11,3,8--11,2,9 & -5.1832 & 84.12 & 0.003$\pm$0.002 \\
\am, $v_{\rm t}$=1 & 87178.4897 & 8,4,5--7,4,4 & -5.1062 & 88.30 & 0.004$\pm$0.0038 \\
\am, $v_{\rm t}$=1 & 87286.7553 & 8,3,5--7,3,4 & -5.0846 & 87.87 & 0.004$\pm$0.003 \\
\am, $v_{\rm t}$=1 & 92337.5385 & 8,1,7--7,1,7 & -5.0641 & 85.46 & 0.004$\pm$0.003 \\
\am, $v_{\rm t}$=1 & 95161.9961 & 9,8,1--9,8,2 & -5.1689 & 92.62 & 0.003$\pm$0.002 \\
\am, $v_{\rm t}$=1 & 98960.6476 & 8,2,6--7,2,5 & -4.9910 & 87.28 & 0.005$\pm$0.003 \\
\am, $v_{\rm t}$=1 & 99191.9445 & 9,1,9--8,0,8 & -4.6669 & 58.50 & 0.010$\pm$0.007 \\
\am, $v_{\rm t}$=1 & 101612.5722 & 12,10,2--12,10,3 & -4.9402 & 121.72 & 0.005$\pm$0.003 \\
\am, $v_{\rm t}$=1 & 108074.6337 & 14,12,3--14,11,4 & -4.8381 & 146.88 & 0.006$\pm$0.004 \\
\am, $v_{\rm t}$=1 & 110000.2604 & 10,1,9--9,2,8 & -4.6286 & 95.56 & 0.010$\pm$0.006 \\
\am, $v_{\rm t}$=1 & 110000.8655 & 13,8,5--13,7,7 & -5.7590 & 118.19 & 0.0007$\pm$0.0005 \\
\hline
\am, $v_{\rm t}$=2 & 85735.8766 & 16,8,8--16,8,9 & -5.1089 & 229.90 & 0.004$\pm$0.002 \\
\am, $v_{\rm t}$=2 & 103232.6376 & 16,9,7--16,9,8 & -4.9038 & 239.40 & 0.005$\pm$0.003 \\
\am, $v_{\rm t}$=2 & 103790.5952 & 9,0,9--8,0,8 & -4.8414 & 139.75 & 0.006$\pm$0.004 \\
\am, $v_{\rm t}$=2 & 106117.5341 & 16,5,11--16,4,12 & -4.8027 & 198.22 & 0.007$\pm$0.004 \\
\am, $v_{\rm t}$=2 & 109284.7294 & 11,5,7--11,4,8 & -4.9046 & 144.84 & 0.005$\pm$0.003 \\
\am, $v_{\rm t}$=2 & 114811.2701 & 20,7,13--20,6,14 & -4.7166 & 264.28 & 0.007$\pm$0.004 \\
\hline
\nm,$v$=0 & 89031.2613 & 8,2,7--7,2,6 & -4.9300 & 22.13 & 0.002$\pm$0.001 \\
\nm,$v$=0 & 89034.2847 & 20,4,16--20,3,17 & -5.0052 & 131.00 & 0.0020$\pm$0.0008 \\
\nm,$v$=0 & 90764.7877 & 8,6,3--7,6,2 & -5.2730 & 44.28 & 0.0011$\pm$0.0004 \\
\nm,$v$=0 & 90764.8105 & 8,6,2--7,6,1 & -5.2730 & 44.28 & 0.0011$\pm$0.0004 \\
\nm,$v$=0 & 91888.7598 & 8,2,6--7,2,5 & -4.9222 & 27.17 & 0.002$\pm$0.001 \\
\nm,$v$=0 & 93406.3797 & 9,0,9--8,0,8 & -4.8163 & 23.10 & 0.0030$\pm$0.0012 \\
\nm,$v$=0 & 102434.8736 & 9,5,4--8,5,3 & -4.9138 & 41.68 & 0.0022$\pm$0.0009 \\
\nm,$v$=0 & 104818.4432 & 28,6,22--28,5,23 & -4.7632 & 251.09 & 0.0029$\pm$0.0012 \\
\nm,$v$=0 & 106004.4774 & 23,4,19--23,3,20 & -4.8749 & 167.88 & 0.0022$\pm$0.0009 \\
\nm,$v$=0 & 107852.5468 & 9,2,7--8,2,6 & -4.7119 & 28.22 & 0.0033$\pm$0.0013 \\
\nm,$v$=0 & 112875.4833 & 11,0,11--10,0,10 & -4.5701 & 37.13 & 0.0044$\pm$0.0018 \\
 \bottomrule
 \end{longtable}
 \textbf{Note.} $^{a}$For all the molecules the levels are indicated as ($J$, $K_{\rm a}$, $K_{\rm c}$), except for HN$^{13}$CO for which also the hyperfine state is indicated, ($J$, $K_{\rm a}$, $K_{\rm c}$, $F$), and for \mi\;and $^{13}$CH$_{3}$NCO for which the levels are labelled as ($J$, $K_{\rm a}$, $K_{\rm c}$, $m$). The quantum numbers $m$ or $K_a$ may carry a sign; in the latter case, $K_c$ is not reported and we indicated it as $K_{\rm c}$=~"--".
 
\twocolumn

 \clearpage

\section{Tentative detections}
\label{figure-tentative}
In this Appendix we show the transitions used for the fit of the $v_{4}$=1 and $v_{5}$=1 states of \ia, of H$^{15}$NCO, and of $^{13}$CH$_{3}$NCO.

\clearpage
\section{Full GUAPOS spectrum}
\label{spectra}
In this Appendix we show the total observed spectra with the best LTE fit of the molecular species studied in this work. Moreover, the synthesised spectrum taking into account the contribution of all the possible molecular species is also shown.

\begin{figure*}
\centering
\includegraphics[width=40pc]{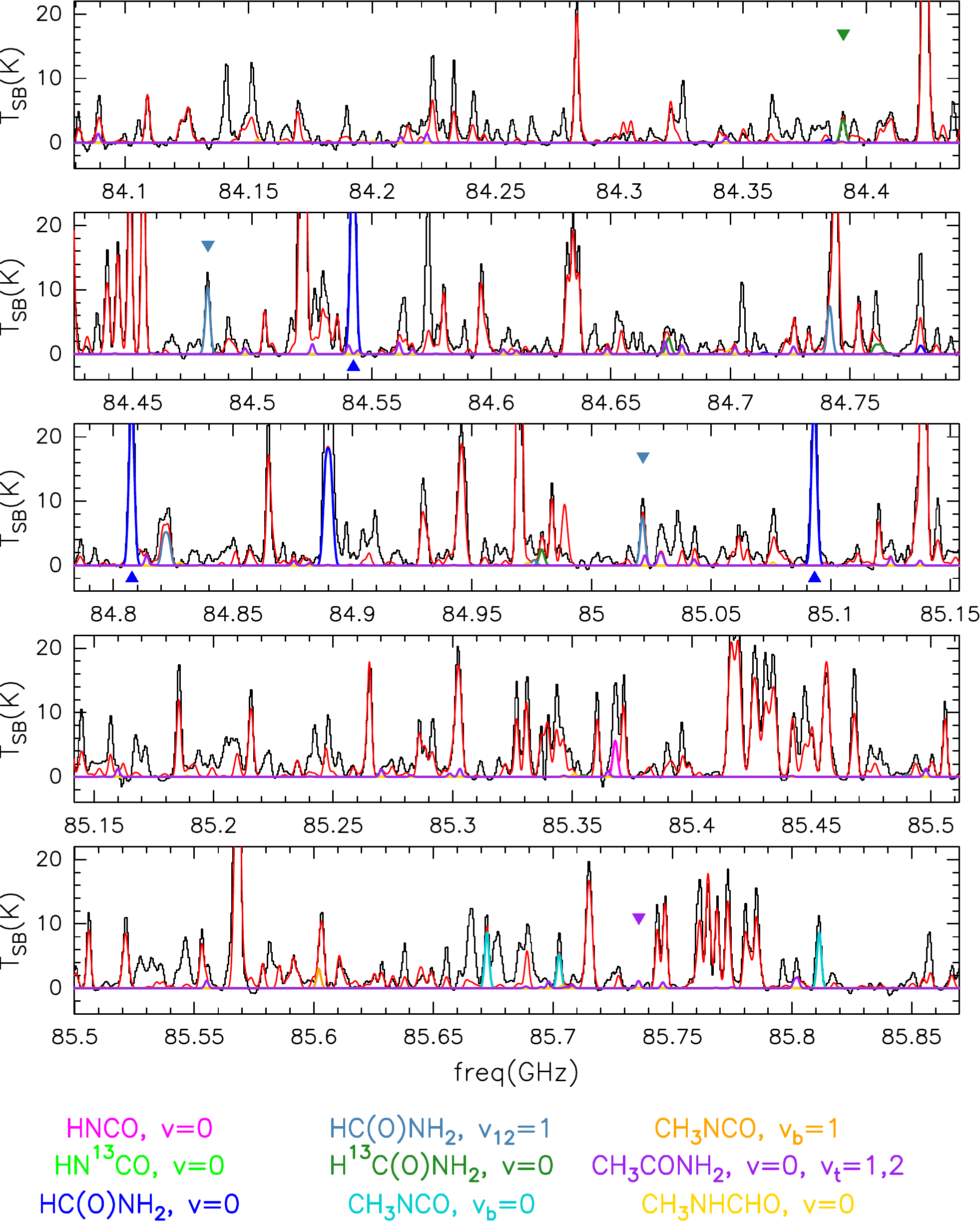}
\caption{Total spectrum of the GUAPOS project in black. The red curve shows the simulated spectrum taking into account all the species identified so far in the region. The best LTE fit of the molecules studied in this work is shown in different colors. The color corresponding to each molecule is shown in the bottom panel. The coloured triangles indicate the transitions used to constrain the fitting procedure. Closer views of those transitions are given in Figs.~\ref{fig-res-ia}--\ref{fig-res-nm}.}
\label{fig-total-spectra-1}
\end{figure*}
 
\begin{figure*}
\centering
\addtocounter{figure}{-1}
\includegraphics[width=40pc]{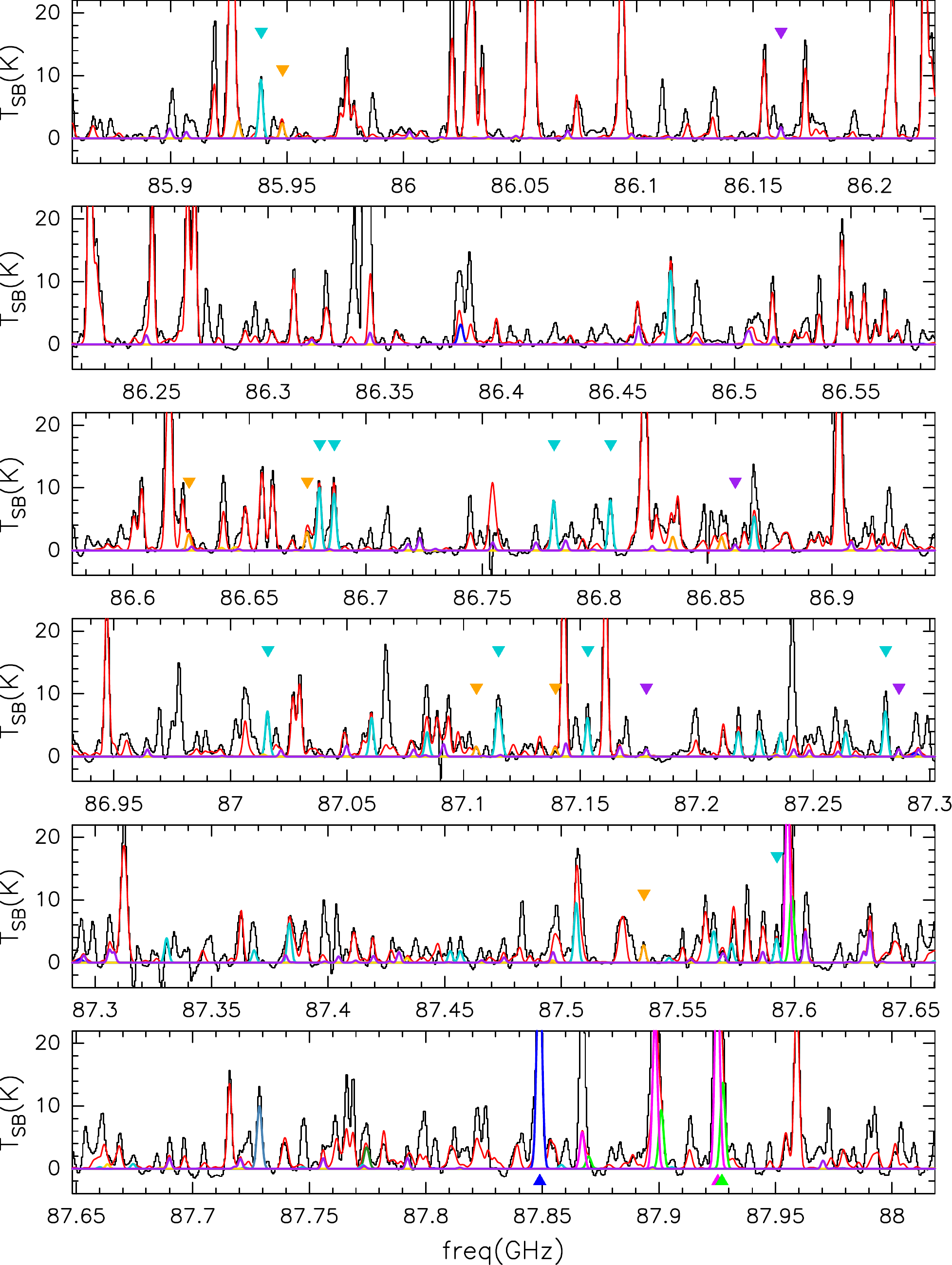}
\caption{Continued.}
\label{fig-total-spectra-2}
\end{figure*}

\begin{figure*}
\centering
\addtocounter{figure}{-1}
\includegraphics[width=40pc]{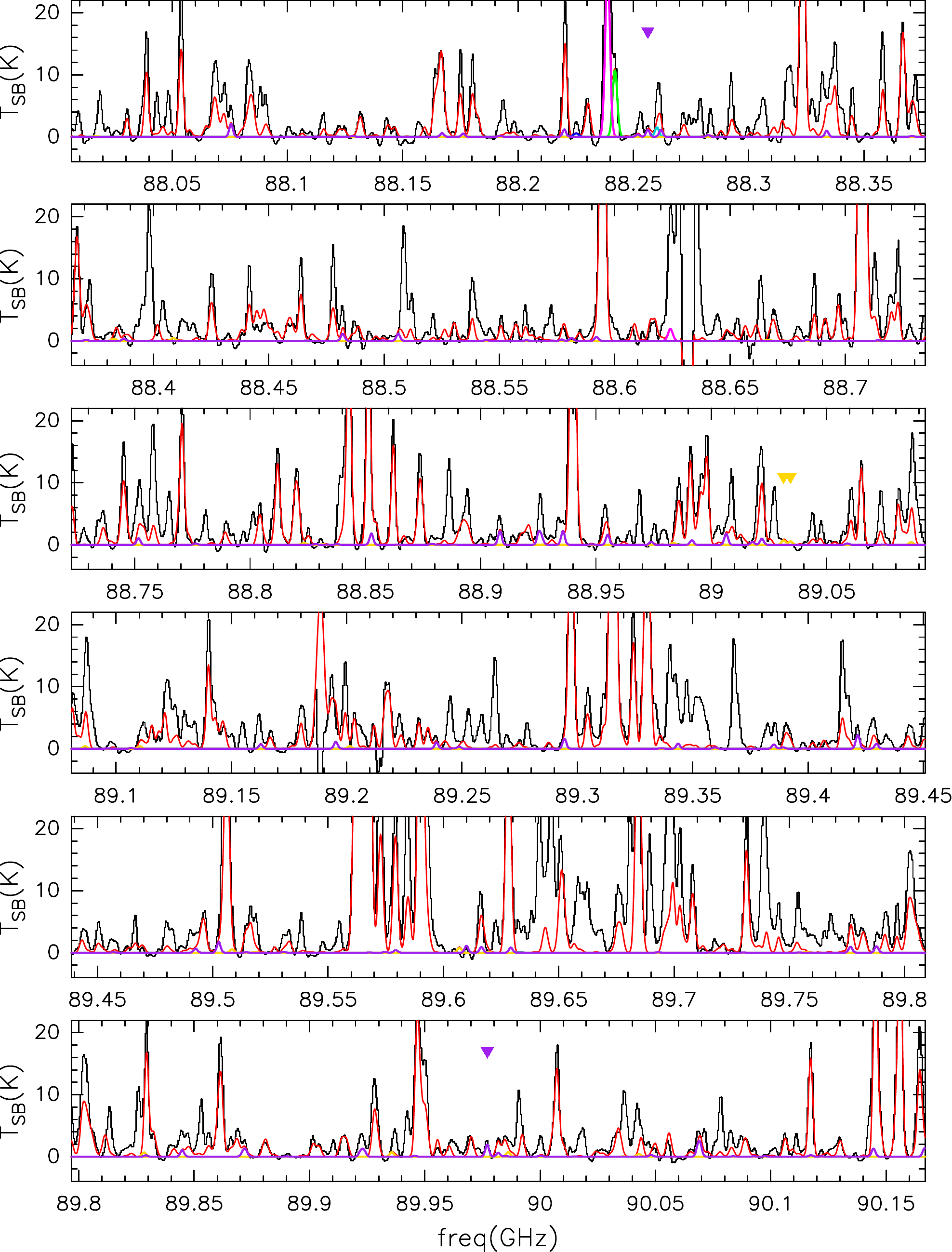}
\caption{Continued.}
\label{fig-total-spectra-3}
\end{figure*}

\begin{figure*}
\centering
\addtocounter{figure}{-1}
\includegraphics[width=40pc]{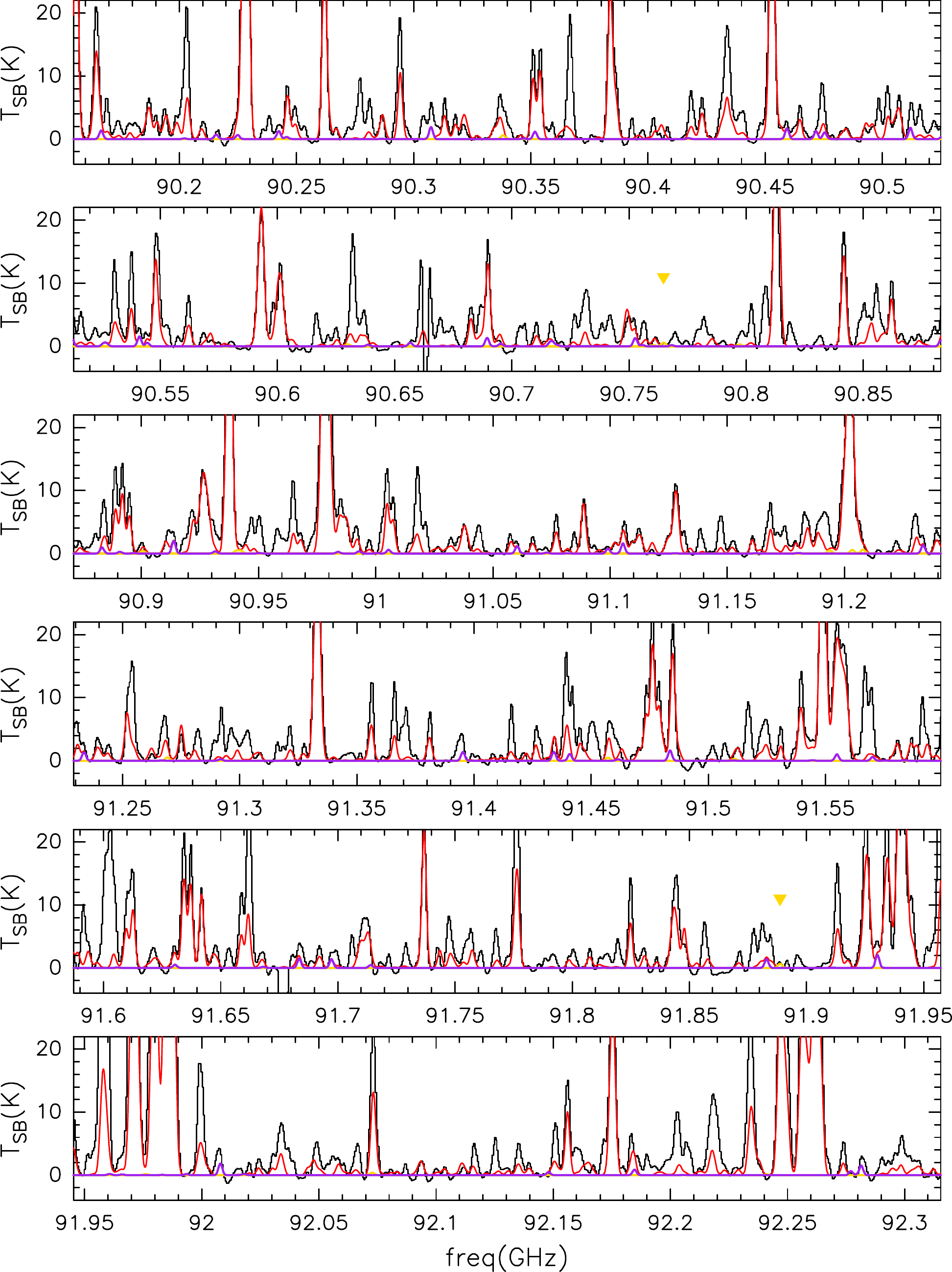}
\caption{Continued.}
\label{fig-total-spectra-4}
\end{figure*} 

\begin{figure*}
\centering
\addtocounter{figure}{-1}
\includegraphics[width=40pc]{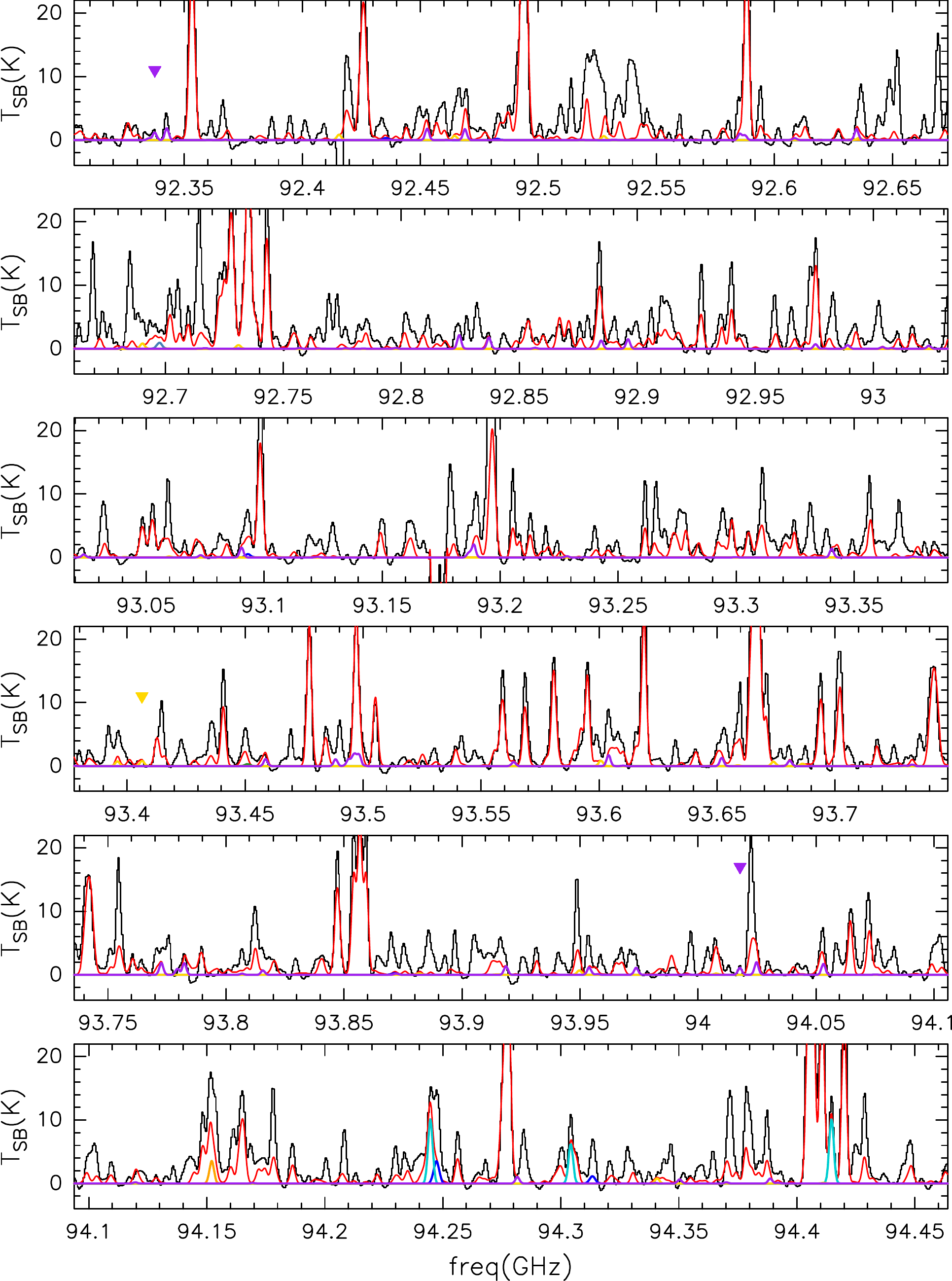}
\caption{Continued.}
\label{fig-total-spectra-5}
\end{figure*} 

\begin{figure*}
\centering
\addtocounter{figure}{-1}
\includegraphics[width=40pc]{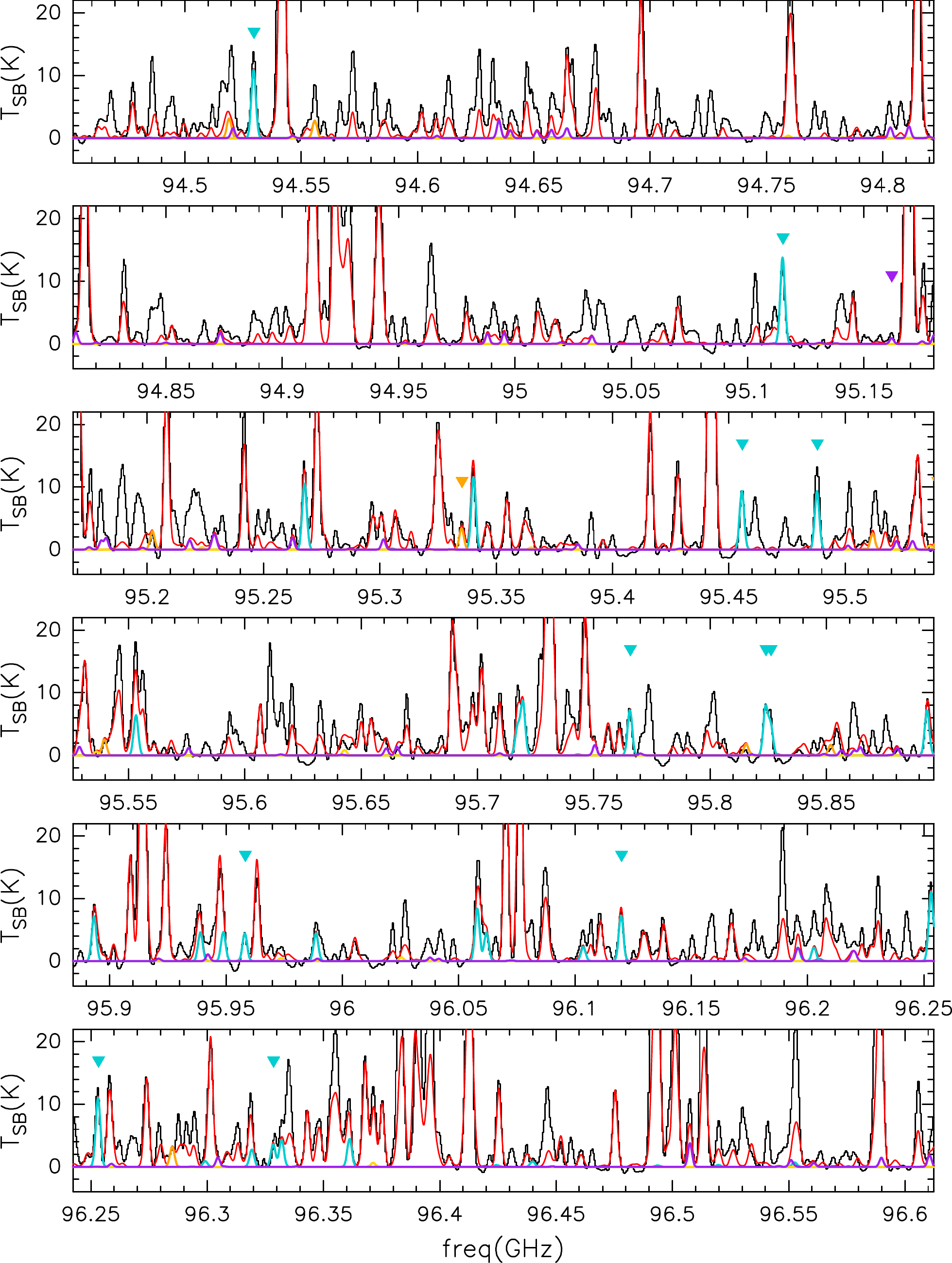}
\caption{Continued.}
\label{fig-total-spectra-6}
\end{figure*}

\begin{figure*}
\centering
\addtocounter{figure}{-1}
\includegraphics[width=40pc]{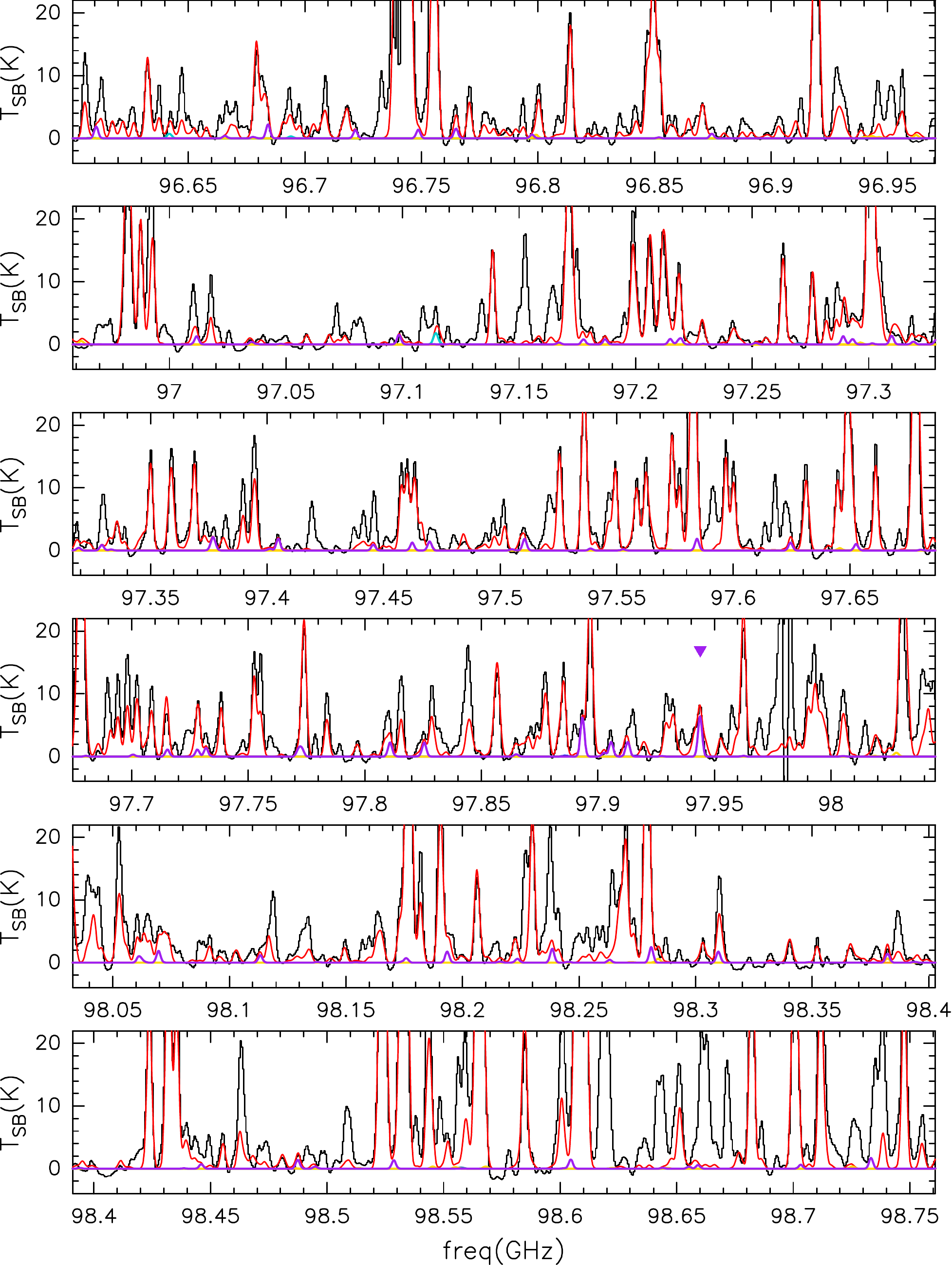}
\caption{Continued.}
\label{fig-total-spectra-7}
\end{figure*}

\begin{figure*}
\centering
\addtocounter{figure}{-1}
\includegraphics[width=40pc]{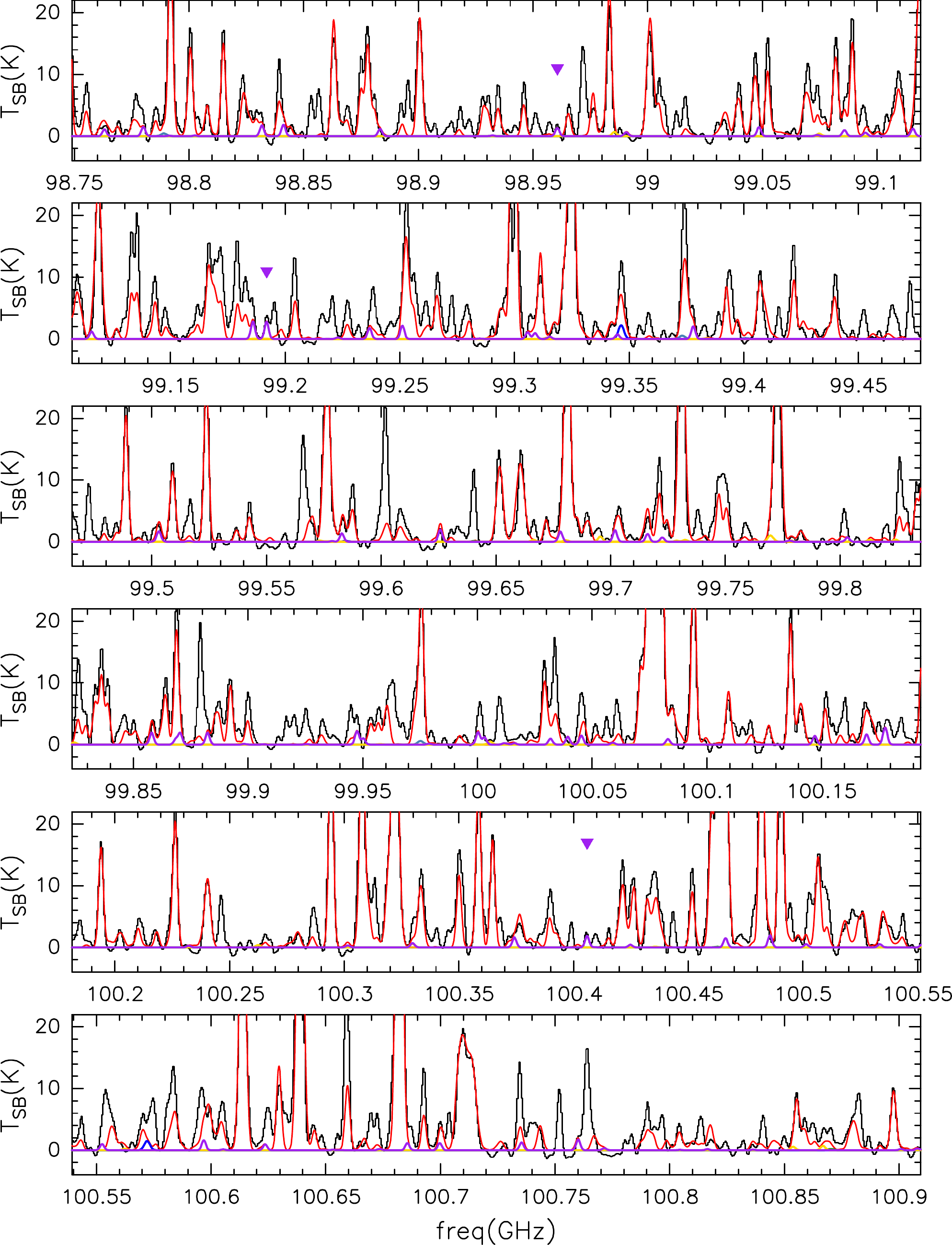}
\caption{Continued.}
\label{fig-total-spectra-8}
\end{figure*}

\begin{figure*}
\centering
\addtocounter{figure}{-1}
\includegraphics[width=40pc]{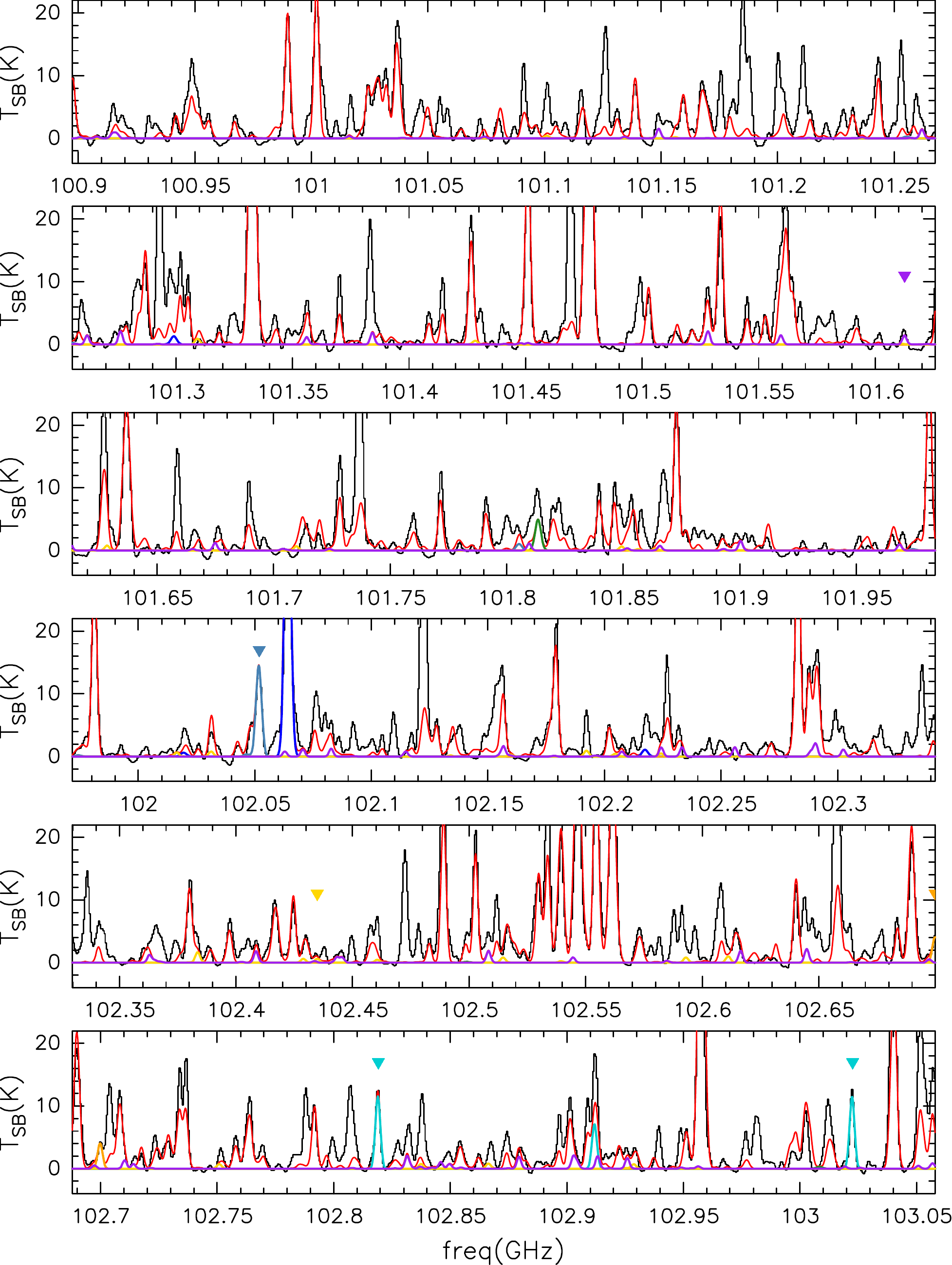}
\caption{Continued.}
\label{fig-total-spectra-9}
\end{figure*} 

\begin{figure*}
\centering
\addtocounter{figure}{-1}
\includegraphics[width=40pc]{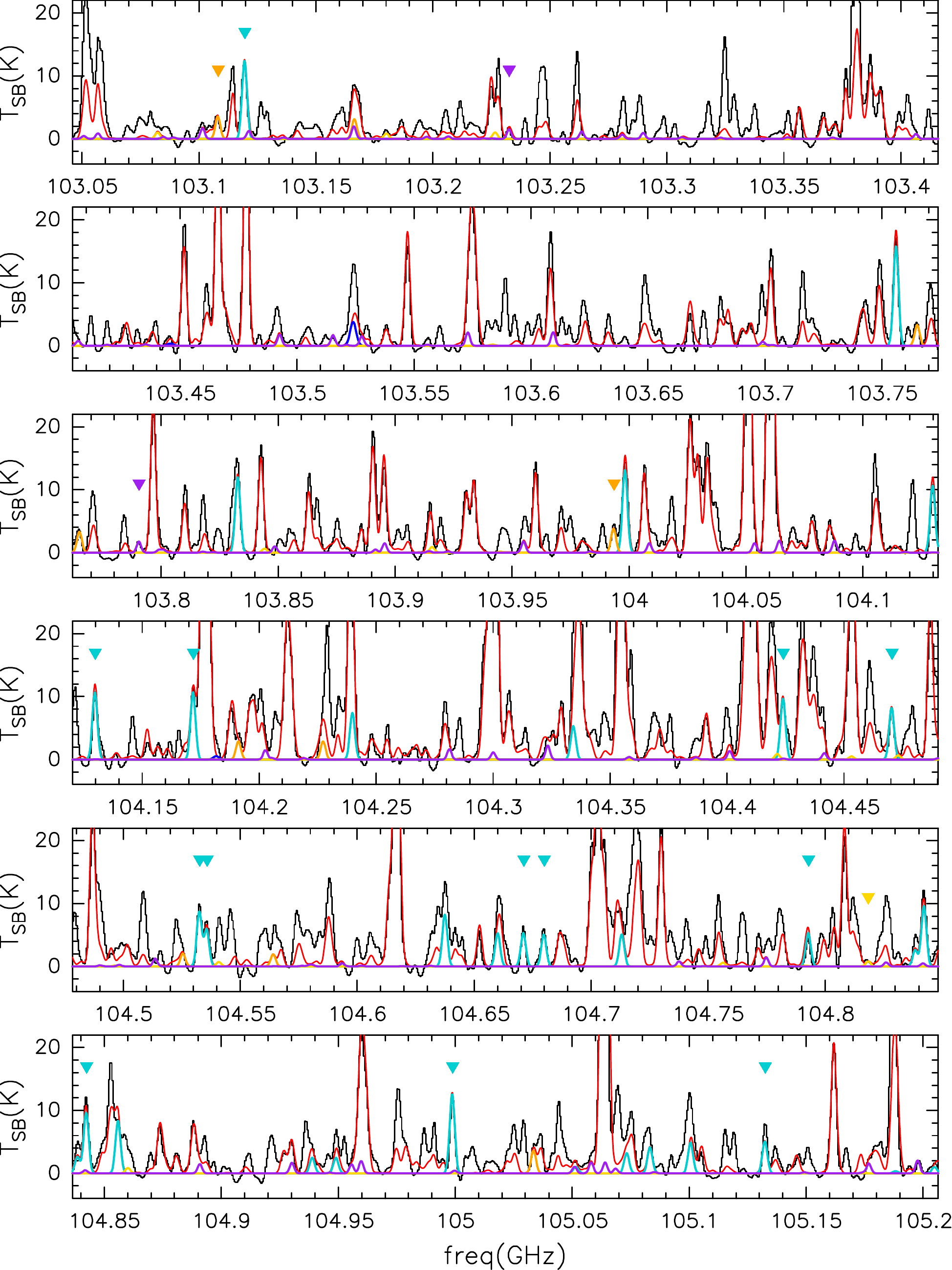}
\caption{Continued.}
\label{fig-total-spectra-10}
\end{figure*} 

\begin{figure*}
\centering
\addtocounter{figure}{-1}
\includegraphics[width=40pc]{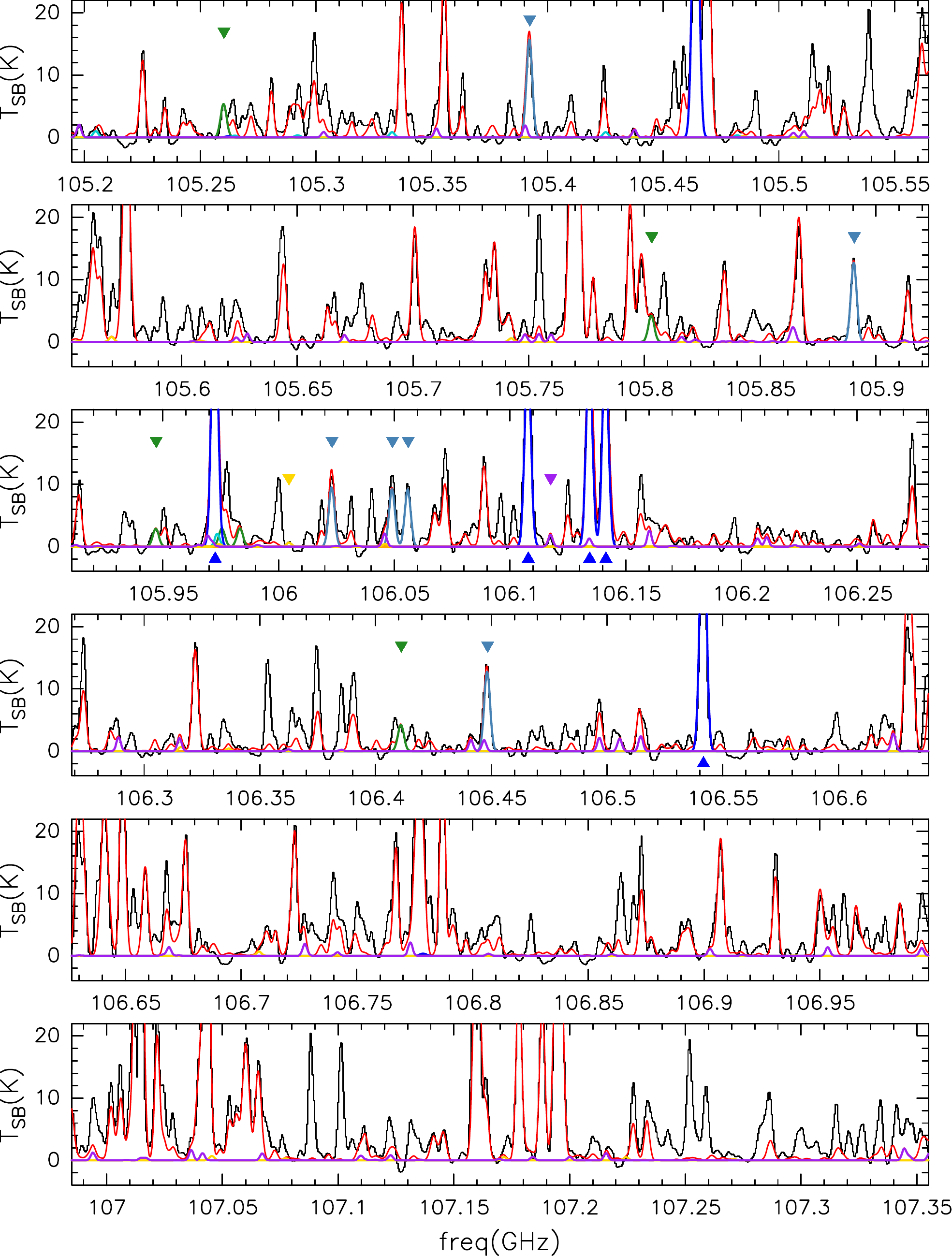}
\caption{Continued.}
\label{fig-total-spectra-11}
\end{figure*} 
 
\begin{figure*}
\centering
\addtocounter{figure}{-1}
\includegraphics[width=40pc]{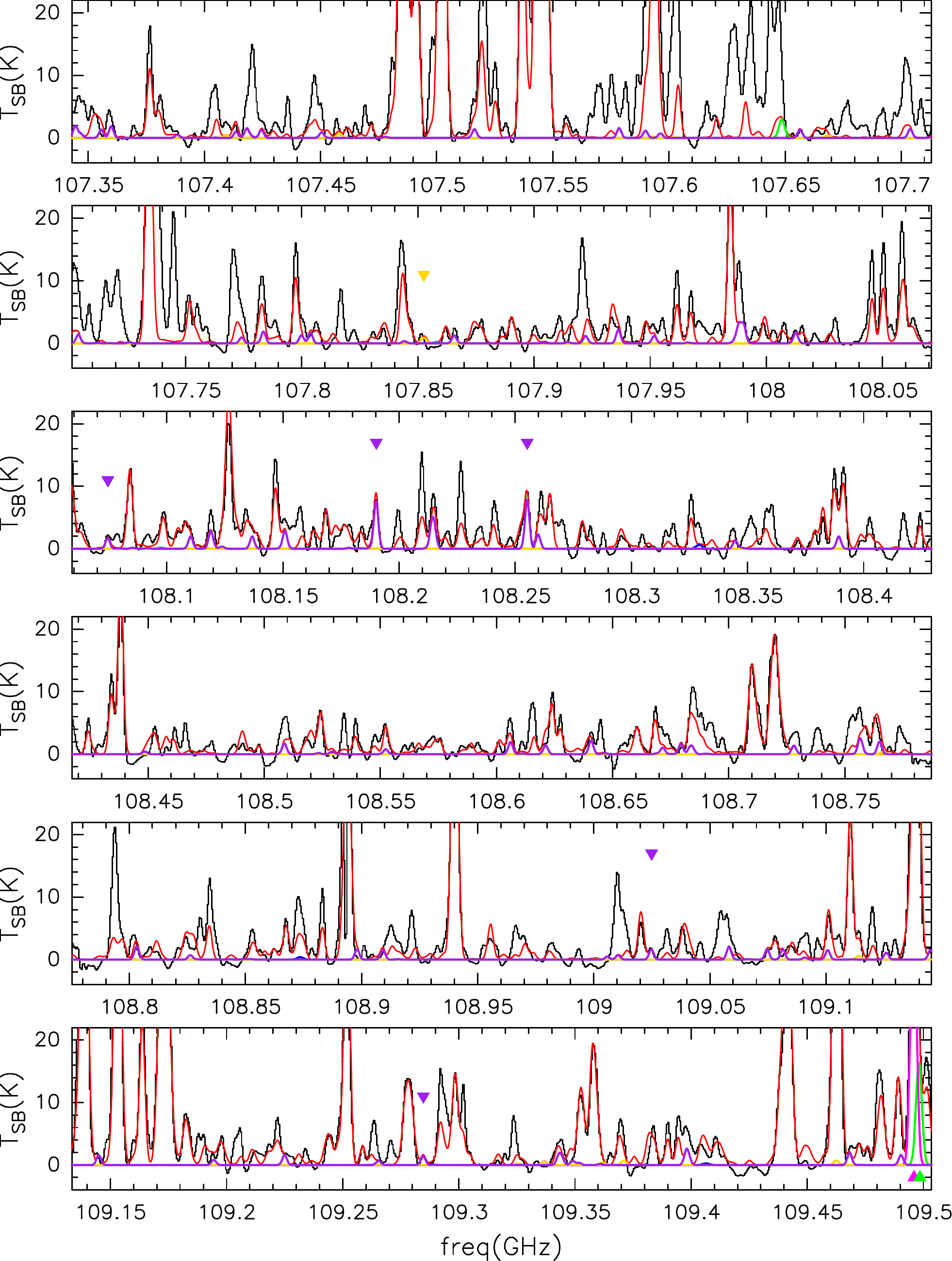}
\caption{Continued.}
\label{fig-total-spectra-12}
\end{figure*} 

\begin{figure*}
\centering
\addtocounter{figure}{-1}
\includegraphics[width=40pc]{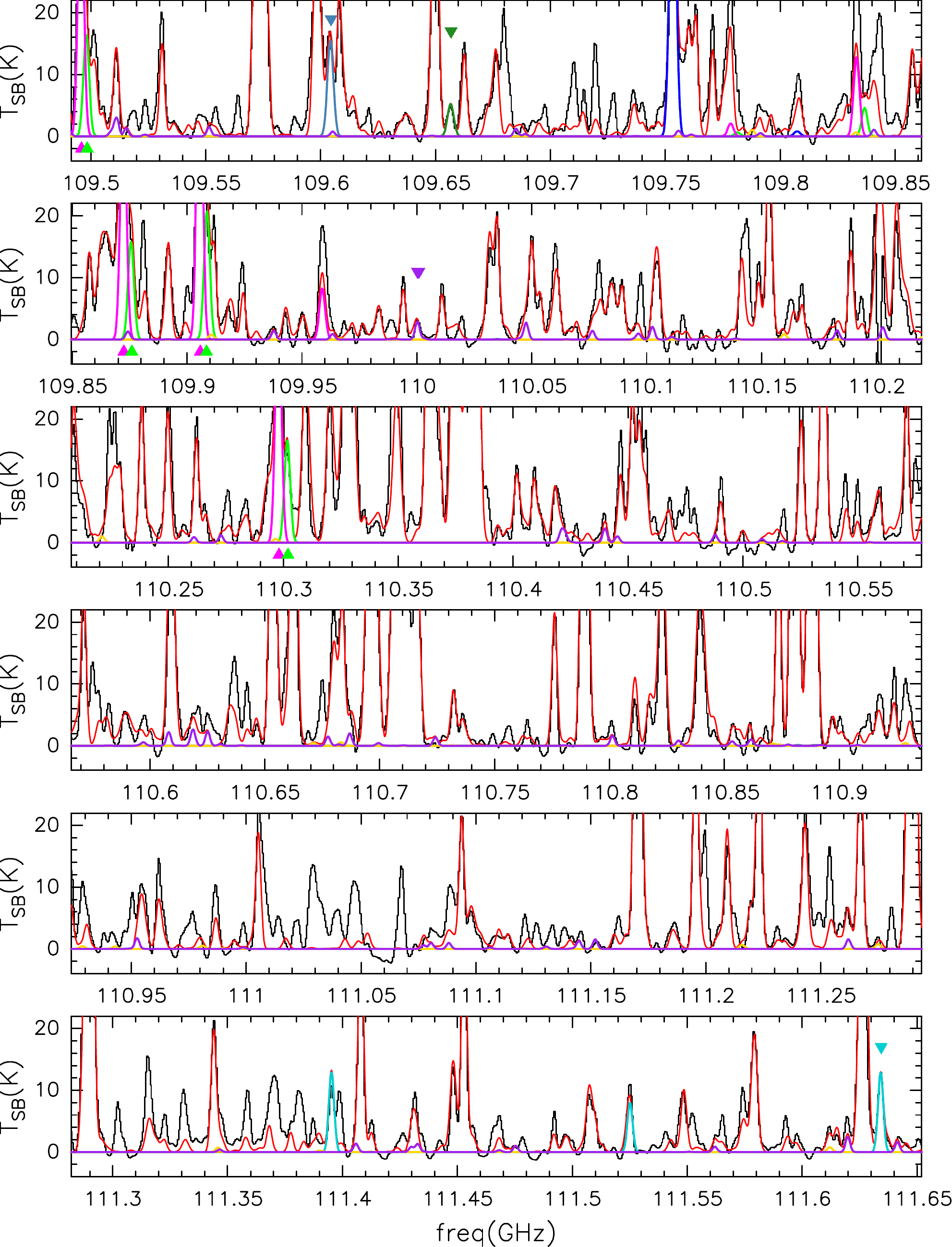}
\caption{Continued.}
\label{fig-total-spectra-13}
\end{figure*} 

\begin{figure*}
\centering
\addtocounter{figure}{-1}
\includegraphics[width=40pc]{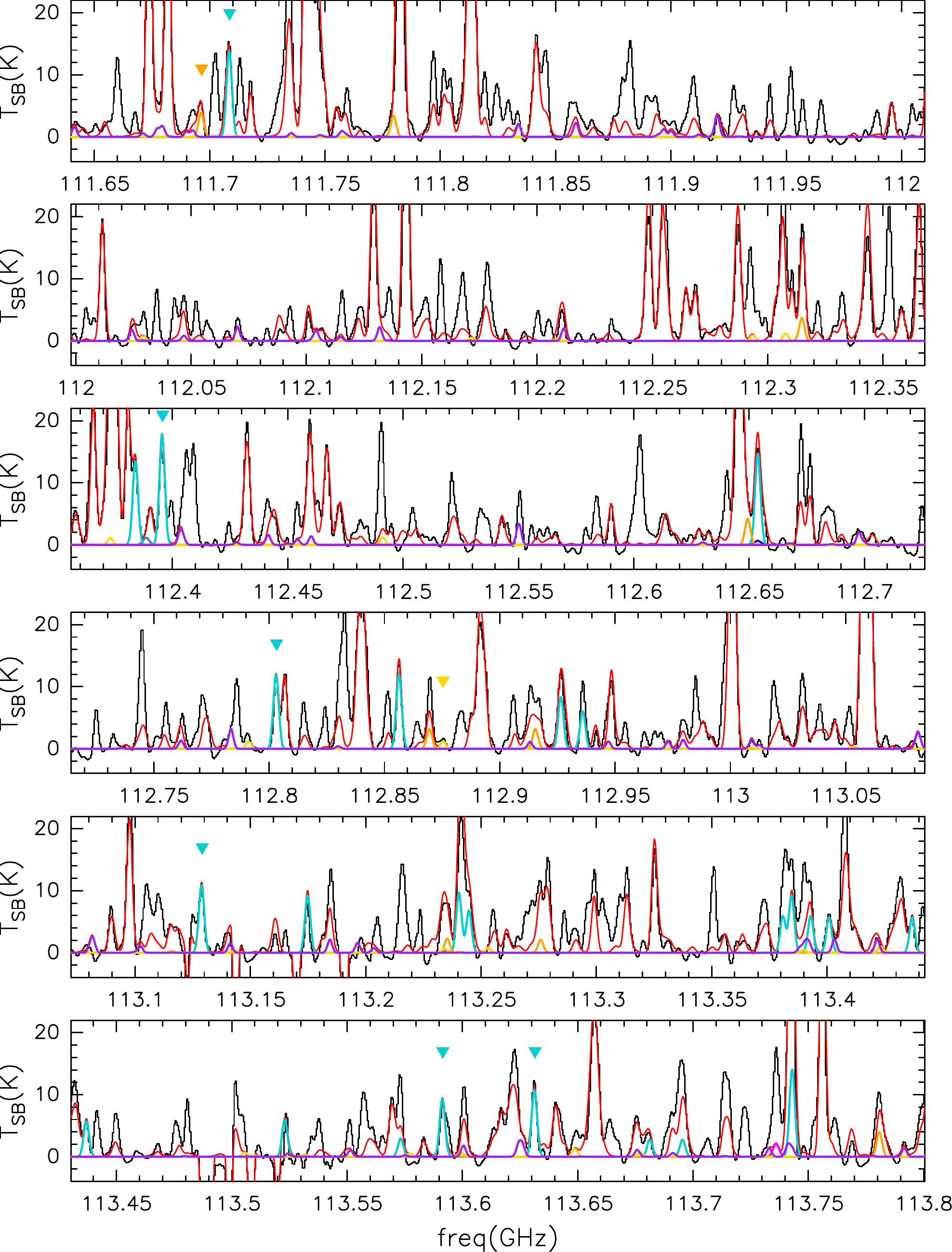}
\caption{Continued.}
\label{fig-total-spectra-14}
\end{figure*}  
 
 \begin{figure*}
\centering
\addtocounter{figure}{-1}
\includegraphics[width=40pc]{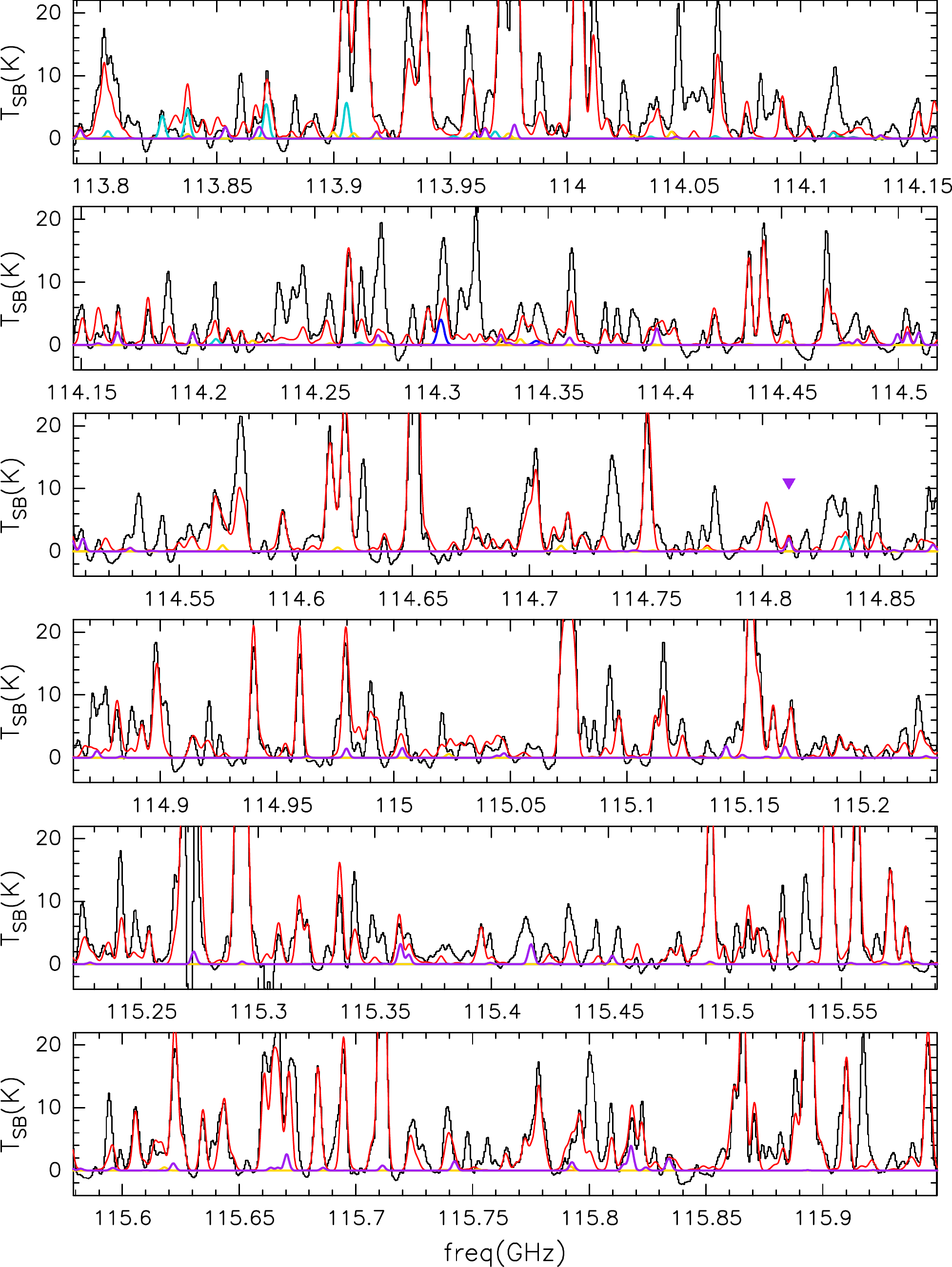}
\caption{Continued.}
\label{fig-total-spectra-15}
\end{figure*}

 \end{appendix}

\end{document}